

\input harvmac
\def\rhob{{\rho\kern-0.465em \rho}}

\def\eps{\epsilon}

\def\ontopss#1#2#3#4{\raise#4ex \hbox{#1}\mkern-#3mu {#2}}

\setbox\strutbox=\hbox{\vrule height12pt depth5pt width0pt}
\def\tablerule{\noalign{\hrule}}
\def\tr{\tablerule}
\def\strut{\relax\ifmmode\copy\strutbox\else\unhcopy\strutbox\fi}

\nref\ritdr{C. Itzykson and J-M. Drouffe, {\it Statistical Field
Theory}, Cambridge University Press, (1991).}
\nref\rsak{B. Sakita, {\it Quantum Theory of Many-Variable Systems and
Fields},
World Scientific (1985).}
\nref\rpol{A.M. Polyakov, {\it Gauge Fields and Strings},
 Harwood Academic (Chur 1987).}
\nref\rpar{G. Parisi, {\it Statistical Field Theory}, Addison-Wesley
(1988).}
\nref\rram{P. Ramond {\it Field Theory: A Modern Primer}, 2 ed.
Addison-Wesley (1989).}
\nref\rlb{M. Le Bellac, {\it Quantum and Statistical Field Theory},
Oxford Univ. Press, (1991).}
\nref\rzj{J. Zinn-Justin, {\it Quantum Field Theory and Critical
Phenomena}, (Oxford Univ. Press 1991).}
\nref\ralice{Lewis Carrol,{\it Through the Looking Glass and what
Alice Found There}, McMillan (1871) chapter 6}
\nref\rramtwo{P. Ramond, {\it op. cit.},p.58}
\nref\rly{C.N. Yang and T.D. Lee, Statistical theory of equations of
state and phase transitions I. Theory of condensation, Phys. Rev. 87
(1952) 404; T.D. Lee and
C.N. Yang, Statistical theory of equations of state and phase
transitions II. Lattice gas and Ising models,
Phys. Rev. 87 (1952) 410 .}
\nref\rons{L. Onsager, Crystal statistics I. A two-dimensional model
with an order disorder transition, Phys. Rev. 65 (1944) 117.}
\nref\rwidom{B. Widom, Equation of state in the neighborhood of the
critical point, J. Chem. Phys. 43 (1965) 3998.}
\nref\rfisher{M.E. Fisher, Theory of equilibrium critical phenomena,
Reports on Progress in Physics, 30 (part 2) (1967) 615.}
\nref\rkad{L.P. Kadanoff, W. Goetze, D. Hamblen, R. Hecht, E.A.S.
Lewis, V.V. Palciuskas, M. Rayl, J. Swift, D. Aspnes, J. Kane, Static
phenomena near critical points: Theory and experiment, Rev. Mod. Phys.
39 (1967) 395.}
\nref\rblo{N. Bloembergen, On the interaction of nuclear spins in a
crystalline lattice, Physica 15 (1949) 386.}
\nref\rdg{P.G. de Gennes, Inelastic magnetic scattering of neutrons at
high temperatures, J. Ohys. Chem. Solids 4 (1958) 233.}
\nref\rsvw{M.Steiner, J. Villain and C. G. Windsor, Theoretical and
experimental studies on one-dimensional magnetic systems, Adv. in
Phys. 25 (1976) 87.}
\nref\rbl{M. B{\"o}hm and H. Leschke, Dynamical aspects of spin chains at
infinite temperature for different spin quantum numbers, Physica A 199
(1993) 116.}
\nref\rmori{H.M. Mori, Transport, collective motion, and Brownian
motion, Prog. Theor. Phys. 33 (1965) 423; and A continued fraction
representation of time correlations functions, Prog. Theor. Phys. 34,
(1966) 399.}
\nref\rfh{R.P. Feynman and A.R. Hibbs, {\it Quantum Mechanics and Path
Integrals}, McGraw-Hill, New York (1965).}
\nref\rgj{J. Glimm and A. Jaffe, {\it Quantum Physics: A Functional
Integral Point of View} Springer-Verlag, New York (1981).}
\nref\rfasl{L.D. Faddeev and A.A. Slavnov, {\it Gauge Fields: Introduction
to Quantum Field Theory}, Benjamin (1980).}
\nref\ritzu{C. Itzykson and J-B. Zuber, {\it Quantum Field Theory},
McGraw Hill, New York (1980).}
\nref\rcol{J.C. Collins, {\it Renormalization}, Cambridge Univ. Press, (1984)}
\nref\rster{G. Sterman, {\it An Introduction to Quantum Field Theory},
Cambridge Univ. Press, Cambridge, (1993).}
\nref\rbd{J.D. Bjorken and S.D. Drell, {\it Relativistic Quantum
Fields}, McGraw-Hill (1965).}
\nref\rds{B. de Wit and J. Smith, {\it Field Theory in Particle
Physics}, North-Holland, Amsterdam (1986).}
\nref\rdirac{P.A.M. Dirac, Physikalische Zeitschrift der
Sowjetunion 3 (1933) 64}
\nref\rfeynman{R.P. Feynman, Space-time approach to non-relativistic
quantum mechanics, Rev. Mod. Phys. 20 (1948) 367.}
\nref\rscha{J. Schwinger, The Theory of Quantized Fields I, Phys. Rev.
82 (1951) 914; II, Phys. Rev. 91 (1953) 713}
\nref\rdysona{F.J. Dyson, The S matrix in quantum electrodynamics,
Phys. Rev. 75 (1949) 1736.}
\nref\rwick{G.C. Wick, Properties of Bethe-Salpeter wave functions,
Phys. Rev. 96 (1954) 1124.}
\nref\rschb{J. Schwinger, On the Euclidean structure of relativistic
field theory, Proc. Nat. Acad. Sci. 44 (1958) 956,
and Euclidean Quantum
Electrodynamics, Phys. Rev. 115 (1959) 721.}
\nref\rwil{K.G. Wilson, Confinement of Quarks, Phys. Rev. D10 (1974) 2445.}
\nref\rgel{M. Gell-Mann, A schematic model of baryons and mesons,
Phys. Letts. 8 (1963) 214.}
\nref\rwk{K.G. Wilson and J. Kogut, The renormalization group and the
$\epsilon$ expansion. Phys. Rep. 12C (1974) 75.}
\nref\rpatpok{A.Z. Patashinskii and V.L. Pokrovskii, Second order
transition in a Bose fluid, Sov. Phys. JETP 19 (1964) 677.}
\nref\rmiga{A.A. Migdal, A diagrammatic technique near the curie point and the
second order phase transition in a Bose liquid, Sov. Phys. JETP 28 (1969)
1036.}
\nref\rgm{V.N. Gribov and A.A. Migdal, Strong coupling in the
Pomeranchuk pole problem, Sov. Phys. JETP 28  (1969) 784.}
\nref\rpola{A.M. Polyakov, Microscopic description of critical
phenomena, Sov. Phys. JETP 28 (1969) 533}
\nref\rmigb{A.A. Migdal, Correlation functions in the theory of phase
transitions: violations of the scaling laws, Sov. Phys. JETP 32 (1971)
552.}
\nref\rpolb{A.M. Polyakov, Properties of long and short
range correlations in the critical region, Sov. Phys. JETP 30 (1971)
151}
\nref\rpolc{A.M. Polyakov, A similarity hypothesis in the
strong interactions I Multiple
hadron production in $e^+e^-$ annihilation, Sov. Phys. JETP 32 (1971)
296.}
\nref\rkauf{B. Kaufman, Crystal statistics II. Partition function
evaluated by spinor analysis, Phys. Rev. 76 (1949) 1232.}
\nref\rwf{K.G. Wilson and M.E. Fisher, Critical
exponents in 3.99 dimensions, Phys. Rev. Letts. 28 (1972) 240.}
\nref\ros{K. Osterwalder and R. Schrader, Axioms for Euclidean Green's
functions, Comm. Math Phys. 31 (1973) 83 and 42 (1975) 281.}
\nref\rwi{A. Wightman, Quantum field theory in terms of vacuum
expectation values, Phys. Rev. 101 (1956) 860.}
\nref\rhk{R. Haag and D. Kastler, An algebraic approach to quantum
field theory,  J. Math. Phys. 5 (1964) 848.}
\nref\rsw{R.F. Streater and A.S. Wightman, {\it PCT, Spin and
Statistics and All That}, Benjamin, New York, (1964).}
\nref\rfr{M.E. Fisher and D. Ruelle, The stability of many-particle
systems, J. Math. Phys. 7(1966) 260.}
\nref\rdy{F.J. Dyson, Ground-state energy of a finite system of
charged particles, J. Math Phys. 8 (1967) 1538.}
\nref\rdl{F.J. Dyson and A. Lenard, Stability of matter I and II, J.
Math Phys. 8 (1967) 423 and 9 (1968) 698.}
\nref\rruelle{D. Ruelle, {\it Statistical Mechanics}, Benjamin,
Amsterdam (1969).}
\nref\rlt{E.H. Lieb and W.E. Thirring, Bound for the kinetic energy of
fermions which proves the stability of matter, Phys. Rev. Letts. 35
(1975) 687.}
\nref\rlieb{E.H. Lieb, The stability of matter, Rev. Mod.
Phys. 48 (1976) 553.}
\nref\rhawk{S.W. Hawking, Quantum gravity and path integrals,
 Phys. Rev. D18 (1978) 1747.}
\nref\rhi{S.W. Hawking, The path integral approach to quantum gravity,
in {\it General Relativity; an Einstein Centenary Survey} ed S.W. Hawking
and W. Isreal, Cambridge Univ. Press. (1979) chapter 15.}
\nref\rgib{G.W. Gibbons, S. W. Hawking and M.J. Perry, Nucl. Phys.
B138 91978) 141.}
\nref\rmm{P.O. Mazur and E. Mottola, The path integral measure,
conformal factor problem and stability of the ground state of quantum
gravity,
Nucl. Phys. B341 (1990) 187.}
\nref\rwald{R.M. Wald, {\it General Relativity}, University of Chicago
Press, Chicago, (1984), chap.14.}
\nref\rmt{C.A. Tracy and B.M. McCoy, An examination of phenomenological
scaling functions used for critical scattering, Phys. Rev. B12 (1975) 368.}

\Title{\vbox{\baselineskip12pt\hbox{ITP-SB-94-07}
\hbox{HEPTH 9403084}}}
{\vbox{\centerline{The Connection Between}
\vskip9pt
\centerline{ Statistical Mechanics and Quantum Field Theory}}}
\vskip9pt

\centerline{ Barry~M.~McCoy~\foot{mccoy@max.physics.sunysb.edu}}

\bigskip\centerline{\it Institute for Theoretical Physics}
\centerline{\it State University of New York}
\centerline{\it  Stony Brook,  NY 11794-3840}

\vskip 13mm

\centerline{\bf Abstract}
\vskip 4mm

A four part series of lectures on the connection of statistical
mechanics and quantum field theory. The general principles relating
statistical mechanics and the
path integral formulation of quantum field theory are presented in the
first lecture. These
principles are then illustrated in lecture 2 by a  presentation of the
theory of the Ising model for $H=0$, where both the homogeneous and
randomly inhomogeneous models are treated and the scaling theory and
the relation with Fredholm determinants and
Painlev{\'e} equations is presented.
In lecture 3 we consider the Ising model with $H\neq 0$, where the
relation with gauge theory is used to discuss the phenomenon of
confinement. We conclude in the last lecture with a discussion of  quantum spin
diffusion in one dimensional chains and a presentation of the chiral
Potts model which illustrates the physical effects that can occur
when the Euclidean and Minkowski regions  are not connected by an
analytic continuation.

\vskip 3mm

\Date{\hfill 3/94}
\vfill\eject

\centerline{\bf Part I. Introduction}

\vskip 13mm

\centerline{\bf Abstract}
\vskip 4mm
The connection between the Euclidean path integral formulation of
quantum field theory and classical statistical mechanics is surveyed
in terms of the theory of critical phenomena and the concept of
renormalization. Quantum statistical mechanics is surveyed with an
emphasis on diffusive phenomena. The particle interpretation of quantum
field theory is discussed in the context of nonperturbative statistical
mechanics.

\vskip 3mm

\newsec{Introduction}
There is probably a theorem somewhere in {\it The Critique of Pure Reason}
which says that it is impossible to use language to
communicate an idea from one
person to the other unless the second person already knows what the
first person is talking about. Consequently it is probably logically
impossible to teach anyone anything.

The practical consequences of this unhappy theorem are all too evident.
If I wish to be clear I must go to great lengths to define and
explain my terms. But by the time I have achieved utmost clarity
I have lost my audience and there is no one left to hear what new things
I have to say. If on the other hand I wish to keep my audience, I must
assume that they will use words the same way that I do and proceed
directly to my results. But the unhappy fact is that since no two people use
words to mean the same thing my conclusion will be lost on the audience
which spends its time in trying to fit my conclusions into their
definitions of the words used.

This basic difficulty in the theory of knowledge is exceptionally
evident in the presentation of the relationship between quantum field theory
and statistical mechanics. This is because the two subjects have
developed from very different roots and have very different languages
and assumptions.

Historically quantum field theory (QFT) is intimately
linked with the classical field theory
of electromagnetism and with particle physics. Experimentally QFT has
been intimately connected to high energy physics
experiments at accelerators starting with cyclotrons in the 30's and
40's and proceeding up to SLAC, CERN, Fermi Lab, KEK, DESY and CEBAF.
Theoretically QFT has gone from QED to QCD, the standard model and
string theory.

The origins of statistical mechanics (SM) are totally and completely
different. Historically SM is linked with the theory of heat,
irreversibility, and the kinetic theory of gases. Experimentally SM
has been intimately connected with calorimeters, specific heats,
magnetic order parameters, phase transitions, and diffusion.
Theoretically SM has gone from ideal gases,  series (virial)
expansions,  and Landau theory to solvable models, Yang--Baxter
equations and holonomic sets of equations for correlation functions.

It would thus seem from the above that QFT and SM have nothing
to do with each other. Indeed in the practical functional sense all
physics departments do indeed act in this fashion. When categories for
recruitment are set up a sharp distinction between a field
theoretician and a statistical mechanician is always made and it is a
career determining decision on the part of a young person whether QFT or
SM is chosen.

Against this background of almost total separation
I have been given the job of  explaining in these lectures why the two subjects
are related. It is not  exactly clear what the organizers  had in
mind when assigning me this task. Perhaps the first answer that comes
to mind is that  the path integral
formulation of Euclidean QFT  is concerned with the computation of path
integral averages that formally look like
\eqn\qf{<O_j>={\int[d\phi ]O_je^{-S_E(\phi )/h}\over Z_E},~~Z_E=
\int[d\phi ]e^{-S_E(\phi)/h}}
where $\int[d\phi ]$ stands for a functional integral over a set of
Euclidean quantum fields in $d$ spatial dimensions
where the distance $s$ is $s^2=x_1^2 +\dots + x_d^2$, $O_j$ is constructed from
those fields, and $S_E(\phi )$ is a function of the fields known as the
action which is expressed in terms of a local Lagrangian
\eqn\elag{S_E(\phi )=\int dx_1\dots dx_d L(\phi ({\vec x})).}
In classical statistical mechanics thermal
expectation values are computed as
\eqn\sm{<O_j>={\sum_{\rm all ~states} O_j e^{-E(\phi )/kT}\over Z},~~Z=
\sum_{\rm all ~states}e^{-E(\phi )/kT}}
where $\sum_{\rm all ~states}$ is a sum over all states of the system
described by the random variables $\phi$, $O_j$ is constructed from
$\phi$, $ E(\phi )$ is a classical  energy functional of $\phi$ , $T$ is
the temperature and $k$ is Boltzmann's constant.
These equations are ``formally the same'' and thus it may be expected that
there is a relation between the subjects.

Of course \qf~ and \sm~ are not the only starting points possible. I
could, for example, choose to start with a Minkowski QFT in $d-1$
spatial and one time dimension defined by
\eqn\mqft{<O_j>={\int [d\phi ] O_je^{iS_M(\phi)/h}\over
Z_M},~Z_M=\int [d\phi ]e^{iS_M(\phi )/h}}
where the metric is $s^2=x_1^2 + \dots +x_{d-1}^2 -t^2$
\eqn\mlag{S_M(\phi )=\int dt \int dx_1 \dots dx_{d-1} L(\phi({\vec x},t))}
and the quantum thermal average
\eqn\qsm{<O_j>={{\rm Tr} O_j e^{-H/kT}\over Z},~Z={\rm Tr}e^{-H/kT}}
where $H$ is a quantum Hamiltonian. Thus from the outset I will
recognize that there is some flexibility in meaning of the words in
the title of these lectures.

But if it is the relation of \qf~ to \sm~ which the organizers
had in mind then I am in
somewhat of a peculiar position because one of the lecturers at this
school is Claude Itzykson whose recent book {\it Statistical Field
Theory}~\ritdr~ is an outstanding discussion of this very relation in a much
more detailed and extensive format than I can possibly match in these
four lectures. Indeed, in the last 10 years many books have
appeared~\ritdr--\rzj~
which exploit the relation between \qf~ and \sm . Their titles range
from gauge theory to string theory to critical phenomena and each has
a distinct point of view. The sheer number of these books demonstrates
the unity of the subjects, a unity which is emphasized by the fact
that almost all these books, regardless of title or orientation of the
author, study the two dimensional Ising model as one of the first
examples of the subject.

These four lectures cannot hope to be a substitute for the one
semester or one year course which can be taught from the existing
literature. Therefore I must be selective in my choice of topics.

In this first lecture I will discuss the general principles which lead
to the unity of \qf~ and \sm . I will approach this from the point
of view of the skeptic who needs to be convinced. This will hopefully
overcome some of the language problems that sometimes obscure
understanding.

In the second and third lectures I will illustrate these general principles
by presenting the field theory construction of simplest of
all systems: the Ising
model in two dimensions. This subject is extremely rich and will illustrate
in terms of exact computations practically all of the general phenomena.
In lecture two I will discuss the Ising model in zero magnetic field and
summarize the field theory insights obtained these past 50 years.

In lecture three I will consider the Ising model in a magnetic field.
In this case I will
present an exact mapping to the simplest model of a two dimensional gauge
field
coupled to a self interacting scalar field. I will then summarize
the analytic computations done on this model in the last 20 years. We will
see that even for this most simple of gauge theories that there are many
interesting unresolved problems.

Finally in lecture four  the considerations are extended to quantum
statistical mechanics where we deal directly with Minkowski space. We
discuss quantum spin diffusion and conclude with a
discussion of the chiral Potts model as
an example of a system where level crossing causes the physics in the
Euclidean and Minkowski regions to be very different.

\bigskip
\newsec{Generalities}

In order to be clear  it is perhaps desirable to make explicit some tacit
assumptions that have been made in writing down \qf~ and \sm.

Perhaps the most basic relation between SM and QFT is not \qf~ and
\sm~ but rather the fact that both deal with systems in an infinite volume
and hence with an infinite number
of degrees of freedom. A major consequence of this is that the formal
definitions \qf~ and \sm~ by themselves have no meaning at all
because they are, at best, evaluated as ${\infty \over \infty}$. There is
always a further definition needed to make sense of these expressions.
In the case of SM that definition is embodied in the thermodynamic
limit which first evaluates \sm~ in a finite volume and then takes
the limit as the size of the box goes to infinity. In the case of QFT
the expression
\qf~ needs the additional definition which is provided by a ``renormalization
scheme'' that usually involves a short distance cutoff as well as a
finite box.

In SM the (infrared) thermodynamic limit is treated
very explicitly.
In QFT the short distance (ultraviolet) cutoff is discussed
extensively. The difference in focus on infrared cutoffs versus
ultraviolet cutoffs is often one of the major barriers to communication
between the two fields and to me seems to constitute a major reason
that QFT and SM are traditionally considered to be different
subjects.

A second tacit assumption is that \qf~ and \sm~ will in the end
have an interpretation in terms of ``particle'' or possibly
``quasi-particles.'' The ``particle'' interpretation of QFT is
fundamental for its relation to particle or high energy physics and
the quasi-particle interpretation of \sm~ is fundamental for such
things as the Landau theory of the Fermi liquid.  This common particle
interpretation has the opposite effect from the limiting procedures
and makes there seem to be great unity between the subjects. Indeed
from at least the late 50's there has been a great industry in the
``application of QFT methods to SM'' which is ultimately based on this
common interpretation.

There is one further point which is related to both the particle description
and the infinite degrees of freedom which should be made explicit;
namely, the distinction between ``bare'' on the one hand and
``renormalized,'' ``collective,'' or ``dressed'' on the other.
The expressions for the fields $\phi$, actions  and interaction
energies which are
used in \qf~ and \sm~ before the various limits are taken are often
referred to as bare quantities. These are the fundamental inputs to the
theory and define the problem we are interested in solving. It is very
common to  give particle names to these bare fields such as electron, neutrino,
quark, or gluon. On the other hand what we observe in
measurements are the expectations defined by \qf~ and \sm~ after the
limits have been taken. These expectation values are also given names
which are often the same at the ``bare'' fields which appear in the
expressions for $O_j$. To distinguish these names apart the words
``renormalized,'' ``dressed'' or ``collective'' are often used when
referring to the expectation values. It is most important to realize
that because we are dealing with systems with an infinite number of
degrees of freedom that
``bare'' and ``renormalized'' quantities may in fact have nothing in
common except the name. The bridging of the gap between these two
constructs is fundamental to the understanding of the connection
between QFT and SM. To attempt to make a proof for a ``renormalized''
quantity by first making a proof for the ``bare'' quantity and
then adding the word ``renormalized'' to it runs the grave danger of
reducing theoretical physics to process of ``proof by means of making
a bad pun.''

By now the difficulty in communication has become abundantly clear.
According to your taste I have either 1) taken a large number of
well known concepts and tried to make them mysterious by putting
quotes around them or 2)  I have introduced a large
number of totally undefined words and pretended that they have an
unambiguous meaning. In this lecture I will adopt the second point of
view and try to explain what I am talking about. If you adopt the
first point of view I leave you with the warning issued by
Humpty-Dumpty~\ralice.
Words mean what I want them to mean, no more-no less.
If they do a good job they come
in on Saturday evening to collect their wages.

It is only fair to close these general remarks with the candid
acknowledgement that there is not universal agreement about what I
will say. As an example it was stated in print as recently as 1989
that the connection between \qf ~and \sm ~was ``not
understood''~\rramtwo. From the very title of these lectures it is clear
that I do not share this view.

\bigskip
\newsec{Classical Statistical Mechanics}

We begin our considerations with classical statistical mechanics and in
particular with a sketch of the theory of phase transitions and
critical phenomena. These
topics have been treated in various ways by other lecturers at this
summer school so I will be concise. A primary intent is to use this
procedure to make contact in sec.~5  with the quantum field theory of \qf.

The  definition of the classical partition function $Z$ and
correlation functions in \sm~ is very
general. However there are a variety of more special cases that are of
great interest. In particular there is more than sufficient richness
in the physics if we consider the following restrictions:

1) The random variables $\phi$ will be chosen to lie on the
intersections of a lattice with $L$ sites in each dimension
($d=1,2,3$).

2) The random variable $\phi$ at each site ${\vec r}$ takes on a finite
number of values $N$.

3) The energy functional $E(\{\phi\})$ is translationally invariant
with periodic boundary conditions. If pressed we will restrict the
interaction between variables to be of finite range.

Some examples of such energy functionals in two dimensions are:

1)The nearest neighbor Ising model in a magnetic field
\eqn\ising{E=-\sum_{j,k=1}^L\{E^v\sigma_{j,k}\sigma_{j+1,k}+
E^h\sigma_{j,k}\sigma_{j,k+1}+H\sigma_{j,k}\}}
where $\sigma_{j,k}=\pm 1.$

2) The $N$ state chiral Potts model

\eqn\cp{E=-\sum_{j,k=1}^L\sum_{n=1}^{N-1}\{E_n^v(\sigma_{j,k}\sigma_{j+1,k}^*)^n
+E_n^h(\sigma_{j,k}\sigma _{j,k+1}^*)^n+H_n\sigma_{j,k}^n\}}
where $\sigma_{j.k}^N=1.$ When $N=2$ we note that this reduces
to~\ising.

The class of restrictions 1--3 is, of course, just one of many. The
variables could lie on the bonds as well as the sites of the lattice.
Continuous variables could be used and a familiar example of such a
system is the $n$ component classical Heisenberg magnet
\eqn\cheis{E=-\sum_{j,k=1}^L\{{\vec  v}_{j,k}\cdot{\vec
v}_{j+1,k}+{\vec  v}_{j,k}\cdot{\vec  v}_{j,k+1}\}}
where ${\vec  v}^2=1$.
A continuum could be considered
instead of a lattice and translational invariance can be broken. But for
general discussion conditions 1--3 may be thought of for concreteness.

We are interested in the thermodynamic limit when
\eqn\thermo{L\rightarrow \infty ,~~~T~{\rm fixed~and ~positive}.}
In this limit we define the free energy per site in dimension $d$ as
\eqn\free{f=-kT\lim_{L \rightarrow \infty} {1\over L^d}\ln Z}
and we define the infinite volume correlations as
\eqn\corr{\lim_{L \rightarrow \infty}<O_j>_L}
where $<O_j>_L$ is computed from \sm~ on the finite lattice.

For a finite system the partition function $Z$ of \sm~ is an entire
function of $1/T$. However, once the infinite volume limit is taken the
free energy \free~can have singularities. When the singularities occur
for real positive $T$ the system has phase transitions.

These singularities in the thermodynamic limit are produced by the
accumulation of zeroes of the finite size partition function as was
first made clear in 1952 by Lee and Yang~\rly. In the general
discussion it is necessary to distinguish between the cases where the
zeroes pinch the real temperature axis at points and when entire
segments of the axis are pinched. The pinching of an entire segment
is known to happen in some cases where the interactions vary randomly
from site to site and thus break translational invariance. We will
consider some of these effects in later lectures. For the present,
however, we will consider the case where the zeroes pinch the real
temperature axis only at isolated points. This  happens at what is
called a critical point or a point of second order phase transition.
If there are other external parameters in the problem, such as the
magnetic field in the Ising model~\ising, the location of these points
of phase transition can depend on these parameters. A plot of the
location of the points of phase transition is called a phase diagram.
One of the important tasks of statistical mechanics is the derivation
of phase diagrams for realistic systems.

Phase diagrams are quite dependent on the specific details of the system
under consideration. Perhaps the most vivid illustration of this is
the transition temperature of a superconductor. There is extreme
interest in finding the material with the highest possible transition
temperature. While in  principle this temperature is contained in the
partition function for the system under consideration, the
practical truth is that theoretical statistical mechanics is far from
being able to tell us how to make high temperature superconductors.

However once the location of the points of phase transition are known
there is a remarkable universality to the behavior of the system near
the critical point. Thus although there are hundreds of superconductors
there are only one or two types of superconductivity.

Superconductivity is of course a quantum phenomenon and thus is not
encompassed by the classical statistical mechanics we are considering here.
But there are plenty of phase transitions in classical statistical
mechanics as well. Familiar examples are the ferromagnetic
transition to the magnetized state and the critical point in a
liquid-gas phase diagram. Indeed, the discussion of~\rly~ on the close relation
of these two phenomena is one of the foundations of the theory of
phase transitions.
Starting in 1944 with the exact work of Onsager on the Ising
model~\rons, extending through the 50's and into the early 60's with
the development of various series expansions and culminating in the
developments of the mid 60's with the invention of scaling theory, we
have developed a very detailed picture of the
physics of these classical critical
phenomena. This is well expounded in the classic articles~\rwidom--
\rkad. This picture is of fundamental importance for our
discussion of the connection with quantum field theory. For
concreteness we will develop this theory in the context of
{}~\ising~and~\cp. However the principles are generally valid.

There are many phenomena which occur at an isolated critical temperature
$T_c$ and the theory of critical phenomena relates them together.
Some of the principal phenomena are a) the singularities in the free
energy, b) the existence of spontaneous symmetry breaking, and c) the
behavior of correlation functions at long distances. These are related
through the construction of the scaling limit and scaling laws. We
will discuss each of these points separately.

\subsec{Singularities in the Free Energy}

To discuss the singularities in the free energy
it is convenient to define the specific
heat $c$ as
\eqn\sheat{c=-T{\partial ^2 f\over \partial T^2}.}
Then the simplest generic singularity the specific heat can have at a
critical temperature $T_c$ is
\eqn\exalpha{c\sim A_{\alpha}|T-T_c|^{-\alpha}.}
The exponent $\alpha$ is referred to as a critical exponent. Somewhat
more generally we recognize that the exponent $\alpha$ could be
different if $T\rightarrow T_c$ from above or below. If the more general
situation holds we let $\alpha$ be the exponent for $T>T_c$ and
$\alpha '$ be the exponent for $T<T_c$ and write
\eqn\sgenheat{c\sim A_{\alpha}^+ (T-T_c)^{-\alpha}~~(T\rightarrow T_c^+)~~{\rm
and}~~c\sim A_{\alpha '}^-(T_c-T)^{-\alpha '} ~~(T\rightarrow T_c^-).}

\subsec{Spontaneous Symmetry Breaking}

In the Ising model~\ising~ the next simplest property to define after
the specific heat at $H=0$ is the magnetization
\eqn\magising{M(H)=\lim_{L\rightarrow \infty}<\sigma_{0,0}>_L}
and for the $N$ state model similar order parameters are defined as
\eqn\ordercp{M^{(n)}({H})=\lim_{L\rightarrow \infty} <\sigma_{0,0}^n>_L.}
If $H=0$ the interaction~\ising~is invariant under
\eqn\sym{\sigma_{j,k}\rightarrow -\sigma_{j,k}}
 and thus if $M(H)$ is
continuous at $H=0$ it follows that $M(0)=0$. If $T>T_c$ this is
indeed the case. However, because we are considering the $L\rightarrow
\infty$ limit in the definition~\magising~or~\ordercp, there is no
reason that $M(H)$ has to be continuous at $H=0$ and, indeed, for
$T<T_c$ the continuity fails. Consequently for $T<T_c$ we define for~\ising~the
spontaneous magnetization $M_-(T)$ as
\eqn\spont{M_-(T)=\lim_{H\rightarrow 0+}\lim_{L\rightarrow \infty}
<\sigma_{0,0}>_L.}
Typically as $T\rightarrow T_c$ the spontaneous magnetization
vanishes. Thus we define a second critical exponent $\beta$ as
\eqn\exbeta{M_-(T)\sim A_{\beta}(T_c-T)^{\beta}~~~
{\rm as}~~T\rightarrow T_c^-.}

Another quantity related to the magnetization
is the magnetic susceptibility $\chi$
\eqn\magsus{\chi={\partial M(H)\over \partial H}|_{H=0}.}
This susceptibility also has singular behavior at $T=T_c$ and we
parametrize this in terms of the exponents $\gamma$ and $\gamma '$ as
\eqn\gammaex{\chi\sim A_{\gamma}^+ (T-T_c)^{-\gamma}~~(T\rightarrow T_c^+)~~
{\rm and}~~\chi\sim A_{\gamma '}^-(T_c-T)^{-\gamma '}~~(T\rightarrow T_c^-).}

Finally we consider the magnetization as a function of $H$ at $T=T_c$.
This also has a singular behavior as $H\sim 0$ which is parametrized as
\eqn\deltaex{M(H,T=T_c)\sim A_{\delta}H^{1/\delta}.}

For the more general $N$ state model~\cp~ the various exponents will depend
on the index $n$ of the order parameters ~\ordercp.
\subsec{Correlations}

Not only do the bulk thermal properties of the system have
singularities at $T_c$ but there are also corresponding phenomena in
the correlation functions as well. Consider, for example, the two point
correlation for~\ising ~ in the infinite volume limit
\eqn\corr{C(m,n)=<\sigma_{0,0}\sigma_{m,n}>.}
We write the coordinates as
\eqn\polar{m=R\sin\theta,~~~n=R\cos\theta.}
When $T<T_c$ we find that as $R\rightarrow \infty$ the correlation
approaches the limiting value of $M_{-}^2$ exponentially as
\eqn\lower{C(m,n)\sim M_-^2(1+{C^-(\theta,T)\over R^{p_{-}}}
e^{-R/\xi^-(\theta,T)}+\dots)}
where $\xi^-(\theta,T)$ is called the correlation length. Similarly
when $T>T_c$
\eqn\upper{C(m,n)\sim {C^+(\theta,T)\over
R^{p_{+}}}e^{-R/\xi^+(\theta,T)}+\dots.}
The correlation lengths depend on $T$ and diverge as $T\rightarrow T_c,$
and thus we define exponents  $\nu$ and $\nu '$ as
\eqn\exnu{\eqalign{&\xi^+(\theta,T)\sim A_{\nu}(\theta) (T-T_c)^{-\nu}
{}~~(T\rightarrow T_c^+)~~{\rm and}\cr
&\xi^-(\theta,T)\sim A_{\nu '}(\theta)(T_c-T)^{-\nu '}~~(T\rightarrow T_c^+).}}

The divergence of the correlation length as $T\rightarrow T_c$ is a
signal that the forms~\lower~ and ~\upper~ break down at $T_c$ and
instead it is found that for $T=T_c$ the correlations decay as a power
law
\eqn\power{C(m,n)\sim{A_c(\theta)\over R^{d-2+\eta}}~~(R\rightarrow \infty)}
where $d$ is the dimensionality of the system and $\eta$ is called the
anomalous dimension.

\subsec{Scaling Limit and Scaling Functions}

Of all the phenomena discussed above that happen at an isolated
critical temperature $T_c$ the most important is the divergence of
the correlation length $\xi$.
The physical meaning of this divergence
is that even though the physics is defined by a nearest neighbor
interaction (on the scale that 1 is defined as the inter atomic
spacing) the physical phenomena occur on a length scale $\xi$ which
is much larger than this scale of definition. At the critical
temperature the physical scale is infinitely large compared to the
scale of definition of the problem and thus it is most natural to
renormalize our length scale from the atomic length to the observed
physical length. Moreover on the atomic length scale the correlation
lengths have an angular dependence on $\theta$ which reflects the
existence of the lattice on which the system was originally defined.
However for many systems as $T\rightarrow T_c$ the angular dependence
approaches that of an ellipse which can be made rotationally invariant by
simply scaling the vertical and horizontal lengths with a different
factor. Thus it is most desirable to define new renormalized lengths
as
\eqn\ren{x=n/\xi(0,T)~~{\rm and}~~y=m/\xi(\pi/2,T)}
and take the limit
\eqn\scalone{T\rightarrow T_c,~~n\rightarrow \infty,~~m\rightarrow
\infty ,~~{\rm with}~~x~~{\rm and}~~y~~{\rm fixed}.}
This particular limit is called mass renormalization and is part of what is
called the scaling limit.

If the limit~\scalone~ were the only step involved the correlation
{}~\lower~ would vanish because the factor of the spontaneous
magnetization vanishes and the factor $C^-(\theta,T)$ is found to go
to a constant independent of T and $\theta$. Consequently we also
divide by $M_-^2$ and define the renormalized Greens function as
\eqn\green{G(r)=\lim_{\rm scaling}M_-^{-2}C(m,n)}
where by $\lim_{\rm scaling}$ we mean ~\scalone~and $r^2=x^2+y^2$.
 The process of dividing
$C(m,n)$ by $M_-^2$ is called wave function renormalization.

When $T>T_c$ there
is a factor called $M_+^2$ similar to $M_-^2$
which ~\upper~may be divided by in order that
the scaling limit exists. More generally we may consider correlations
of $n$ $\sigma$'s. Thus we formulate the extension of the
definition~\green~to the general case of the scaling functions of the
theory
\eqn\ngreen{G({\vec r}_1,\dots,{\vec r}_n)_{\pm}=\lim_{\rm
scaling}M^{-n}_{\pm}<\sigma_{m_1,n_1}\dots\sigma_{m_n,n_n}>}
where the subscript $\pm$ makes explicit that there are two different
field theories to be constructed, one from above and one from below $T_c$.

\subsec{Scaling Laws}
Thus far the theory discussed may be considered as descriptive and all
the exponents and functions introduced may be considered as independent
subject only to the general requirements of thermodynamic stability.
However if we make one additional assumption we find that this theory
makes predictions about the relation of critical exponents to each
other. These relations are known as scaling laws.

The additional requirement is the assumption that there are no other
length scales in the problem other that the atomic length scale of
definition and the physical length scale of the correlation length and
that these two scales join together smoothly.

As a first example of the relations between exponents which this new
assumption leads to we consider the relation between the scaling
function~\green ~for $T<T_c$ and the lattice correlation~\power~ at $T=T_c.$

Considering the isotropic case for convenience and using
the assumption of one length scale we
extend ~\polar ~and ~\exnu~ as
\eqn\newr{r=R(T_c-T)^{\nu '}/{A_{\nu '}}.}
Then using again the assumption of one length scale we write
\exbeta~and~{\green} for $T\sim T_c$ as
\eqn\newgreen{C(m,n) \sim A_{\beta}^2(T_c-T)^{2\beta}G(R(T_c-T)^{\nu
'}/A_{\nu '}).}
The assumption of one length scale further says  for large, but
fixed $R$, that as $T\rightarrow T_c$ this form of $C(m,n)$ must agree
with the $T=T_c$ behavior ~\power. Thus by comparing these two
expressions we find not only that it is required for the $R$ dependence
to match that for $r\sim 0$
\eqn\gzero{G(r)\sim {G_0 \over r^{d-2+\eta}}}
where $G_0$ is a constant
but furthermore in order for the dependence on $T_c-T$ to match we need
\eqn\scalingeta{2\beta=\nu ' (d-2+\eta).}
This relation between critical exponents is called a scaling law.

We may also use the one length scale assumption to find a relation
with the exponent $\gamma '$ and an expression for $A_{\gamma '}^-$ in
terms of $G(r).$ From ~\magising~ and ~\magsus~ we may write
\eqn\chisum{\chi (T)={1\over kT}\sum_{m,n}(<\sigma_{0,0}\sigma_{m,n}>-M_-^2).}
In order to study the divergence in $\chi$ as $T\rightarrow T_c$ it
is sufficient to consider only the large $R$ contributions to \chisum.
In this region the sum is approximated by an integral as
\eqn\intspprox{\eqalign{\sum_{m,n}&\sim \int_{-\infty}^{\infty}A_{\nu
'}(T_c-T)^{- \nu '}dx \int_{- \infty}^{\infty}A_{\nu '}(T_c-T)^{- \nu
'}dy\cr
& = A_{\nu '}^2(T_c-T)^{-2\nu '}\int_{0}^{2 \pi }d\theta
\int_{0}^{\infty}rdr.}}
Thus using~\green~ we find that as $T\rightarrow T_c-$
\eqn\chidiv{\chi (T)\sim (kT_c)^{-1} A_{\nu '}^2(T_c-T)^{2(\beta - \nu ')}
A_{\beta}^2 2\pi \int _{0}^{\infty} r(G(r)-1)dr}
and hence comparison with ~\gammaex~ gives
(written more generally in d dimensions)
\eqn\gammascaling{\gamma ' =d\nu ' -2\beta}
and (for d=2)
\eqn\amp{A_{\gamma '}^-=(kT_c)^{-1}A_{\nu '}^2 A_{\beta}^2 2 \pi
\int_{0}^{\infty}r(G(r)-1)dr.}
This expression for $\gamma '$ is another example of a scaling law.

There are several other scaling law relations between exponents which
are all obtained following the same line of reasoning using an
assumption of one length scale. We refer the reader to
{}~\rwidom--\rkad~ for the details and here quote the final results
valid for any dimension
\eqn\allphaascaling{\alpha ' +2\beta +\gamma '=2,}
\eqn\nuscaling{d\nu' =2-\alpha'}
and
\eqn\etadelta{\eta=2-{d(\delta -1)\over \delta +1}.}
The corresponding scaling laws for $T>T_c$ are obtained by replacing
all primed exponents with unprimed exponents and replacing $\beta$
with the appropriate exponent needed for the existence of the scaling
limit~\ngreen~for $T>T_c.$

\subsec{Universality}

The final concept in the theory of critical phenomena that must be
discussed is the concept of universality. In the examples of the Ising
model and the chiral Potts model the defining formulae~\ising~and ~\cp~
were written down with nearest neighbor interactions. But nothing in
the subsequent discussion appears to depend on this assumption. Indeed
it is thought that at the very least the interactions can be
generalized to an arbitrary number of finite range (ferromagnetic)
interactions and
the only change in the above discussion will be that the amplitude
factors and $T_c$ become functions of all of the interaction constants.
However the exponents $\alpha, \beta, \gamma, \delta, \eta, \nu$ and
the corresponding primed exponents are independent of all of these
interaction constants. More generally all of the scaling functions are
independent of the interaction constants. This independence is
referred to as universality.

\subsec{Discussion}

This general picture of the scaling theory of the critical point was
developed over 25 years ago. At the time it was originally put
forward the proponents were exceedingly cautious. The nature of the
one length scale assumption was clearly investigated and the
consequences of singularities more complicated  than the pure power
laws were examined. But a quarter century is a long time and it is
appropriate to close this section by surveying the subsequent progress
made in verifying this theory.

The first very significant piece of information is that in the last
25 years it has been verified that the two dimensional
Ising model at $H=0$
totally confirms all aspects of the theory. Thus the class of scaling
theories is definitely not empty. The full explanation of these
results will be given in lecture two.

There is no other theory as well studied as the  two dimensional
Ising model and in
three dimensions there are only numerical confirmations of the scaling
laws available. However, the current status here is that after 25 years
of most serious endeavor no known counter example has been found.

In two dimensions there exist a plethora of models for which partial
results on the scaling behavior are known. In all known cases the simple
power law forms hold (unless the exponent is an integer in which case
a logarithm may appear). Moreover whenever sufficient exponents have
been computed to test a scaling law the prediction has always been
verified. However for no model other than the Ising model have the
scaling functions been fully characterized.

The verification of the concept of universality is less studied. Even
for the Ising model exact computations have never been done if the
interaction bonds do not form a planar graph. However there is
absolutely no evidence that universality fails. This is one of those
cases in physics where it would be most desirable to have a precise
mathematical theorem.

\bigskip
\newsec{Quantum Statistical Mechanics}

 We next turn to quantum statistical mechanics as defined by ~\qsm.
As with classical statistical mechanics it helps to consider
specific systems in order to illustrate the problems involved.
An example that will possess more that sufficient richness is the XYZ
chain in an external field
\eqn\hxyz{H=-\sum_{j}(J^x S^x_j S^x_{j+1}+J^yS^y_j
S^y_{j+1}+ J^zS_j^zS_{j+1}^z+HS_j^z)}
where $S^i$ can be either the three Pauli spin matrices $\sigma^i$ or
the spin $S$ representation of the rotation group.
When $J^x=J^y=J^z$ the resulting chain is called the spin $S$ isotropic
Heisenberg magnet.

As in the case of classical statistical mechanics, phase transitions exist in
quantum systems. Unfortunately our knowledge of the details of these
transitions is much less than that of their classical counterparts. Thus
although the theory of critical phenomena developed in the previous
section is just as applicable to the quantum as to the classical case
and even though there is series expansion data for critical exponents
of several quantum magnets in three dimensions we will not pursue the
topic further in this lecture.

However there is one new feature in quantum statistical mechanics
which is very important to consider. Namely the quantum dynamics
which is obtained from time dependent correlation functions such as
\eqn\timecorr{C^j(n,t;T)=
{\rm Tr}(e^{-itH}S_0^j e^{itH}S_n^j e^{-H/kT})/{{\rm Tr} e^{-H/kT}}}

Just as the theory of equilibrium critical phenomena starts from the
behavior of the correlations for large spatial separations, so the
theory of transport properties starts from the behavior of these time
dependent correlations for large times. This theory originates in the
work of Bloembergen~\rblo~ and de Gennes~\rdg. We here follow the
treatment of ~\rsvw~ and~\rbl.

It will make a great deal of difference in the  asymptotic behavior
whether or not there is a conservation law. We will concentrate here on
the case where there is such a conservation law of the form
\eqn\cons{\sum_{n=-\infty}^{\infty}C^z(n,t)=1.}
Such a conservation law holds for the XYZ model~\hxyz~ if $J^x=J^y.$
For this case we define the spatial variance
\eqn\svar{\sigma^2(t)=\sum_{n=-\infty}^{\infty}C^z(n,t)n^2}
and the spatial Fourier transform
\eqn\four{I(k,t)=\sum_{n=-\infty}^{\infty}C^z(n,t)\cos(kn).}
The process of spin diffusion is said to occur if for small $k$ and
large $t$  we have
\eqn\spindif{I(k,t)\sim Ae^{-Dk^2t}}
where $0<D<\infty$. This asymptotic behavior implies  the  following
long time behaviors for the autocorrelation function
\eqn\cass{C^z(0,t)=\int_{-\pi}^{\pi}{dk\over 2 \pi}I(k,t)\sim (4 \pi D
t)^{-1/2}}
and the spatial variance
\eqn\cvar{\sigma^2(t)=-{\partial^2 I(k,t)\over \partial
k^2}|_{k=0}\sim 2Dt.}

These long time behaviors play the same role as does the exponential
decay of the spatial correlations discussed in the previous section and
as before we need to know the validity of~\spindif.
In 1976 it was stated~\rsvw~ that ``this results rather naturally from
the phenomenological theories of irreversible    processes, and can
be more or less proved by modern statistical mechanics, for instance
by Mori's methods~\rmori''. However, despite the optimism of this claim
the fact is that this spin diffusion form has never been demonstrated
for any system and the only exact computations which do exist have a
different asymptotic behavior altogether. This is not to say that the
spin diffusion form does not occur, but it does mean that  our
knowledge of the subject is not nearly refined as our knowledge of
critical phenomena in classical systems. Consequently, even though
these quantum time dependent correlation functions are naturally
connected with finite temperature field theories, we will not pursue
the connection further in this lecture. We will consider these
phenomena at length in lecture four.

\bigskip
\newsec{Quantum Field Theory}

The formula for thermal averages \sm , which was the basis for sec. 3,
was found by Gibbs and Boltzmann in the 19th century. It forms the
first chapter of any course on statistical mechanics and all our
treatment of thermal phenomena follows from \sm . There
is never any argument about the starting point of equilibrium statistical
mechanics.

On the other hand the   formula \qf~ has no such universal status in
quantum field theory. Consequently it may be useful to discuss where
\qf ~comes from.

It is here, surely, that the warning of Humpty-Dumpty must be kept in
mind because the words quantum field theory mean many things to many
people. Some of the different usages are exemplified by the following
list:

{}~1) Canonical QFT~~~~~~~~~~~~~~2) Lagrangian QFT

{}~3) Relativistic QFT~~~~~~~~~~~~4) Perturbative QFT

{}~5) Renormalized QFT~~~~~~~~~6) Finite Temperature QFT

{}~7) Conformal QFT~~~~~~~~~~~~~8) Topological QFT

{}~9) Lattice QFT~~~~~~~~~~~~~~~~10) Gauge QFT

11) Axiomatic QFT~~~~~~~~~~~12) S Matrix Bootstrap QFT

{}~~~~~~~~~~~13) Path Integral QFT

The formula (1.1) may be taken as the definition of the last item on
the list, but there are literally tens of thousands of papers written
about the other twelve items  and this list is far from complete. In
fact, there are so many different usages of the words ``quantum field
theory''  that
it often useful to regard the words QFT by themselves as not
specifying anything at all until the qualifying adjective is added.
This warning saves untold hours of discussion which communicates nothing
because the participants are all talking at cross purposes. It is certainly
not the case that all books on quantum field theory start with (1.1)
as the definition. Some books~\rfh~\rgj~ put (1.1) in the very title, some
make use of it in the text in places ranging from the very beginning
to near the end~\rfasl--\rster~ and some do not use it at
all~\rbd~\rds.  This lack of a common starting point
marks a fundamental distinction between
statistical mechanics and quantum field theory.

With this being said, however, it is certainly true that the first twelve items
are not totally unrelated to the thirteenth. Therefore I will begin
this section by sketching the history of the development of formula (1.1).
As in the previous two sections it is useful to have
several examples in mind to make the considerations concrete. Three
such examples of Lagrangians are~\ritzu

1) The $n$ vector model
\eqn\nlsigma{L(\phi)=\sum_{k=1}^d({\partial {\vec \phi}\over \partial
x_k})^2 +{m^2\over 2} {\vec \phi }^2 +{g\over4}({\vec \phi}^2)^2}
This model can be viewed as a generalization of the $n$ component
Heisenberg model~{\cheis } (which in quantum field theory is often
called the non-linear sigma model).

2) Scalar Electrodynamics
\eqn\scalar{\eqalign{L&={1\over 4} \sum_{j,k=1}^d  (\partial_j
A_k-\partial_k A_j)^2+{\lambda\over 2}(\partial \cdot
A)^2\cr
&+\sum_{j=1}^d[(\partial_j+ieA_j)\phi]^{\dagger}[(\partial_{j}+ieA_{j})\phi
]+m^2 \phi^{\dagger} \phi +{g\over 4}(\phi^{\dagger}\phi)^2}}

3) Quantum Electrodynamics
\eqn\qed{L={1\over 4}\sum_{j,k=1}^d
(\partial_jA_k-\partial_kA_j)^2+{\lambda \over
2}(\partial \cdot A)^2 +{\bar \psi}(\partial \cdot \gamma
-eiA \cdot \gamma -m){\psi}}
where $\gamma_j$ are the Dirac gamma matrices.

Quantum field theory did not originate with the discovery of (1.1).
Instead quantum field theory developed from classical field theory by
taking the equations of motion of a classical field (of which the
Maxwell's equations coupled to electrons
was by far and away the most important
application) and replacing classical fields by quantum operators
which obey commutation or anti-commutation relations. From these field
equations Green's functions were computed and from this such triumphs
as the anomalous magnetic moment of the electron came.

Classical field theory can be obtained from a classical action and
as early as the work of Dirac in  1933~\rdirac~ and
the work of Feynman~\rfeynman~
in 1948 it was understood how to extend this to quantum mechanics
using a path integral of the Minkowski form ~\mqft.~The action principle was
extended from quantum mechanics to quantum field theory shortly
thereafter by Schwinger~\rscha. At almost the same time it was realized
by Dyson~\rdysona, and later by Wick~\rwick, that great advantages could be
obtained if one
considered the analytic continuation to Euclidean space and the
Euclidean action approach of (1.1) was considered by Schwinger~\rschb~
in the late 50's.

It may now be asked what is the origin of  the action principle. This can be
answered by first forming the partition function $Z$ of (1.1). If $h$ is
considered to go to zero and if we choose to evaluate the
functional integral for $Z$ by the method of steepest descents where the
action is expanded about classical configurations which satisfy
\eqn\stataction{{\delta L\over \delta \phi({\vec x})}=0}
 then
the results of Schwinger's action principle and of canonical
quantization are the same. If in addition external sources are
considered then the partition function can be regarded as the
generating function for the Greens functions of the quantum theory. In
this construction $h$ sets the scale of the commutation relations and
this scale is identified with the experimentally determined Planck's constant.

Thus far the path integral (1.1) has been used as a device for doing
computations which could and indeed were first performed by other methods.

However in this procedure we have been rather sloppy because
questions of existence and convergence have been ignored. In
order to make direct sense of (1.1) some sort of additional
definitions need to be made.

For the scalar field theory 1) it is
sufficient to replace the fields $\phi (x)$ defined on continuum space
by fields defined only on the lattice sites and to replace derivatives
by differences which amounts to the replacement
\eqn\replace{(\partial_k\phi )^2\rightarrow -\phi (x) \phi(x')}
where $x$ and $x'$ are nearest neighbors.

For the gauge field theories 2) and 3) the lattice definition was
first given in~\rwil. The vector potential $iA$ is replaced by a
compact group element such as $e^{iA}$ which is defined on the links of
the lattice and the matter fields are $\phi$ or $\psi$ are defined on
the sites of the lattice. Then, calling the gauge group element on
the link between sites $j$ and $k$ $U_{j,k}$ the Lagrangian 2) is
replaced by
\eqn\wil{{1\over 2 g_0}{\rm Tr}U_{j,k}U_{k,l}U_{l,m}U_{m,j} +e\phi^{\dagger}
(x_j)U_{j,k}\phi (x_k) +{m\over 2}\phi^{\dagger}\phi +{\lambda \over
4}(\phi^{\dagger}\phi )^2}
where in the first term the sites $j,k,l,m$ lie on the vertices of a
square (plaquette).

But now that we have reproduced the statistical mechanical starting
point we are forced, if we want to regain a continuous result, to follow
the procedure that leads to the scaling limit in statistical mechanics
and find those points in the phase diagram, including the variable
$h$, where the system undergoes a phase transition and at those points
to construct a scaling limit. If the point $h=0$ is a point of phase
transition then we will regain the situation we started from with
canonical quantization. But in statistical mechanics there are a host
of physically interesting systems where the critical temperature is
not at zero. The corresponding field theories will be outside the
realm of those considered by canonical quantization.

We may now examine the concept of perturbative QFT as derived from
(1.1). In principle one obtains a field theory from (1.1) by finding
the points of phase transition and constructing a scaling theory in
the vicinity. In perturbative QFT the action is written as
$S_0+\lambda S_1$ where $S_0$ has a phase transition at $h=0$. An
expansion in $\lambda$ is then made assuming that the full theory
still has a phase transition at $h=0$. Actions for which this
procedure works are called renormalizible and the resulting expansion
is called a renormalized perturbation expansion. These renormalized
perturbation expansions form a large part of the literature on
quantum field theory.

In these renormalized perturbation expansions the physical content of
particles of the full theory is the same as the content of the
``free'' theory $S_0$ (up to possible questions of bound states). The
particle spectrum of $S_0$ is called bare as opposed to the genuine
physical particle content of the full $S$ which are called
dressed or renormalized in particle physics.

 But starting in the mid 60's, certainly with the seminal paper of
Gell-Mann~\rgel, it has been realized that for the gauge theory of
strong interactions (QCD) that neither the gluons (gauge bosons) nor
the quarks (bare fermions) which are the fields in $S$ actually show up
as physical particles of the full theory. The word for this is
confinement and the lattice gauge theory of~\rwil~ is one way of
addressing the problem. (Any reader who has not in fact read~\rgel~
is strongly urged to read the last sentence and to marvel at the
subtlety of the English language. Even Humpty-Dumpty must stand in awe
of an author who can introduce into a physics paper a word whose
origin is explained by a footnote to {\it Finnegan's Wake}~and then
suggest an experiment to determine if these objects are real and if
they exist.)

The need to consider field theories with confinement is a most
compelling reason to extend the notion of QFT beyond the perturbation
theory fixed point at $h=0$.

Consequently there has been a movement in the past 30 years to change
our point of view and to take (1.1) as the starting point for QFT and
to allow for all possible points of phase transition, not just $h=0$.

This change of starting point for QFT now makes QFT and SM into
identical subjects. This is  natural for a statistical
mechanician but from the historical perspective of QFT we must accept
the fact that if the point of phase transition is not $h=0,$ there will
no longer be a classical limit of the QFT. Such a field theory may
rightly be called fully quantum mechanical.

In more old fashioned terms we are accepting the fact that for field
theories with points of phase transition not at $h=0$ there is no
correspondence principle.

In the past 20 years we have had great success in two dimensions in
finding, constructing and in many cases solving these quantum field
theories with no classical limit. Indeed, most of the conformal and
topological field theories discussed by other lecturers at this summer
school have this property.

Moreover once it is understood that fully quantum mechanical quantum
field theories exist and that at least in two dimensions some of them can be
explicitly seen to have confined degrees of freedom, it is not
possible to accept as dogma the predictions about high energy
scattering made solely on the basis of renormalized perturbation
theory. The development of QFT from quark confinement to string
theory is the attempt to understand physics beyond the perturbation
theory of the 60's.

At this point it is most helpful if a very candid admission is made.
If one reads the papers on the path integral foundation of quantum
field theory just discussed above~\rdirac--\rschb~ it is very clear that
this procedure of looking for points of phase transition in the
Euclidean path integral is NOT what these authors had in mind. We have
quite deliberately extended the meaning of the words quantum field
theory beyond what the founders of quantum electrodynamics had in mind
by equating quantum field theory with phase transitions in statistical
mechanics. The reader has every reason to ask who it was that first
made this new extension.

It is one of the curious mysteries of physics that there does not seem
to be any one person or paper to whom this idea can be unambiguously
attributed. In his 1974 paper Wilson~\rwil~ writes as if it were well
known and refers the reader to the paper of Wilson and Kogut~\rwk. But
in~\rwk ~ the authors also do not claim that they are inventing a new
principle. The statement of many of the ideas relating  field theory and
critical phenomena in a perturbative setting
is given several years earlier~\rpatpok--~\rpolc~where
the focus seems not to be on the invention of a new principle  but on
rather heavy diagrammatic computations  and the relation with
field theory is treated as rather obvious. Indeed in \rpola~ the Ising
model is discussed and the Ising model had been discussed in terms of
a field theory of fermions as early as 1949~\rkauf.

My conclusion from this is that the emergence of the identity of QFT
as defined by (1.1) and statistical mechanics as defined by \sm~
emerged from the community as a whole as something that once you saw
what was going on you truly decided that it was so obvious and basic
that it was implied in all that had gone before. It is perhaps fitting
that a thermodynamic principle has a collective origin.

However, the lack of a definitive statement of origin
leads to confusion even to
the present day. If you ask many physicists if there can be a quantum
field theory without a corresponding classical limit you will often
get a blank stare of incomprehension. And yet it tortures the brain to
force the Ising model \ising~ into the language of classical field
theory and a stationary action principle. In particular, as explained
in sec. 3 there are two field theories that can be constructed near
the $T_c$ of the Ising model: one from above $T_c$ and one from below
$T_c$. Even worse is the fact that there are actually an infinite
number of field theories that can be constructed as $H\rightarrow 0$
and $T\rightarrow T_c$ depending on the value of $h=H/|T-T_c|^{15/8}$.
One way in which it has been attempted to preserve the notion of a
classical limit is to consider the Ising model as an $n=1$ non linear
sigma model~\nlsigma~where the interaction term is $\lambda(\phi
^2-1)^2$ in the limit where $\lambda \rightarrow  \infty$ and to
develop a perturbation scheme in which $\epsilon =4-d$ is a small
parameter. This was originally suggested in~\rwf~ and is reviewed in
detail in ~\ritdr. However the $\epsilon$ expansion does not converge
and thus far it has not been possible to use it to regain the exact
results known in two dimensions. Consequently I will here adopt the
view that it is far better to
abandon the notion of a correspondence principle rather than to turn
an exactly solvable model into an insolvable mess merely to preserve a
fiction that indeed may not represent all the physics.

It finally remains to consider how to obtain results in the physical
Minkowski metric from the Euclidean path integral (1.1). Formally this
is done by analytically continuing one of the Euclidean variables, say
$x_0,$ to $it$. The mathematical questions that need to be addressed are
1) the circumstances under which this continuation is possible
and 2) what the relation is of the resulting continuation with
correlations computed directly from the Minkowski path integral \mqft.
Fortunately these questions have been extensively investigated in
axiomatic quantum field theory and are well discussed in the
literature~\rgj,\ros. Consequently we can be brief.

For a Euclidean QFT to be acceptable it is not enough that the limits
exist but that the limiting correlations satisfy several properties
called the axioms of Osterwalder and Schrader~\ros. In non technical
language these axioms are:
E0) Temperedness; E1) Rotational invariance;
E2) Reflection (Osterwalder-Schrader) positivity; E3) Symmetry;
E4) Cluster property;

The virtue of the OS axioms is that they hold for the Euclidean QFT if
and only if
the following Wightman axioms~\rwi--\rsw~ hold for the Minkowski correlation
functions:
M0) Temperedness; M1) Lorentz Invariance; M2) Positivity;
M3) Local commutativity; M4) Cluster property;
M5) The existence of a lower bound on the energy spectrum.

For further discussion of these axioms (especially for the somewhat
technical explanation of temperedness) see ref ~\ros and \rgj.

The verification that the correlations of a Euclidean QFT satisfy
the OS axioms can be taken as guaranteeing that the analytic
continuation from Euclidean to Minkowski space is possible. Thus for
our purposes we will consider that whenever they are both defined that
(1.1) and~\mqft~ are equivalent. It may, however, be that there are
cases where
{}~\mqft~ is not necessarily well defined. In that case we will use (1.1)
and analytic continuation as the definition of what is meant by a
Minkowski QFT.

\bigskip
\newsec{The Need to Exist}

One of the enduring contributions made by Kant in {\it The Critique
of Pure Reason} is the demonstration that the mere making of a definition
does not by itself ensure the existence of the thing defined. Kant
raises this most important distinction in his discussion of the
ontological proof given by  Descartes in the {\it Discourse on
Method}
for the existence of God.
However the truth of this distinction is universal and since Kant's
time the existence proof has become a very important part of
mathematics.

Consequently the definitions  of path integral quantum field theory
given above will only be useful if we can demonstrate that the
category of such field theories is not empty.

In statistical mechanics a great deal of attention has been given to
the determination of restrictions which must be placed on the
classical energy functional $E(\phi)$ or the quantum Hamiltonian $H$
in order for the thermodynamic
limit of the correlation functions \sm~ or \qsm~to exist \rfr--\rlieb.
To prove existence it
is necessary that the interactions are not so attractive at short
distances that the system collapses and that at long distances the
interactions are not so repulsive that the system explodes. Somewhat
more precisely the requirement of no short distance collapse will be
ensured for a quantum system if the ground state energy  has the lower bound
of
\eqn\lbound{E_{GS}>-AN}
where $N$ is the number of particles in the system and $A$ is a positive
constant.

It is one of the triumphs of mathematical physics that the lower bound
\lbound~ has been proven for the system of electrons and protons
interacting via the non-relativistic Coulomb
interaction~\rdl,\rlt~and\rlieb.  For the
existence of the bound it is essential for the electrons and protons
to obey Fermi statistics. A system consisting of only charged bosons
is known to have an upper bound on the ground state energy of~\rdy~\rdl
\eqn\ubound{E_{GS}<-A_1 N^{7/5}}
which manifestly violates~\lbound. Such a system is unstable against
collapse and the laws of statistical mechanics \sm~ are not directly
useful in its description.

There is an identical need for the demonstration of existence for
quantum field theory as defined by (1.1). However similar detailed
existence proofs have not been carried out for all actions commonly
believed to be interesting. But just as in the case of statistical
mechanics the need for existence places restrictions on the actions
which can be considered and in particular a lower bound on the action
similar to \lbound ~is required.

If it were the case that all the actions which are considered
interesting in quantum field theory
had such lower bounds then it would not be particularly
appropriate to dwell on this point. However, this is the not
case. The most famous case in point is the  attempt to
quantize the Euclidean version of the Einstein action of general relativity
\eqn\ein{S_E=-{1\over 16 \pi G}\int R g^{1/2}d^4x}
where $R$ is the curvature and $g$ is the determinant of the
metric tensor $g_{\mu , \nu}$.
This classical action is indeed not bounded below and thus
the definition of QFT given above based on (1.1) will not apply.
 There
have been attempts~\rhawk--\rgib~to introduce extra restrictions into the
Euclidean path integral to exclude the regions of path space which
make (1.1) diverge. There have also been proposals~\rmm~for different
Euclidean actions obtained from a more sophisticated analytic
continuation procedure than used in non gravitational field theories.
Some of these quantization procedures seem to lead~\rwald~
to further difficulties. Moreover, even if such a
prescription could be found there is a further problem of how to pass
from the Euclidean to the Minkowski metric. In a general relativistic
theory the correlations are certainly not expected be to be rotationally or
Lorentz invariant and thus there is no known equivalent of the
Osterwalder-Schrader
axioms~\ros. Consequently there is no reason to believe that there is
enough analyticity to continue from a Euclidean to a Minkowski region.
Indeed the global question of what is meant by Minkowski is obscure
for quantum gravity~\rwald.

Of course it is formally possible to construct a perturbation
expansion about the point $h=0$ if (1.1) is used in an uncritical
fashion and the classical solution is asymptotically flat~\rwald~\rmm, but the
resulting perturbation theory is readily shown to be
nonrenormalizible and is thus not particularly suitable to be taken as
a definition.

One physical effect that lies at the bottom of the difficulty of using (1.1) to
provide a theory of pure quantum gravity  is that gravity is a
purely attractive bosonic theory
and this attraction at short distances leads to precisely the sort of
collapse that occurred in the system of nonrelativistic charged bosons.

Because gravity is always with us and has been studied for literally
3000 years it is no small criticism that this interaction is not well
incorporated into the definition (1.1). I am also well aware that there
is a school of thought that argues that the only way to obtain a
consistent theory of elementary particles is to find a theory which
does incorporate gravity. Many attempts have made in addition to
{}~\rhawk--~\rmm~ to either
1) modify the classical action to make the path integral exist, 2) add
sufficient fermionic fields (usually of spin 3/2) to provide enough
repulsion to produce existence of the path integral or 3) abandon
(1.1) altogether and to develop new principles. Some of
these attempts are summarized in ~\rwald~ but it is fair to  say that
none has been successful on all counts.

It is thus an open problem to find a definition of QFT that will
incorporate gravity. But following the philosophy of Humpty-Dumpty I
will not let this stop me from defining what {\it I} mean by quantum
field theory. Consequently in these lectures the words
quantum field theory will always mean the limiting theory constructed from a
lattice regularized path integral (1.1). No more--No less.

\bigskip
\newsec{Particles}

I have now given some sort of meaning to the formal connection between
quantum field theory and statistical mechanics based on \qf~and \sm~
and I have sketched many of the physical phenomena measured in the
statistical mechanical part of the connection. But in terms of quantum field
theory I have said nothing at all about the experiments that quantum
field theory was invented to study in the first place; namely the study
of the properties of particles.

The word particle is one that has many connotations. Particles are
conceived of as having properties such as momentum, charge, mass, spin
and a variety of other quantum numbers given in the particle data
tables. There is the distinction between stable and unstable particles
and there is the question of how unstable a particle can be and still
be dignified by that name. Particle may be ``elementary'' such as
neutrinos and electrons or ``composite'' such as nuclei and we may
argue over which of the two categories neutrons and protons are
supposed to be in. Particles such as neutrons and protons have some
sort of intrinsic size or extension. To complete the relation of
quantum field theory and statistical mechanics we must say something
about the statistical mechanical counterparts of these objects.

In this lecture I will not address all of these different properties
connected with the word particle. In particular I will not consider
what should be meant either by the word unstable or the word extended.
Thus the discussion of this section will only begin to study this most
fundamental of concepts.

I will begin   with the
concept of a particle like energy spectrum of a Hamiltonian.
Such a spectrum has the
property that in infinite space  all the excited state energies
are given by
\eqn\espectrum{E_{ex}-E_{GS}=\sum_{\alpha =1}^{S}\sum_{j_{\alpha}
=1,\rm rules}^{m_{\alpha}}e_{\alpha}({\vec p}_{j_{\alpha}})}
 and all the momenta of the states are given by
\eqn\pspectrum{{\vec P}_{ex}=\sum_{\alpha=1}^S \sum_{j_{\alpha}=1,\rm
rules}^{m_{\alpha}}{\vec p}_{j_{\alpha}}}
where $S$ is the number of different species of particles, $m_{\alpha}$
is the number of particles of type $\alpha$ in the state,
$e_{\alpha}(P)$  is the single particle energy of the particle of type
$\alpha$, and the subscript rules indicates the rules with which the
excitations can be combined. One such example of the combination rules
is
\eqn\fermi{{\vec p}_{j_{\alpha}}\neq{\vec p}_{l_{\alpha}}~~
{\rm if}~~j_{\alpha}\neq
l_{\alpha}.}
This is a called a Pauli exclusion rule and particles which satisfy it
are called fermions and said to satisfy Fermi statistics. Another
example of such
combination a rule is
\eqn\pair{m_1+m_2=0 ~~({\rm mod}~2)}
which expresses the fact that particles of type 1 and 2 must be created
in pairs. When this type of rule exists there is no one particle state
for either type of particle and the single particle energy must be
inferred out of the observed multiparticle energy levels.

In a relativistic quantum field theory the single particle energies
are of the form
\eqn\relen{e_{\alpha}({\vec p})=(M_{\alpha}^2 + {\vec p}^2)^{1/2}}
and the statement that all the states of the system
satisfy \espectrum~ is referred to as asymptotic completeness~\rsw.
It is an open question as to whether the Wightman axioms are
sufficient to guarantee this form.

In statistical mechanics this type of spectrum is referred to as a
quasi-particle spectrum The prefix ``quasi'' being attached to
indicate the feeling that somehow magnons and phonons are different
from protons and electrons and that single particle energy levels
other than \relen~ are allowed.

This particle spectrum has a major impact on the Greens functions.
This is best described by considering the Fourier transform of the
correlation functions \qf~ or ~\mqft. For concreteness consider the
Fourier transform of a Euclidean two point function
\eqn\ft{G({\vec k}) =\int d{\vec x}~C({\vec x}) e^{i {\vec k}\cdot
{\vec x}}}
where $C({\vec x})$ is obtained from the  Minkowski correlation
function by Wick rotation. Then from \espectrum~ and \relen~ it
follows that $G({\vec k})$ depends only on ${\vec k}^2$ and has poles at
\eqn\poles{{\vec k}^2=-M_{\alpha}^2}
and has multi-particle threshold singularities at all those
combinations
\eqn\multi{{\vec k}^2=- (\sum_{\alpha}n_{\alpha}M_{\alpha})^2}
where $n_{\alpha}$ takes on all integer values consistent with the
rules of combination.

We will take the spectral condition~\espectrum~ and the corresponding
statement on the location of singularities in the Greens functions
{}~\poles--\multi~ as the definition of the existence of particles in a
relativistic quantum field theory.

To explain further the relation of particles to Greens functions in
statistical mechanics we introduce the concept of the transfer matrix $T$
{}~(considered in two dimensions for concreteness).
Consider a system with nearest neighbor interactions on a square lattice
where the energy is of the form
\eqn\esquare{E=\sum_{l=1}^{L_h}\sum_{j=1}^{L_v}(E^v(\phi_{j,l},\phi_{j+1,l})
+E^h(\phi_{j,l},\phi_{j,l+1}))}
where $E^v(\phi_{j,l},\phi_{j+1,l})$ $(E^h(\phi_{j,l},\phi_{j,l+1}))$ are
the vertical (horizontal) interactions.
Then letting $\{\phi\}$ stand for the collection of variables $\phi_{j,l}$ in
row $j$ and defining
\eqn\thor{T^h_{\{ \phi \},\{ \phi ' \}}=\delta_{\{\phi \},\{\phi ' \}} e^{-
\sum_{l=1}^{L_h}E^h(\phi_l , \phi_{l+1})/kT}}
and
\eqn\tver{T^v_{\{\phi\},\{\phi ' \}}
=e^{-\sum_{l=1}^{L_h} E^v(\phi _l, \phi_l ')/kT}}
a row to row transfer matrix may be defined either as
\eqn\trow{T^R=T^v T^h}
or more symmetrically as
\eqn\trowsym{{T^R} '=(T^h)^{1/2}T^v(T^h)^{1/2}.}
The partition function \sm~ with periodic boundary conditions in both the
vertical and horizontal directions is then written in terms of
either of these transfer  matrices $T$ as
\eqn\ztran{Z={\rm Tr} T^{L_v}.}

It is also useful to consider a diagonal to diagonal transfer matrix
\eqn\tdiag{T^D_{\{\phi\},\{\phi '\}}=\prod_{l=1}^{L_h}
e^{-E^v(\phi_l,\phi'_l)/kT}
e^{-E^h(\phi_l,\phi '_{l+1})/kT}.}
Then if the boundary conditions on the lattice are suitably shifted
the partition function is still given by \ztran.

In general transfer matrices are diagonalizable. Thus calling $\lambda_m$
the eigenvalues of $T$ we may write ~\ztran~ as
\eqn\zlamb{Z=\sum_{m}\lambda_m^{L_v}}
and  in the thermodynamic limit the free energy~\free~
is written in terms of the maximum eigenvalue $\lambda _0$ of $T$ as
\eqn\flam{f=-kT\lim_{L_h \rightarrow \infty}{1\over L_h}\ln \lambda_0.}

The correlation functions may also be expressed in terms of the transfer
matrix.
As an example consider the two point correlation of the operator
$\sigma ^n_{j,l}$ of the chiral Potts model \cp. By definition
\eqn\corrcp{<\sigma^n_{j,l}\sigma^{*n}_{j',l'}>={1\over Z}\sum_{\rm states}
\sigma^n_{j,l}\sigma_{j',l'}^{*n}e^{-E/kT}}
which is written in terms of the transfer matrix as
\eqn\corcptr{<\sigma^n_{j,l}\sigma^{*n}_{j', l'}>=
{1\over Z}\Tr \sigma^n_l T^{j'-j}
\sigma_{l'}^{*n} T^{L_v-(j'-j)}}
(where for definiteness we have chosen $j'\geq j$).
In the basis where the spins $\sigma_{j,l}$ take on the values
$e^{2 \pi i r/N}$ with $r=0,1,\dots ,N-1$
 the operator $\sigma_l$ is
\eqn\opsigma{\sigma_l=I\otimes\cdots \otimes \underbrace{ S^z}_{{\rm site}~l}
 \otimes \cdots \otimes}
where $S^z$ is the $N \times N$ matrix with elements
$\delta_{r,r'} e^{2 \pi i r/N}$
Thus, writing $T$ in terms of its eigenvectors $|m>$ and  eigenvalues
$\lambda_m$ we find in the thermodynamic limit
\eqn\corrther{<\sigma_{j,l}^n\sigma_{j',l'}^{*n}>=\sum_m<0|\sigma^n_l|m>
(\lambda_m/\lambda_0)^{(j'-j)}<m|\sigma_{l'}^{*n}|0>.}

Corresponding to the quasi-particle property~\espectrum~ of
the energies there is a corresponding quasi-particle form for
the eigenvalues of the transfer matrix
\eqn\traneig{\lambda_{ex}/\lambda_0=\prod_{\alpha=1}^S \prod_{j_{\alpha}=1,
{\rm rules}}^{m_{\alpha}}\lambda_{\alpha}(p_{j_{\alpha}}).}
Moreover, for a translationally invariant system the operator
$\sigma_k$ is obtained from the operator $\sigma_0$ by a
lattice translation operator whose eigenvalues are $e^{i k P_{ex}}$
where $P_{ex}$ is given by ~\pspectrum.
Thus the correlation ~\corrther~ is expressed as
\eqn\klcp{\eqalign{<\sigma^n_{j,l} \sigma^{*n}_{j',l'}>=&
\sum_{\alpha =1}^S\sum_{m_{\alpha}=0}^{\infty}
\sum_{\{p_{j_{\alpha}}\}}
<0|\sigma_0^n|\{p_j{_{\alpha}}\}>
<\{p_{j_{\alpha}}\}|\sigma_0^{*n}|0>\cr
&e^{\sum_{\alpha=1}^S\sum_{j_{\alpha}=1}^{m_{\alpha}}[i(l'-l)p_{j_{\alpha}}
+(j'-j)\ln(\lambda_{\alpha}(p_{j_{\alpha}}))]}\cr}}
This representation of the correlation function is the statistical mechanical
analogue of the K{\"a}llen-Lehmann representation of
quantum field theory~\ritzu~
where $\ln \lambda_{\alpha}(p)$ plays the role of the single particle
energy levels. If the scaling limit exists and is relativistically
invariant with $S$ particles of masses $M_{\alpha}$ we obtain the two
point Greens function as
\eqn\green{\eqalign{ G_n({\vec r})=&\lim_{\rm scaling}M_n^{-2}<\sigma_{j,l}^n
\sigma_{j',l'}^{*n}>\cr=
&\sum_{\alpha=1}^S\sum_{m_{\alpha=0}}^{\infty}<0|\sigma_0^n|\{p_{j_{\alpha}}\}>
<\{p_{j_{\alpha}}\}|\sigma_0^{*n}|0>\cr
&e^{\sum_{\alpha=1}^S\sum_{j_{\alpha}=1}^
{m_\alpha}
[ir_x p_{j_\alpha}+r_y(M_{\alpha}^2+p_{j_\alpha}^2)^{1/2}]}\cr}}
Here the matrix elements $<0|\sigma_0^n|\{p_{j_\alpha}\}>$ are referred to as
form factors. In particular for the correlations to be rotationally
invariant the single particle form factors $<0|\sigma^n_0|p_{j_\alpha}>$
must be constants. Then the Fourier transform ~\ft~ will have poles
at ~\poles~ and we see that the two notions we have introduced
for the masses $M_{\alpha}$ of the particles coincide.

But in applications of field theory to particle physics the notion of
particle spectrum and masses of particles are
not sufficient to describe the experiments. We must further be able to
describe the scattering of particles. Particle scattering has been
at the core
of elementary particle physics experiments for most of this century
and I will presume the reader has some acquaintance with the subject.
For concreteness I will follow the treatment of ~\ritzu.

The fundamental  object in scattering theory ~\ritzu~ is the S matrix
which represents the transition amplitude from an ``in'' state with
$n_{\rm in}$ particles of momenta $\{p_\alpha\}$ to an ``out''
state with $n_{\rm out}$ particles of momenta $\{p_{\alpha}'\}$.
One of the crucial results is that the S matrix elements,
which are supposed to be what are measured in  laboratory experiments,
are expressed as the residues of Greens functions when the external
momenta satisfy the mass shell condition ~\poles. This result is
the LSZ reduction theorem~\ritzu.

However, when the attempt is made to relate the LSZ reduction theorem
to the discussion of Greens functions given above there are two problems
which cannot be avoided.

The first problem is that there may be many poles  in a two point function
and thus there does not need to be a one to one relation between particles and
fields. This is an old observation~\rsw~ and for such systems
the notions of ``creation operators'' and ``free fields'' which are used in
the derivation of the LSZ theorem need a more careful treatment. To make a
``particle'' to be observable we would like, in some theoretical sense at
least, to be able to find an operator that produces  only that single particle.
The definition of mass and spectrum made above do not provide an answer to
this question. This phenomenon of many  poles in a
two point function occurs in
the field theories constructed from the Ising model in the limit
$H\rightarrow 0$, $T\rightarrow T_c$ with $H/|T-T_c|^{15/8}$ held fixed and
will be discussed in detail in the third lecture.

The second problem is the question of
which Greens functions are to be used in the LSZ reduction formula. In the $N$
state chiral Potts model explicitly discussed above there were
$N-1$ different two point functions which were considered,
all of which are constructed from the same energy spectrum.
The same is true for the four and higher point functions. It would thus
appear that I can construct several different S matrices for the same
 particles  by use of the LSZ reduction. This would seem not to be
compatible with the
notion of a particle as a thing in itself and consequently it can be argued
that the consistency of the particle interpretation will
force an identity between these several different constructions. However
I know of no proof of this for any model and, in fact, do not even know
where the point is discussed.

If in fact there are different S matrices obtained from different Greens
functions we would be forced to somewhat modify our interpretations.
 However it does seem to
reflect reality to adopt the point of view that in any experiment
we will only say that a particle is observed if it is detected in some way.
The detection process must refer to the   interaction with some classical
external apparatus. The word ``observable'' is used indiscriminately  in
speaking of the operators $O_j$ used in the correlation functions, but in
actual practice we can only observe those    operators for which apparatus
can be manufactured that will couple to it. It would seem that at least
in terms of Greens functions that it is an act of some philosophic subtlety
to speak of a particle in the abstract without considering a concrete
measurement and it does not  seem excluded that different measuring devices
could in fact detect different objects.

Particles are not things of direct sensory perception but are rather
philosophic abstractions. Their existence is the existence of the abstract and
no matter how appealing it is at times to think of them as small billiard balls
such a  concept is not adequate. We will return to this in subsequent lectures.

\bigskip
\newsec{Statistical Mechanics and Quantum Field Theory Compared}

I have now provided one answer to the question of the connection of
statistical mechanics and quantum field theory via the path integral
construction (1.1) and \sm. However to make completely clear the
relation between the two subjects I will close this lecture by pointing
out that the scaling limit has a very different status in statistical
mechanics than it does in quantum field theory.

In the theory of magnets and critical fluids the notion of the atomic
length scale is very real and physical. Moreover the critical
temperature $T_c$ is directly observable. There are many measurements
of specific heats, order parameters, and magnetic susceptibilities for
arbitrary temperatures away from $T_c.$ These non critical (and
nonuniversal) properties of real statistical mechanical systems are
very familiar. What is difficult to measure in a magnet or a fluid are
the properties in the critical region. Because critical theory is by
definition a limiting theory the question of how close to $T_c$ you
have to go before the formulae of the theory may be applied is of very
great practical concern. The scaling functions are almost never
measured in detail because of the difficulties in precisely
controlling $T-T_c$. For an in depth discussion of these problems
the reader is referred to ~\rmt.

In quantum field theory, on the other hand the situation is exactly reversed.
Here the scaling functions are the measured scattering amplitudes whereas
the order parameters and the correlation lengths are the
renormalization constants which can never be measured. Indeed because
the bare theory and the physical theory are cut off by an infinite
renormalization of scale it is not totally clear how unique the
concept of the bare fields are. Even if in some sense it is unique it
is most doubtful that it can ever be reconstructed from a finite
number of measurements. We will return to this question of a
practical reconstruction theory in the next lecture.

The spirit of the relation of the scaling to the
bare theory in both SM and QFT is
captured by the old children's rhyme:

{}~~~~~I met a man upon the stairs.

{}~~~~~The little man who was not there.

{}~~~~~He wasn't there again today.

{}~~~~~My God, I wish he'd go away.

The difference between the subjects is that in statistical mechanics
the critical theory is the little man whereas in quantum field theory
he is the bare theory.


\bigskip
\bigskip



\vfill

\eject
\listrefs

\vfill\eject

\global\secno=0

\global\meqno=1

\global\subsecno=0

\global\refno=1


%


\nref\rlenz{W. Lenz, Phys, ZS. 21 (1920) 613.}
\nref\rising{E. Ising, Beitrag zur theorie des ferromagnetismus, Z.
Physik 31 (1925) 253.}
\nref\rkw{H.A. Kramers and G.H. Wannier, Statistics of the
two dimensional ferromagnet, I and II, Phys. Rev. 60 (1941) 252 and
263.}
\nref\ronsagera{L. Onsager, Crystal statistics I. A two-dimensional
model with an order disorder transition, Phys. Rev. 65 (1944) 117.}
\nref\rwan{G.H. Wannier, The statistical problem in cooperative
phenomena, Rev. Mod. Phys. 17 (1945) 50.}
\nref\rkauf{B. Kaufman,  Crystal statistics II. Partition function
evaluated by spinor analysis, Phys. Rev. 76 (1949) 1232.}
\nref\rkafons{B. Kaufman and L. Onsager, Crystal statistics III.
Short-range order in a binary Ising lattice, Phys. Rev. 76 (1949)
1244.}
\nref\ronsagerb{L. Onsager, discussion, Nuovo Cimento 6 Suppl.(1949)
261.}
\nref\rhou{R.M.F. Houtappel, Order-disorder in hexagonal lattices,
Physica 16 (1950) 425.}
\nref\ryang{C.N. Yang, The spontaneous magnetization of the
two-dimensional Ising model, Phys. Rev. 85 (1952) 808.}
\nref\rchang{H.Chang, The spontaneous magnetization of a two
dimensional rectangular Ising model, Phys. Rev. 88 (1952) 1422.}
\nref\rlya{C.N. Yang and T.D. Lee, Statistical theory of equations of
state and phase transitions I. Theory of condensation,
Phys. Rev. 87 (1952) 404.}
\nref\rlyb{T.D. Lee and C.N. Yang, Statistical theory of equations of
state and phase transitions II. Lattice gas and Ising models, Phys.
Rev. 87 (1952) 410.}
\nref\rsz{ G. Szeg{\"o}, On certain hermitian forms associated with the Fourier
series of a positive function, {\it Communications du S{\'e}minaire
math{\'e}matique de l'Universit{\'e} de Lund}, Tome Suppl{\'e}mentaire (1952)
d{\'e}di{\'e} a Marcel Reisz, 228.}
\nref\rkw{M. Kac and J.C. Ward, Combinatorial solution of the
2-dimensional Ising model, Phys. Rev. 88 (1952) 1332.}
\nref\rnm{G.F. Newell and E.W. Montroll, On the theory of the Ising model of
ferromagnetism, Rev. Mod. Phys. 25 (1953) 353.}
\nref\rkac{M. Kac, Toeplitz matrices, transition kernels and a related
problem in probability theory, Duke Math.  J. 21 (1954) 501.}
\nref \rpw{R.B. Potts and J.C. Ward, The combinatorial method and the
two-dimensional Ising model, Prog. Theor. Phys. (Kyoto) 13 (1955) 38.}
\nref\rhg{C.A. Hurst and H.S. Green, New solution of the Ising problem
for a rectangular lattice, J. Chem. Phys. 33 (1960) 1059.}
\nref\rkasa{P.W. Kasteleyn, The statistics of dimers on a lattice, the
number of dimer arrangements on a quadratic    lattice, Physica 27
(1961) 1209.}
\nref\rkasb{P.W. Kasteleyn, Dimer statistics and phase transitions, J.
Math. Phys. 4 (1963) 287.}
\nref\rmpwaa{E.W. Montroll, R.B. Potts and J.C. Ward, Correlations and
spontaneous magnetization of the two-dimensional Ising model, J.
Math. Phys. 4 (1963) 308.}
\nref\rsml{T.D. Schultz, D.C. Mattis and E.H. Lieb, Two-dimensional
Ising model as a soluble problem of many fermions, Rev. Mod. Phys. 36
(1964) 856}
\nref\rgh{H.S. Green and C.A. Hurst, {\it Order disorder phenomena},
Interscience, London (1964)}
\nref\rstepa{J. Stephenson, Ising model spin correlations on
triangular lattices,
J. Math. Phys. 5 (1964) 1009.}
\nref\rwu{T.T. Wu, Theory of Toeplitz determinants and the spin
correlations of the two-dimensional Ising model, Phys. Rev. 149 (1966)
380.}
\nref\rkada{L.P. Kadanoff, Scaling laws for Ising models near $T_c$,
Physics 2 (1966) 267.}
\nref\rkadb{L.P. Kadanoff, Spin-spin correlations in the
two-dimensional Ising model, Nouvo Cim. 44B (1966) 276.}
\nref\rstepb{J. Stephenson, Ising model spin correlations on
triangular lattices II: Fourth order correlations,
J. Math. Phys. 7 (1966) 1123.}
\nref\rmwa{B.M. McCoy and T.T. Wu, Theory of Toeplitz determinants and the spin
correlations of the two dimensional Ising model II, Phys. Rev. 155
(1967) 438.}
\nref\rcw{H. Cheng and T.T. Wu, Theory of Toeplitz determinants and
the   spin correlations in the two-dimensional Ising model III, Phys. Rev.
162 (1967) 436.}
\nref\rmwb{ B.M.McCoy and T.T. Wu, Theory of Toeplitz determinants and spin
correlations of the two-dimensional Ising model IV, Phys. Rev. 162
(1967) 436.}
\nref\rff{M.E. Fisher and A.E. Ferdinand, Interfacial, boundary and
size effects at critical points, Phys. Rev. Letts. 19 (1967) 169.}
\nref\rhecht{R. Hecht, Correlation functions for the two-dimensional
Ising model, Phys. Rev. 158 (1967) 557.}
\nref\rmwc{B.M. McCoy and T.T. Wu, Theory of Toeplitz determinants and
the spin correlations of the two-dimensional Ising model V, Phys.
Rev. 174 (1968) 546.}
\nref\rmwd{B.M. McCoy and T.T. Wu, Random impurities as the cause of
smooth specific heats near the critical temperature, Phys. Rev. Letts.
21 (1968) 631.}
\nref\rmwe{B.M. McCoy and T.T. Wu, Theory of a two dimensional Ising
model with random impurities, Phys. Rev. 176 (1968) 631.}
\nref\pinka{D.A. Pink, Three--site correlation function of the two--dimensional
Ising model, Can. J. Phys. 46 (1968) 2399.}
\nref\rgri{R.B. Griffiths, Nonanalytic behavior above the critical
point in a random Ising ferromagnet, Phys. Rev. Letts. 23 (1969) 17.}
\nref\rmwf{B.M. McCoy and T.T. Wu, Theory of a two-dimensional Ising
model with random impurities II. Spin correlation functions, Phys. Rev.
188 (1969) 982.}
\nref\rbma{B.M. McCoy, Incompleteness of the critical exponent
description for ferromagnetic systems containing random impurities,
Phys. Rev. Letts. 23 (1969) 383.}
\nref\rbmb{B.M. McCoy, Theory of a two--dimensional Ising model with
random impurities III. Boundary effects, Phys. Rev. 188 (1969) 1014.}
\nref\rffb{A.E. Ferdinand and M.E. Fisher, Bounded and inhomogeneous
Ising models I. Specific heat anomaly of a finite lattice, Phys. Rev.
185 (1969) 832.}
\nref\rfis{M.E. Fisher, Aspects of critical phenomena,
  J. Phys. Soc. Jpn. Suppl. 26 (1969) 87.}
\nref\rkadd{L.P. Kadanoff, Correlations along a line in the
two--dimensional Ising model, Phys. Rev. 188 (1969)
859.}\nref\rpinkb{D.A. Pink, Row correlation functions of the
two-dimensional Ising model, Phys. Rev. 188 (1969) 1032.}
\nref\rbmc{B.M. McCoy, Theory of a two--dimensional Ising model with
random impurities IV, Phys. Rev. B2 (1970) 2795.}
\nref\rstepc{J. Stephenson, Ising model spin correlations on
triangular lattices III: Isotropic antiferromagnetic lattices,
J. Math. Phys.11 (1970) 413.}
\nref\rstepd{J. Stephenson, Ising model spin correlations on
triangular lattices IV: Anisotropic ferromagnetic and antiferromagnetc
lattices,
J. Math. Phys. 11 (1970) 420.}
\nref\ronsagertwo{L. Onsager, in {\it Critical Phenomena in Alloys,
Magnets and Superconductors}, edited by R. Mills, E. Ascher and R. Jaffee
(McGraw Hill, New York 1971) 3.}
\nref\rkc{L.P. Kadanoff and H. Ceva, Determination of an operator
algebra for the two-dimensional Ising model, Phys. Rev. B3 (1971)
3918.}
\nref\rweg{F.J. Wegner, Duality in generalized Ising models and phase
transitions without local order parameters, J. Math. Phys. 12 (1971)
2259.}
\nref\rsuz{M. Suzuki, Equivalence of the two-dimensional Ising model
to the ground state of the linear XY model, Phys. Lett. A34 (1971)
94.}
\nref\reytan{E. Barouch, On the two-dimensional Ising model with random
impurities, J. Math. Phys. 12 (1971) 1577.}
\nref\rsm{M.J. Stephen and L. Mittag, A new representation of the solution
of the Ising model, J. Math Phys. 13 (1972) 1944.}
\nref\rbook{B.M. McCoy and T.T. Wu, {\it The Two dimensional Ising
Model}, Harvard Univ. Press, Cambridge, (1973).}
\nref\rbmw{E. Barouch, B.M. McCoy, and T.T. Wu, Zero-field
susceptibility of the two-dimensional Ising model near $T_c$, Phys.
Rev. Letts. 31 (1973) 1409.}
\nref\rmta{C.A. Tracy and B.M. McCoy, Neutron scattering and the
correlation functions of the Ising model near $T_c$, Phys. Rev. Letts.
31 (1973) 1500.}
\nref\raya{H. Au-Yang,  Thermodynamics of an anisotropic boundary of a
two-dimensional Ising model, J. Math. Phys. 14 (1973) 937.}
\nref\rayma{H. Au-Yang and B.M. McCoy, Theory of layered Ising models
I. Thermodynamics, Phys. Rev. B10 (1974) 886.}
\nref\raym2{H. Au-Yang and B.M. McCoy, Theory of layered Ising models
II. Spin correlation functions, Phys. Rev. B10 (1974) 3885.}
\nref\rmtb{C.A. Tracy and B.M. McCoy, An examination of
phenomenological scaling functions used for critical scattering, Phys.
Rev. B12 (1975) 368.}
\nref\rbdi{R. Balian, J.M. Drouffe and C. Itzykson, Gauge fields on a
lattice II. Gauge invariant Ising model, Phys. Rev. D11
(1975) 2098.}
\nref\rayf{H. Au-Yang and M.E. Fisher, Bounded and inhomogeneous Ising
models II. Specific-heat scaling for a strip, Phys. Rev. 11 (1975) 3469.}
\nref\rfisau{M.E. Fisher and H. Au-Yang, Regularly spaced defects in
Ising models, J. Phys. C8 (1975) L418.}
\nref\rwmtb{T.T. Wu, B.M. McCoy, C.A. Tracy and E. Barouch,  Spin-spin
correlation functions for the two-dimensional Ising model: exact
theory in the scaling region, Phys. Rev.
B13 (1976) 315.}
\nref\rayff{H. Au-Yang, M.E. Fisher and A.E. Ferdinand, Bounded and
inhomogeneous Ising models. III. Regularly spaced point defects, Phys.
Rev. B13 (1976) 1238.}
\nref\ray{H. Au-Yang, Bounded and inhomogeneous Ising models. IV.
Specific-heat amplitude for regular defects, Phys. Rev. B13 (1976) 1266.}
\nref\rbara{R. Z. Bariev, Higher order correlation
functions for the  planar Ising model,
Physica 83A (1976) 388.}
\nref\rmtwa{B.M. McCoy, C.A. Tracy and T.T. Wu, Painlev{\'e} equations of
the third kind, J. Math. Phys. 18 (1977) 1058.}
\nref\rmtwb{B.M. McCoy, C.A. Tracy and T.T. Wu, Two-dimensional Ising
model as an exactly relativistic quantum field theory. Explicit
formulas for the $n$-point functions, Phys. Rev. Letts. 38 (1977) 793.}
\nref\rsmjaa {M. Sato, T. Miwa and M. Jimbo, Studies on holonomic
quantum fields I, Proc. Jpn. Acad. 53A (1977) 6}
\nref\rsmjab{M. Sato, T. Miwa and M. Jimbo, Studies on holonomic
quantum fields II, Proc. Jpn. Acad. 53A (1977) 147.}
\nref\rsmjac{M. Sato, T. Miwa and M. Jimbo, Studies on holonomic
quantum fields III--XII, Proc. Jpn. Acad. 53A (1977) 153, 183, 219; 54A
(1978) 36, 221, 263, 309; 55A (1979) 6, 73.}
\nref\rhamm{J.R. Hamm, Regularly spaced blocks of impurities in the
Ising model; critical temperature and specific heat, Phys. Rev. B15
(1977) 5391.}
\nref\rayj{H. Au--Yang,  Multispin correlation functions for Ising
models, Phys. Rev. B15 (1977) 2704.}
\nref\rayk{H. Au--Yang, Four--spin correlation function along the
diagonal,  Phys. Rev. B16 (1977) 5016.}
\nref\rbanitz{M. Bander and C. Itzykson, Quantum--field theory
calculation of the two--dimensional Ising model correlation function,
Phys. Rev. D15 (1977) 463.}
\nref\rzuIt{J.B. Zuber and C. Itzykson, Quantum field theory and the
two--dimensional Ising model, Phys. Rev. D15 (1977) 2874.}
\nref\rwil{D. Wilkinson, Continuum derivation of the Ising model
two-point function, Phys. Rev. D17 (1978) 1629.}
\nref\rmwg{B.M. McCoy and T.T. Wu, Two-dimensional Ising field theory
for $T<T_c$: String structure of the three point function, Phys. Rev.
D18 (1978) 1243.}
\nref\rmwh{B.M. McCoy and T.T. Wu, Two-dimensional Ising field theory
for $T<T_c$: Green's function strings in $n$--point functions, Phys. Rev.
D18 (1978) 1253}
\nref\rmwi{B.M. McCoy and T.T. Wu, Two-dimensional Ising field theory in
a magnetic field:
breakup of the cut in the 2--point function, Phys. Rev. D128 (1978)
1259.}
\nref\rmwj{B.M. McCoy and T.T. Wu, Two-dimensional Ising model near $T_c$:
Approximation for small magnetic field, Phys. Rev. B18 (1978) 4886.}
\nref\rsmjb{M. Sato, T. Miwa and M. Jimbo, Holonomic
quantum fields I., Pub. Res. Math. Sci. 14 (1978) 223.}
\nref\rbaxc{R.J. Baxter, Solvable eight-vertex model on an arbitrary
planar lattice,  Phil. Trans. Roy. Soc. (London) 289 (1978) 315.}
\nref\rbe{R.J. Baxter and I.G. Enting, 399th solution of the Ising
model, J. Phys. A 11 (1978) 2463.}
\nref\rbarb{R.Z. Bariev, Higher order correlation functions of the
planar Ising model II, Physica 93A (1978) 354.}
\nref\rrichban{J.L. Richardson and M. Bander, $n$--spin correlation
functions for the two--dimensional Ising model, Phys.
Rev. B17 (1978) 1464.}
\nref\rjmlett{M. Jimbo and T. Miwa, Studies on holonomic quantum
fields XIII,XIV, Proc. Acad. Jpn. (1979) 115, 157.}
\nref\rsmjc{M. Sato, T. Miwa, and M. Jimbo, Holonomic quantum
fields II. The Riemann--Hilbert problem, Pub. Res. Math. Sci. 15
(1979) 201}
\nref\rsmjd{M. Sato. T. Miwa and M. Jimbo, Holonomic fields
III, Pub. Res. Math. Sci. 15 (1979) 577.}
\nref\rsmje{M. Sato, T. Miwa and M. Jimbo, Holonomic fields
IV, Pub. Res. Math. Sci. 15 (1979) 871.}
\nref\rbard{R.Z. Bariev, Effect of linear defects on the local
magnetization of a plane Ising lattice, Sov. Phys. JETP, 50 (1979) 613.}
\nref\rbarc{R.Z. Bariev, Correlation functions of the semi-infinite
two-dimensional Ising model I. Local magnetization, Theo. and Math.
Phys. 40 (1979) 623.}
\nref\rmwk{B.M. McCoy and T.T.Wu, Lattice renormalization of non
perturbative quantum field theory, in Bifurcation Phenomena in
Mathematical Physics, ed C. Bardos and D. Bessis, (1980) 69.}
\nref\rmwl{B.M. McCoy and T.T. Wu, Nonlinear partial difference
equations for the
two-dimensional Ising model, Phys. Rev. Letts. 45 (1980) 675.}
\nref\rperka{J.H.H. Perk, Equations of motion for the transverse
correlations of the one dimensional $XY$ model at finite temperature,
Phys. Letts.A79 (1980) 1.}
\nref\rperkb{J.H.H. Perk, Quadratic identities for the Ising model,
Phys. Letts. A79 (1980) 3.}
\nref\rmp{B.M. McCoy and J.H.H. Perk, Two spin correlation functions
of an Ising model with continuous exponents, Phys. Rev. Letts. 44
(1980) 840.}
\nref\rmwm{B.M. McCoy and T.T. Wu, Some properties of the non-linear
partial difference equation for the Ising model correlation function, Phys.
Lett. A80 (1980) 159.}
\nref\rsmjf{M. Sato, T. Miwa and M. Jimbo, Holonomic fields
V, Pub. Res. Math. Sci. 16 (1980) 531.}
\nref\rbarf{R.Z. Bariev, Correlation functions of the semi--infinite
two--dimensional Ising model II. Two--point correlation functions,
Theo. Math. Phys. 42 (1980) 173.}
\nref\rjm{M. Jimbo and T. Miwa, Studies on holonomic quantum fields
XVII, Proc. Japan Acad. 56A (1980) 405; 57A
(1981) 347.}
\nref\rayfisd{H. Au-Yang and M.E. Fisher, Wall effects in critical
systems: Scaling in Ising model strips, Phys. Rev. B 21 (1980) 3956.}
\nref\rmwn{B.M. McCoy and T.T. Wu, Non-linear partial difference
equations for the two-spin correlation function of the two dimensional
Ising model, Nucl. Phys. B 180[FS2] (1981) 89.}
\nref\rmpw{B.M. McCoy,
J.H.H. Perk and T.T. Wu, Ising field theory: quadratic difference
equations for the $n$-point Green's functions on the lattice, Phys. Rev.
Letts. 46 (1981) 757.}
\nref\rpt{J. Palmer and C. Tracy, Two-dimensional Ising correlation
functions: convergence of the scaling limit, Adv. in Appl. Math. 2
(1981) 329.}
\nref\rsal{S. Salihoglu, Two-dimensional Ising model and Ising field theory
for $T>T_c$ in a small magnetic field: shift in the pole in the two
point function, Phys. Rev. D24 (1981) 3255.}
\nref\rhilvl{H.J. Hilhorst and J.M.S. van Leeuwen, Nonuniversal and
anomalous decay of boundary spin correlation in inhomogeneous ising
systems, Phys. Rev. Letts. 47 (1981) 1188.}
\nref\rayl{H. Au--Yang, Four--spin correlation function of two
parallel nonaligned pairs in the planar Ising model,
Phys. Rev. B 23 (1981) 4749.}
\nref\rbax{R.J. Baxter, {\it Exactly Solved Models in Statistical
Mechanics}, Academic Press, London (1982).}
\nref\rit{C. Itzykson, Ising fermions (I), Two dimensions,
Nucl. Phys. B210(FS6) (1982) 448.}
\nref\rmya{B.M. McCoy and M-L. Yan, Gauge Invariant correlation
functions for the Ising-gauge Ising-Higgs system in 2-dimensions,
Nucl. Phys. B215[FS7] (1983) 278.}
\nref\rrptb{J. Palmer and C. Tracy, Two-dimensional Ising
correlation functions: the SMJ analysis, Adv. in Appl. Math. 4 (1983) 46.}
\nref\raypa{H. Au-Yang and J.H.H. Perk, Ising correlations at the
critical temperature, Phys. Lett. A104 (1984) 131.}
\nref\rpcqn{J.H.H. Perk, H.L. Capel, G.R.W. Quispel and F.W. Nijhoff,
Finite temperature correlations for the Ising chain in a transverse
field, Physica 123A (1984) 1.}
\nref\rbuguhivl{T.W. Burkhardt, I. Guim, H.J. Hilhorst and J.M.J. van
Leeuwen, Boundary magnetization and spin correlations in inhomogeneous
two-dimensional Ising systems, Phys. Rev. B 30 (1984) 1486.}
\nref\rblhi{H.W.J. Bl{\"o}te and H.J. Hilhorst, Surface critical
behavior of the smoothly inhomogeneous planar Ising model: The
Pfaffian method, J. Phys. A18 (1985) 3030.}
\nref\rmyb{B.M. McCoy and M--L. Yan, Energy density correlation
function of the Ising model in a
small magnetic field for $T>T_c$, Nucl. Phys. B257[FS14](1985) 303.}
\nref\rkaup{L--F Ko, H. Au-Yang and J.H.H. Perk, Energy density
correlation functions in the two-dimensional Ising model with a line
defect, Phys. Rev. Letts. 54 (1985) 1091.}
\nref\raypb{H. Au-Yang and J.H.H. Perk, Toda lattice equation and
Wronskians in the 2d Ising model, Physica A18 (1986) 365.}
\nref\rkaypa{X.P. Kong, H. Au-Yang and J.H.H. Perk, New Results for
the susceptibility of the two-dimensional Ising model at criticality,
Phys. Letts. A 116 (1986) 54.}
\nref\rkaupb{X.P. Kong, H. Au--Yang and J.H.H. Perk, Logarithmic singularities
of $Q$--dependent
susceptibility of $2$--D Ising model, Phys. Letts. A 118 (1986)
336.}
\nref\rkaupc{X.P. Kong, H. Au-Yang and J.H.H. Perk, Comment on a paper
by Yamada and Suzuki, Prog. Theor. Phys. 77 (1987) 514.}
\nref\rmpb{B.M. McCoy and J.H.H. Perk, The relation of conformal field
theory and deformation theory for the Ising model, Nucl. Phys.
B285[FS19] (1987) 279.}
\nref\raypc{H. Au-Yang and J.H.H. Perk, Critical correlations in a Z
invariant inhomogeneous Ising model, Physica A 144 (1987) 44.}
\nref\rshanmur{R. Shankar and G. Murthy, Nearest--neighbor frustrated
random bond model in $d=2$: Some exact results, Phys. Rev. B36 (1987) 536.}
\nref\rtra{C.A. Tracy, Universality class of a Fibonacci Ising model, J. Stat.
Phys. 51 (1988) 481.}
\nref\rtrb{C.A. Tracy, Universality classes of some aperiodic Ising
models, J. Phys. A 21 (1988) L603.}
\nref\rayptheta{H. Au--Yang and J.H.H. Perk, Solutions of Hirota's
discrete--time Toda lattice equation and the critical correlations in
the Z--invariant Ising model, Proc. of Symp. in Pure Math. 49 (1989)
partI, 287.}
\nref\nzam{A.B. Zamolodchikov, Integrable field theory from conformal
field theory, Advanced Studies in Pure Mathematics 19 (1989) 641.}
\nref\rbard{R.Z. Bariev, Correlation functions in a two-dimensional
Ising model with a large scale inhomogeneity caused by a line defect,
J. Phys. A 22 (1989) L397.}
\nref\rtrc{C.A. Tracy, Asymptotics of $\tau$-function arising in the
two-dimensional Ising model, Comm. Math. Phys. 142 (1991)
297.}
\nref\rbt{E.L. Basor and C.A. Tracy, Asymptotics of a tau-function and
Toeplitz determinants with a singular generating function, Int. J. Mod.
Phys. A7, Suppl 1A (1991) 83.}
\nref\rfis{D. Fisher, Random transverse field Ising spin chains,
Phys. Rev. Letts. 69 (1992) 534.}
\nref\rjbt{J. Palmer, M. Beatty and C.A. Tracy, Tau-functions for the
Dirac operator on the Poincar{\'e} disk, to appear in Comm. Math. Phys.}
\nref\rvgri{G. von Gehlen and V. Rittenberg, $Z_n$-symmetric quantum chain
with an infinite set of conserved charges and $Z_n$ zero modes,
Nuclear Physics B257 [FS14] (1985) 351.}
\nref\rdogr{L. Dolan and M. Grady, Conserved charges from self
duality,  Phys. Rev. D 25 (1982) 1587.}
\nref\rptheta{J.H.H. Perk, Star--triangle equations, quantum Lax pairs
and higher genus curves, Proc. of Symp. in Pure Math. 49, part 1 (1989) 341.}
\nref\rdaviesa{B. Davies, Onsager's algebra and superintegrability,
 J. Phys. A23 (1990) 2245.}
\nref\rdaviesb{B. Davies, Onsager's algebra and the Dolan-Grady condition
in the non-self-dual case, J. Math. Phys. 32 (1991) 298.}
\nref\rjimi{M. Jimbo, K, Miki, T. Miwa and A. Nakayashiki,
Correlation functions of the XXZ model for $\Delta<-1,$ Phys. Letts A170
1992) 255. }
\nref\rbaxcp{R.J. Baxter, Free energy of the solvable chiral Potts model,
J. Stat. Phys. 52 (1988) 639.}
\nref\rpain{P. Painlev{\'e}, Sur les {\'e}quations diff{\'e}rentielles
du second ordre et d'ordre superieur dont l'integral g{\'e}n{\'e}ral
est uniform, Acta Math. 25 (1902) 1.}
\nref\rince{E.L. Ince, {\it Ordinary Differential Equations}, Dover
New York (1956).}
\nref\roka{K. Okamoto, Polynomial Hamiltonians associated with
Painlev{\'e} equations I, Proc. Jpn. Acad. 56 (1980) 264.}
\nref\rokb{K. Okamoto, Polynomial Hamiltonians associated with
Painlev{\'e} equations II.
Differential Equations satisfied by polynomial Hamiltonians, Proc.
Jpn. Acad. 56 (1980) 367.}
\nref\rjmu{M. Jimbo, T. Miwa and K. Ueno, Monodromy preserving
deformation of linear ordinary differential equations with rational
coefficients I, General theory and $tau$--function,
Physica 2D (1981) 306.}
\nref\rjmb{M. Jimbo and T. Miwa, Monodromy preserving deformation of
ordinary differential equations with rational coefficients II,
 Physica 2D (1981) 407.}
\nref\rjmc{M. Jimbo and T. Miwa, Monodromy preserving deformation of
ordinary differential equations with rational coefficients III,
 Physica 4D (1981) 26 .}
\nref\rokc{K. Okamoto, On the $\tau$--function of the Painlev{\'e}
equations,  Physica 2D (1981) 525.}
\nref\rbpza{A.A. Belavin, A.M. Polyakov and A.B. Zamolodchikov,
Infinite conformal field theory of critical fluctuations in two dimensions,
J. Stat. Phys. 34 (1984) 763.}
\nref\rbpzb{A.A. Belavin, A.M. Polyakov and A.B. Zamolodchikov,
Infinite conformal symmetry in two--dimensional quantum field theory,
Nucl. Phys. B241 (1984) 333.}
\nref\rtoda{M. Toda, Vibration of a chain with nonlinear interaction,
J. Phys. Soc. Jpn. 22 (1967) 431.}
\nref\rhir{R. Hirota, Nonlinear partial difference equations. II.
Discrete-time Toda equation, J. Phys. Soc. Jpn. 43 (1977) 2074.}
\nref\rjmms{M. Jimbo, T. Miwa, Y. Mori and M. Sato, Density matrix of
an impenetrable Bose gas and the fifth Painlev{\'e} transcendent,
Physica D (1980) 80.}
\nref\rctwa{D.B. Creamer, H.B. Thacker and D. Wilkinson, Some exact
results for the two--point function of an integrable quantum field
theory, Phys. Rev. D23 (1981) 3081.}
\nref\rctwa{D.B. Creamer, H.B. Thacker and D. Wilkinson, A study of
correlation functions for the delta--function Bose gas, Physica 20D
(1986) 155.}
\nref\rmtang{B.M. McCoy and S. Tang, Connection formulae for
Painlev{\'e} V functions. II The $\delta$ function Bose gas problem,
Physica 20D (1986) 187.}
\nref\riiks{A.R. Its, A.G. Izergin, V.E. Korepin and N.A. Slavnov,
Differential equations for quantum correlation functions, Int. J. Mod.
Phys. S4 (1990) 1003.}
\nref\rksa{V.E. Korepin and N.A. Slavnov, The time dependent
correlation function of an impenetrable Bose gas as a Fredholm minor.I,
Comm. Math. Phys. 129 (1990) 103.}
\nref\rksb{V.E. Korepin and N.A. Slavnov, Time dependence of the
density--density temperature correlation function of a
one--dimensional Bose gas, Nucl. Phys. B340 (1990) 759.}
\nref\iika{A.R. Its, A.G. Izergin and V.E. Korepin, Space correlations
in the one--dimensional impenetrable Bose gas at finite temperature,
Physica D 53 (1991) 187.}
\nref\rksc{V.E. Korepin and N.A. Slavnov, Correlation function of
fields in one--dimensional Bose--gas, Comm. Math. Phys. 136 (1991)
633.}
\nref\riikb{A.R. Its, A.G. Izergin and V.E. Korepin, Large time and
distance asymptotics of the temperature field correlator in the
impenetrable Bose gas, Nucl. Phys. B348 (1991) 757.}
\nref\riikc{A.R. Its, A.G. Izergin and V.E. Korepin, Time and
temperature dependent correlation function of impenetrable Bose gas
field correlator in the impenetrable Bose gas, Contp. Math. 122, (1991) 61.}
\nref\rkbi{V.E. Korepin, N.M. Bogoliubov and A.G. Izergin, {\it
Quantum Inverse Scattering Method and Correlation Functions}, Camb.
Univ. Press, (1993).}
\nref\rberka{A. Berkovich and J.H. Lowenstein, Correlation function of
the one-dimensional Fermi gas in the infinite--coupling limit
(repulsive case), Nucl. Phys. B 285 (1987) 70.}
\nref\rberkb{A. Berkovich, Temperature and magnetic
field--dependent correlation of the exactly integrable
$(1+1)$--dimensional gas of impenetrable fermions, J. Phys. A 24 (1991) 1543.}
\nref\rlsm{E. Lieb, T. Schultz and D. Mattis,  Two soluble models of
an antiferromagnetic chain, Ann. Phys. 16 (1961)
407.}
\nref\rcp{H.W. Capel and J.H.H. Perk, Autocorrelation function of the
$x$--component of the magnetization in the one--dimensional
$XY$--model, Physica 87A (1977) 211.}
\nref\rpeca{J.H.H. Perk and H.W. Capel,  Time--dependent
$xx$--correlation functions in the one--dimensional $XY$ model,
Physica 89A (1977) 265.}
\nref\rpecb{J.H.H. Perk and H.W. Capel, Transverse correlations in the
inhomogeneous one dimensional $XY$--model at infinite temperature,
 Physica 92A (1978) 163.}
\nref\rpecc{J.H.H. Perk and H.W. Capel, Time-- and
frequency--dependent correlation functions for the homogeneous and
alternating $XY$--models, Physica 100A (1990) 1.}
\nref\rmpsa{B.M. McCoy, J.H.H.Perk and R.E. Shrock, Correlation
functions of the transverse Ising chain at the critical field, Nucl.
Phys. B220[FS8] (1983) 35.}
\nref\rmpsb{B.M. McCoy. J.H.H.Perk and R.E. Schrock, Correlation
functions of the transverse Ising chain at the critical temperature
for large temporal and spatial separations, Nucl. Phys. B220[FS8]
(1983).}
\nref\rmtangb{B.M. McCoy and S. Tang, Connection formulae for
Painlev{\'e} V functions, Physica 19A (1986) 42.}
\nref\iiks{A.R. Its, A.G. Izergin, V.E. Korepin and V.Ju. Novokshenov,
temperature autocorrelations of the transverse Ising chain at the
critical magnetic field, Nucl. Phys. B340 (1990) 752.}
\nref\rcikt{F. Colombo, A.G. Izergin, V.E. Korepin and V. Tognetti,
Correlators in the Heisenberg $XX0$ chain as Fredholm determinants,
Phys. Letts. A 169 (1992) 243.}
\nref\rdezh{P. Deift and X. Zhou, Courant Math. Inst. Report (1992).}
\nref\riiksc{A.R. Its, A.G. Izergin, V.E. Korepin and N.A. Slavnov,
Temperature correlations of quantum spins, Phys. Rev. Letts. 70 (1993)
1704.}
\nref\rikt{H. Itoyama, V.E. Korepin and H.B. Thacker, Fredholm
determinant representation of quantum correlationfunction for
sine--Gordon at special value of coupling constant, Mod. Phys. Letts.
B6 (1992) 1405.}
\nref\rbele{D. Bernard and A. Le Clair, Differential equations for
Sine--Gordon correlation functions at the free fermion point, CLNS
94/1276,SPhT--94--021 (1994).}
\nref\rmehta{M.L. Mehta, {\it Random Matrices}, second ed. Academic
Press, 1991.}
\nref\rbk{E. Br{\'e}zin and V. Kazakov,  Theories of closed strings,
Phys. Letts. B 236 (1990) 144.}
\nref\rds{M. Douglas and S. Shenker, String theory in less that one
dimension, Nucl. Phys. B335 (1990) 635.}
\nref\rgma{D.J. Gross and A.A. Migdal, Nonperturbative
two--dimensional quantum gravity, Phys. Rev. Letts. 64 (1990) 127.}
\nref\rgmb{D.J. Gross and A.A. Migdal, A nonperturbative treatment of
two--dimensional quantum gravity, Nucl. Phys. B340 (1990) 333.}
\nref\rmoore{G. Moore, Matrix models of 2D gravity and isomonodromic
deformation, Prog. Theo. Phys. Suppl. 102 (1990) 255.}
\nref\rbobr{M.J. Bowick and E. Br{\'e}zin, Universal scaling of the tail of
the density of eigenvalues in random matrix models, Phys. Letts. B268
(1991) 21.}
\nref\rbtw{E.L. Basor, C.A. Tracy and H. Widom, Asymptotics of level
spacing distributions for random matrices, Phys. Rev. Letts. 69 (1992) 5.}
\nref\rmehtab{M.L. Mehta, A non--linear differential equation and a
Fredholm determinant, J. de Phys. I France 2 (1992) 1721.}
\nref\rtrwia{C.A. Tracy and H. Widom, Introduction to random matrices
in {\it Geometric and Quantum Aspects of Integrable Systems}, G.F.
Helminck, ed., Lecture Notes in Physics, 424, Springer--Verlag (Berlin,
1993) 103,}
\nref\rmame{G. Mahoux and M.L. Mehta, Level spacing functions and
nonlinear differential equations, J. Phys. I France 3 (1993) 697.}
\nref\rtrwib{C.A. Tracy and H. Widom, Level--spacing distributions and
the Airy kernel, Phys. Letts. B305 (1993) 115.}
\nref\rtrwinew{C.A. Tracy and H. Widom, Level--spacing distrubutions
and the Airy kernel, Comm. Math. Phys. 159 (1994) 151.}
\nref\rtrwic{C.A. Tracy and H. Widom, Level--spacing distrubutions and
the Bessel kernel, Comm. Math. Phys. (in press).}
\nref\rtrwid{C.A. Tracy and H. Widom, Fredholm determinants,
differential equations and matrix models, Comm. Math. Phys. (in
press).}
\nref\rdysb{F.J. Dyson, The Coulomb fluid and the fifth Painlev{\'e}
transcendent, IASSNSS-HEP--92/43 preprint.}
\nref\rceva{S. Cecotti and C. Vafa, Ising model and $N=2$ supersymmetric
theories, Comm. Math. Phys. 157 (1993) 139.}
\nref\rzinn{J. Zinn--Justin, {\it Quantum Field Theory and Critical Phenomena},
Oxford Univ. Press (1991).}
\nref\rdodo{V.S. Dotsenko and V.S. Dotsenko, Critical behavior of the phase
transition in the 2 dimensional Ising model with impurities,
Adv. Phys. 32 (1983) 129.}
\nref\rzieone{K. Ziegler, Quenched thermodynamics of the random bond
Ising model on the square lattice, Nucl. Phys. B344 (1990) 499.}
\nref\rzietwo{K. Ziegler, Disorder-induced phase in the random bond
Ising lattice on the square lattice, Europhys. Lett. 14 (1991) 415.}
\nref\rziethree{K. Ziegler, The energy-energy correlation function of
the random bond Ising model in two dimensions, Phys. Rev. B (in press).}
\centerline{\bf Part II. Ising Field Theory for $H=0$}



\vskip 13mm

\centerline{\bf Abstract}
\vskip 4mm
We discuss the theory of the two dimensional Ising model at zero magnetic
field in both the homogeneous and inhomogeneous cases. For the homogeneous
case
we discuss in detail the scaling limit construction of the field theory
limit given in the previous lecture and present the relation of
the two spin correlation to Painlev{\'e} functions.
For the inhomogeneous case
we discuss the breakdown of this scaling picture which is caused by the
spreading out of the critical point into a critical region.
\vskip 3mm

%

\newsec{Introduction}

The oldest and most completely studied system in statistical
mechanics which has finite range forces and has a critical point is the
two dimensional Ising model in zero magnetic field. Consequently, by
means of the connection between statistical mechanics presented in the
previous lecture, the Ising model is also the most extensively studied
quantum field theory. The principal purpose of this lecture is the
presentation of the theory of this model. This will concretely illustrate
many of the general principles previously presented.

However the most surprising feature of the Ising model is that there
are many connections with quantum field theory that go beyond the
theory of the first lecture. The Ising model has led to developments
in mathematics which have been widely applied to the theory of random
matrices, to quantum gravity, and the Ising correlations themselves
are directly related to $N=2$ supersymmetric quantum field theory in
two dimensions. Therefore a second purpose of this lecture will be a to
briefly indicate some of these further connections between SM and QFT.

The  two dimensional Ising models we will consider in this lecture
will be various special cases of the general nearest neighbor
inhomogeneous model in a magnetic field specified by the interaction
energy
\eqn\ising{E=-\sum_{j=1}^{L_v}\sum
_{k=1}^{L_h}(E^v(j,k)\sigma_{j,k}\sigma_{j+1,k}+
E^h(j,k)\sigma_{j,k}\sigma_{j,k+1})}
where $\sigma_{j,k}=\pm1$,~j(k) specifies the row (column) of the
lattice, periodic boundary conditions are assumed
 and in the most general inhomogeneous case the interaction
constants $E^v(j,k)$ and $E^h(j,k)$ are allowed to vary from site
to site. We will be particularly concerned with 1) the homogeneous case
where $E^v$ and $E^h$ are independent of position and 2) the case of
layered impurities where $E^v$ and $E^h$ depend only on the row index
$m$ and $H=0$.

For the case of the homogeneous Ising model the discussion of the previous
lecture has given an unambiguous prescription to obtain a field theory by
constructing a scaling limit. The primary goal of this lecture is to carry
out this construction in detail for the Ising model.
In sec. 2 we summarize
briefly the various methods used to solve the Ising model at $H=0$ on the
lattice and in
sec. 3 we present the results of these studies. In sec. 4 we use these results
to
construct the field theory limit and to present the important
relations the spin correlations  have with  Painlev{\'e} functions.
We also indicate how these considerations arise in other field theory
contexts
and we discuss the particle content of the theory. We
conclude in sec. 5 with a discussion of the inhomogeneous model and the
new physics which emerges in this situation.

The Ising  model was first solved in one dimension by E. Ising in 1925 and
for that reason it now bears his name. Since then its study has
involved some of the most creative developments in mathematical
physics. The papers on this model are many starting with the original
work of Ising and continuing to the present day. I have attempted to
list the papers most relevant to the present considerations as items
\rlenz--\rjbt~of the reference section.

In any bibliography there is inevitably some selection criteria that
is applied and I should make explicit the criteria used here. Firstly,
with possibly the exception of~\rbook, I have included only original
material and excluded the large amount of pedagogic writing on the
subject. Secondly I have excluded the many applications of Ising model
theory to physical systems.

There is one further selection that has been made which is more subtle
and needs a bit of explanation. There are two important relations between the
homogeneous Ising model and the XY spin chain in a magnetic field defined by
\eqn\hxy{H_{XY}=-\sum_m({1\over 2}(1+\gamma)\sigma^{x}_m\sigma^{x}_{m+1}+
{1\over 2}(1-\gamma)\sigma^{y}_m\sigma^{y}_{m+1}+h\sigma^{z}_m)}
where $\sigma^x,~\sigma^y$ and $\sigma^z$ are the three Pauli spin matrices.
Both of these relations involve a commutation relation of $H_{XY}$ with a
transfer matrix of the Ising model.

 The first relation is obtained if we consider a row to row transfer matrix
of sec. 7 of part I which may be explicitly written in terms   of
\eqn\tv{T^v=(2 \sinh 2K^v)^{L_h/2}\exp (K^{v*}\sum_{k=1}^{L_h}\sigma^z_k)~~
{\rm and}~~T^h=\exp(K^h\sum_{k=1}^{L_h}\sigma_k^x\sigma_{k+1}^x)}
where $K^{v,h}=E^{v,h}/kT~~~{\rm and}~~\tanh K^{v*}=e^{-2K^v}$
as
\eqn\tr{T^{R''}=(T^v)^{1/2}T^h (T^{v})^{1/2}.}
Then~\rsuz
\eqn\comone{[T^{R''},H_{XY}]=0}
if
\eqn\relation{\cosh 2K^{v*}={1\over \gamma} ~~{\rm and}~~\tanh 2K^h=
(1-\gamma)^{1/2}/h.}
{}From this commutation relation it follows that the correlation of two
Ising spins in the same row are given in terms of XY correlations as
\eqn\isxy{<\sigma_{n,m}\sigma_{n,m'}>=
\cosh^{2}K^{v*}<\sigma^x_m\sigma^x_{m'}>_{XY}
-\sinh^{2}K^{v*}<\sigma^y_{m}\sigma^y_{m'}>_{XY}.}

The second relation is obtained if we consider the diagonal transfer matrix
defined by (I.7.15) as
\eqn\tdig{T^D_{\{\sigma\},\{\sigma '\}}=
\prod_{k=1}^{L_h}e^{K^v\sigma_k\sigma_k'} e^{K^h\sigma^k \sigma_{k+1}'}.}
This commutes with the special case of the XY Hamiltonian~\hxy~ with
\eqn\tising{\gamma=1~~{\rm and}~~h=(\sinh 2K^h \sinh 2K^v)^{-1}}
and thus we find that the correlation of two spins on a diagonal in the Ising
model are given in terms of the XY correlations as
\eqn\istis{<\sigma_{0,0}\sigma_{n,n}>=<\sigma_0^x \sigma_n^x>_{XY}.}

Because of this intimate relation of the XY spin chain with the
two-dimensional Ising model there are some results for the XY model
which will also give results for the Ising model. In this lecture we
make no attempt to be complete in referencing the literature of the XY
model and give only those papers for which there is no analogous Ising
result in the literature. We will return to the XY model particularly
considered as a dynamical system at finite temperature in the fourth
lecture and will give  further XY references there.

\bigskip
\newsec{Methods of Solution for $H=0$}

In the 50 years since Onsager~\ronsagera~ first computed the free energy,
the Ising model at $H=0$ has been solved by many different methods. Some
methods are powerful enough to compute correlation functions	 and
some  can be generalized to solve many other models in two dimensions
but there is no method known at the present time which is both general
and powerful enough to compute all correlation functions of all models.
Because our focus here is on the Ising model itself we will
naturally discuss the most
powerful of the methods. However for the purpose of orientation we
will begin our considerations with a brief sketch of five of the
principal methods of solution in their historical order:

1) Onsager's algebra \ronsagera

2) Fermionization~\rkauf~\rkafons

3) Combinatorial ~\rkw~\rpw~\rhg~\rkasa~\rkasb

4) Star triangle equation and functional equations \rwan~\rbax

5) The 399 Solution~\rbe

\subsec{Onsager's algebra}

Onsager in his original 1944 paper~\ronsagera~ begins his considerations
with the XY model Hamiltonian  ~\tising~ written as
\eqn\xyising{H_{XY}=A_0+hA_1}
where
\eqn\azero{A_0=-\sum_{m=1}^{L_h}\sigma_m^x\sigma_{m+1}^x~~{\rm and}~~
A_1=-\sum_{m=1}^{L_h}\sigma_m^z}
and demonstrates that these two operators are part of a much larger algebra
\eqn\onal{\eqalign{[A_l,A_m]=&4G_{l-m}\cr
[G_l,A_m]=&2A_{m+l}-2A_{m-l}\cr
[G_l,G_m]=&0\cr}}
He also finds the relation which incorporates the fact that the
lattice has $L_h$ sites
\eqn\closing{G_{m+L_h}=-CG_m=-G_mC}
where C is the spin reversal operator
and from these relations alone is able to compute all the eigenvalues
of $H_{XY}$ and the spectrum of the transfer matrix $T^vT^h$ which is seen
to be of the form $e^{a_0 A_0}e^{a_1 A_1}.$ This algebra and method
of solution remained unused and unexplored until it was
found~\rvgri~in 1985 to be related to the special case of the chiral
Potts model which is now called superintegrable which is a generalization
of ~\azero~ from two to $N$ states per site.
The algebra follows from the two special cases~\rdogr~\rptheta
\eqn\aaaa{[A_0,[A_0,[A_0,A_1]]]=16[A_0,A_1]~~{\rm and}~~
[A_1,[,A_1,[A_1,A_0]]]=16[A_1,A_0].}
In greatest generality it has been shown~\rdaviesa~\rdaviesb~ that
the algebra~\onal~is sufficient to guarantee that all the eigenvalues of the
Hamiltonian are in sets containing $(2s)^{j_{max}}$ elements of the form
\eqn\eigen{E=a+hb+\sum_j 4m_j(1+h^2+ha_j)^{1/2}~~{\rm where}~~m_j=-s,-s+1,
\cdots,+s} where $s$ is either integer or half integer
and $j_{max}$ and the $j_{max}$ numbers $a_j$  vary from set to set.
So far the only known representations of the algebra have $s=1/2.$

\subsec{Fermionization}

In order to compute correlations in the Ising model the algebra of
Onsager does not seem to be enough. To compute these correlations
we need the exceptionally powerful method of fermionization
introduced by Kaufman~\rkauf~in 1949. This method is so powerful
that, unlike Onsager's original method, it can be used to obtain
results  on the completely inhomogeneous lattice~\ising~with $H=0$
and $E^v(j,k)$ and $E^h(j,k)$ allowed to vary in an arbitrary
random fashion from point to point. The Ising model is the only
model which has been studied in this enormous degree of generality.

The transfer matrix~\tv~ written in terms of $\sigma_k^i$ is not in a
fermionic form because the $\sigma_k^i$ considered as operators commute
on different sites whereas fermions anticommute. The key to the fermionization
of the Ising model by Kaufman~\rkauf~ is the introduction of anticommuting
operators $\gamma _j$~by means of the Jordan-Wigner
transformation
\eqn\jw{\sigma_k^x=2^{-1/2}(\prod _{l=1}^{k-1} P_l)\gamma_{2j-1},~~
\sigma_k^y=2^{-1/2}(\prod_{l=1}^{k-1} P_l)\gamma_{2k},~~
\sigma^z_k=-i\gamma_{2k-1}\gamma_{2k}=-{1\over 2}P_k}
where
\eqn\anti{\gamma_k \gamma_l+\gamma_l \gamma_k =\delta_{k,l}.}
If we write the completely inhomogeneous transfer matrices
\eqn\intm{T^h_j=\exp(\sum_{k=1}^{L_h}K^h(j,k)\sigma^x_k\sigma^x_{k+1})~~
{\rm and}~~
T^v_{j+1/2}=\exp(\sum_{k=1}^{L_h}K^{v*}(j,k)\sigma^z_k)}
in terms of the $\gamma_j,$ the resulting expressions are still quadratic
in the $\gamma_j$. Consequently the diagonalizing of the transfer
matrix is reduced to the study of free fermions on
an (inhomogeneous ) lattice. Furthermore
since the local correlations of Ising spins are now
expressible in terms of the
correlation of Jordan-Wigner strings of free Fermi operators and since
all free Fermi correlations can be reduced by Wick's theorem to Pfaffians
of two point correlations, it is possible in principle to study all of
the correlations even on completely inhomogeneous lattices.

\subsec{Combinatorial}

The combinatorial method differs from the two preceding methods in that it
does not make use of a transfer matrix and treats the vertical and horizontal
interactions on an equal footing. It starts with the observation that
because $\sigma ^2=1$  we have the identity
\eqn\htem{e^{K\sigma \sigma '}=\cosh K +\sigma \sigma '\sinh K .}
 which may be used to write the partition function as
\eqn\part{\eqalign{Z=&\sum_{\{\sigma\}}e^{\sum_{j=1}^{L_v}\sum_{k=1}^{L_h}
[K^v(j,k)\sigma_{j,k}\sigma_{j+1,k}+K^h(j,k)\sigma_{j,k}\sigma_{j,k+1}]}\cr
=&\prod_{j=1}^{L_v}\prod_{k=1}^{L_h}(\cosh K^v(j,k) \cosh K^h(j,k))\cr
&\sum_{\{\sigma\}}\prod_{j=1}^{L_v}\prod_{k=1}^{L_h}
(1+z_v(j,k)\sigma_{j,k}\sigma_{j+1,k})(1+z_h(j,k)\sigma_{j,k}\sigma_{j,k+1})}}
with $z_i=\tanh K^i(j,k).$ The product on the RHS of~\part~ may be expanded
as a sum of graphs on the square lattice where on each bond either the factor
$1$ or $z_i\sigma \sigma '$ is used. The sum over $\{ \sigma \}$ thus reduces
the partition function to the sum of all closed graphs with  weights of
$z_i(j,k)$ for each occupied bond. The combinatorial method then shows that
this sum may be expressed as the Pfaffian of an antisymmetric matrix and when
the lattice is homogeneous this Pfaffian is easily computed by Fourier
transformation.

Correlation functions are treated in a similar fashion. We start
with the observation that again since $\sigma ^2=1$ any
correlation function such as $<\sigma_{m,n}\sigma_{m',n'}>$ may be written as
\eqn\path{<\sigma_{m,n}\sigma_{m',n'}>=<\prod_{\rm path}\sigma \sigma '>}
where $\sigma \sigma '$ are nearest neighbors and the path is any path which
connects the sites $(m,n)$ and $(m',n')$. Then using~\htem~ we write each
factor on the path as
\eqn\newfac{\sigma \sigma ' e^{K^i\sigma \sigma '}=
\sinh K^i +\sigma \sigma ' \cosh K^i =\sinh K^i (1+ z_i^{-1}\sigma \sigma ')}
and thus correlations are expressed as quotients of a partition function
of a lattice with the bonds $z_i$ on the path replaced by $z_i^{-1}$ divided by
the original partition function. The numerator partition function is
expressed as a Pfaffian exactly as was the original partition function and
hence all correlation functions can in principle be reduced to the
evaluation of determinants.

Even though no explicit use has been made of
fermions, operators, or transfer matrices
this method is actually very close to the fermionization method and if
proper choice
of the arbitrary path is made the two methods can be shown to coincide.

\subsec{Commuting transfer matrices and functional equations}

The diagonal transfer matrix~\tdig~ possesses the very remarkable property
that if we consider two such transfer matrices
$T^D(u)$ and $T^D(u')$ which are obtained from ~\tdig~ by allowing
$K^v$ and $K^h$ to depend on a variable $u$ such that the $h$ of ~\tising~
is held fixed that
\eqn\tcom{[T^D(u), T^D(u')]=0.}
This relation seems to have been known to Onsager and follows from a relation
on the Boltzmann weights~\rwan~\rhou~\rbax~known as the star-triangle equation
that
\eqn\star{\sum_{\sigma_d=\pm
1}e^{-\sigma_d(L_1\sigma_a+L_2\sigma_b+L_3\sigma_c)}=
Re^{-(K_1\sigma_b\sigma_c+K_2\sigma_c\sigma_a+K_3\sigma_a\sigma_b)}}
if
\eqn\cond{\sinh 2K_j \sinh2 L_j =k^{-1}~~{\rm and}~~
R^2=2k\sinh 2L_1 \sinh 2L_2 \sinh 2 L_3}
 Using this star triangle equation
Baxter~\rbax~ demonstrates that the transfer matrix satisfies a simple
functional equation and he solves that functional equation to obtain all
the eigenvalues of the transfer matrix for the homogeneous Ising model.
This method is quite unlike the previous three methods in that it has been
applied to a very large number of two dimensional models.
However it is more restrictive, in a sense, than the fermionization
and combinatorial method, in that it is not able to deal with the
completely inhomogeneous lattice.

This technique has been extended by Baxter to compute the spontaneous
magnetization of the Ising model by means of the introduction of a new
object called the corner transfer matrix.
Moreover, with the exception of
some very recent work~\rjimi~ on the $XXZ$ chain, these techniques have
not yet been  extended to the computation of other  correlation functions.
The hope is that eventually for the homogeneous lattice
 all the results of the fermionization method can be reproduced by
these more general techniques. However further discussion of these
developments is outside the scope of these lectures.

\subsec{The 399th Solution}

The final method of solution~\rbe~ is an ingenious use of the star triangle
equation
to compute the free energy of the Ising model without recourse to
functional equations. This method has been extended to compute the free
energy of the chiral Potts model~\rbaxcp~but thus far it has not been extended
to obtain all the other eigenvalues of the transfer matrix.

\bigskip
\newsec{The homogeneous Ising model at $H=0$}

After this brief discussion of the methods used to obtain results for the
Ising model at $H=0$ we are now able to summarize the results of the
computations and use them to illustrate the scaling theory and field theory
construction outlined in the previous lecture. It should be kept in mind,
however, that in fact much of the general picture of the first lecture was
developed after many of these Ising results were found and that the general
theory
was developed as an attempt to understand the Ising model in a much
wider context. It is fair to say that most of the general theory of critical
phenomena originates in the results we will present below.
\subsec{Free energy}

The first analytic result is the famous computation of Onsager~\ronsagera~ of
the free energy
\eqn\free{\eqalign{F=-kT\{\ln 2 +
{1\over 2}(2\pi)^{-2}\int_0^{2\pi}
&d\theta_1\int_{0}^{2 \pi}d\theta_2 \ln[\cosh2\beta E^h \cosh 2\beta E^v\cr
-&\sinh2 \beta E^h \cos \theta_1 -
\sinh 2 \beta E^v \cos \theta_2]\}}}
where we use $\beta=1/kT.$
This free energy is singular at the unique value of $T=T_c$
determined from the condition
\eqn\tc{\sinh 2E^v/kT_c \sinh2E^h/kT_c =1}
which had been previously found by Kramers and Wannier~\rkw.

The internal energy $u$ is defined as
${\partial \beta F\over \partial \beta}$ and from the definition
of $F$ is written in terms of the nearest neighbor correlations  as
\eqn\inen{u={\partial \beta F\over \partial \beta}=
-E^h<\sigma_{0,0}\sigma_{0,1}>-E^v<\sigma_{0,0}\sigma_{1,0}>.}
Hence from ~\free~ we obtain
\eqn\epsans{<\sigma_{0,0}\sigma_{0,1}>={1\over 2 \pi}\int_0^{2 \pi}d\theta
\left[ {(1-\alpha_1 e^{i\theta})(1-\alpha _2e^{-i\theta})\over
(1-\alpha_1 e^{-i \theta})(1-\alpha_2 e^{i \theta})}\right] ^{1/2}}
with
\eqn\al{\alpha_1=e^{-2K^v}\tanh K^h~~{\rm and}~~\alpha_2= e^{-2K^v}\coth K^h}
and $<\sigma_{0,0}\sigma_{0,1}>=<\sigma_{0,0}\sigma_{1,0}>$
with $K^h$ and $K^v$ interchanged.
At $T_c$ we find
\eqn\ecrit{<\sigma_{0,0}\sigma_{0,1}>=
{2\over \pi}\coth 2\beta _c E^h {\rm gd}2\beta_c E^h}
where ${\rm gd}x=\arctan \sinh x$ is the Gudermannian of $x.$ {}From the
internal energy we obtain the specific heat and find that near $T_c$
\eqn\sheat{c={\partial u\over \partial T}\sim-{ 2k\beta_c^2\over \pi}
(E^{h2}\sinh^2 2\beta_c E^v +2E^v E^h +E^{v2}\sinh^2 2\beta_c E^h)
\ln|1-T/T_c|.}
Thus the specific heat has a logarithmic divergence at $T_c$ which corresponds
to the critical index $\alpha =0$ of the previous lecture.
\subsec{Transfer matrix spectrum }

Onsager~\ronsagera~ also computed the complete excitation spectrum  of
the transfer matrix. He found that the full set of $2^{L^h}$ eigenvalues
is composed of 2 sets given as (for $L^h$ even)
\eqn\eigone{\ln\lambda^{+}={L^h\over 2}\ln(2\sinh 2K^h)+
{1\over 2}\sum_{r=1}^{L^h}\pm\gamma_{2r-1}}
and
\eqn\eigtwo{\ln\lambda^-={L^h\over 2}\ln(2\sinh 2K^h)\pm(K^{v*}-K^h)+
{1\over 2}\sum_{r=1}^{L^h-1}\pm\gamma_{2r}}
where $\gamma _r$ is the positive solution of
\eqn\gamms{\cosh \gamma_r=\cosh 2K^{v*}\cosh 2K^h -
\sinh 2K^{v*}\sinh 2K^h\cos{r\pi\over L^h}.}
The eigenvectors corresponding to $\lambda^+$ ($\lambda^-$) are even (odd)
under spin reflection
and the $\pm$ are chosen independently with the restriction that the
number of $-$ signs is even (odd).
When $T<T_c$ ($K^h>K^{v*}$) the maximum $\lambda^{\pm}$ are asymptotically
degenerate as $L^h\rightarrow \infty$ whereas for $T>T_c$ $\lambda^+$ is the
maximum eigenvalue and is non degenerate.

\subsec{Spontaneous magnetization}

The spontaneous magnetization was announced by Onsager~\ronsagerb~in 1949
and a derivation was given by Yang~\ryang~in 1952. The remarkably simple
result is
\eqn\mag{M_-(T)=(1-(\sinh 2K^h \sinh 2 K^v)^{-2})^{1/8}.}
Consequently the critical exponent $\beta =1/8.$

\subsec{Spin-spin correlations}

The simplest result for the spin-spin correlation is for the diagonal
correlation at $T=T_c$ where we have the remarkably simple expression
\eqn\diag{<\sigma_{0,0}\sigma_{N,N}>=({2\over\pi})^N
\prod_{l=1}^{N-1}[1-{1\over 4l^2}]^{l-N}}
which as $N\rightarrow \infty$ behaves as
\eqn\tccorr{<\sigma_{0,0}\sigma_{N,N}>\sim AN^{-1/4}}
where $A$ is a transcendental constant $\sim .6450\cdots.$ Thus the
critical exponent $\eta =1/4.$

More generally all spin correlation functions may be expressed as
finite determinants~\rmpwaa~\rbook~with a size that depends on the
separation of the spins. The simplest of these determinental expressions are
\eqn\rowcor{<\sigma_{0,0}\sigma_{0,N}>=\left|\,\matrix{
a_0\hfill& a_{-1}\hfill\cdots&a_{-N+1}\cr
a_1\hfill &a_0\hfill\cdots&a_{-N+2}\cr
\,\vdots\hfill&\,\vdots\hfill&\,\vdots\hfill\cr
a_{N-1}&a_{N-2}\cdots&a_{0}\cr}\right|}
with
\eqn\an{a_n={1\over 2 \pi}\int _{-\pi}^{\pi}d\theta e^{-in\theta}
\left[ {(1-\alpha_1 e^{i\theta})(1-\alpha_2e^{-i\theta})\over
(1-\alpha_1e^{-i\theta})(1-\alpha_2e^{i\theta})}\right] ^{1/2}}
and
\eqn\diacorr{<\sigma_{0,0}\sigma_{N,N}>=\left|\,\matrix{
{\tilde a}_0\hfill& {\tilde a}_{-1}\hfill\cdots&{\tilde a}_{-N+1}\cr
{\tilde a}_1\hfill &{\tilde a}_0\hfill\cdots&{\tilde a}_{-N+2}\cr
\,\vdots\hfill&\,\vdots\hfill&\,\vdots\hfill\cr
{\tilde a}_{N-1}&{\tilde a}_{N-2}\cdots&{\tilde a}_{0}\cr}\right|}
with
\eqn\antild{{\tilde a}_n={1\over 2\pi}\int_{-\pi}^{\pi}d\theta e^{-in
\theta}
\left[ {\sinh 2K^v \sinh 2K^h-e^{-i \theta}\over \sinh 2K^v \sinh 2K^h
-e^{i \theta}}\right] ^{1/2}.}
These finite determinental expressions are particularly effective in
studying the correlations for small separations.

The study of the large distance expansions of the two point
function for $T\neq T_c$ is significantly more involved than the
$T=T_c$ expansion. To effectively study this case it is useful to
recognize that there are not one but many determinental expressions for
the correlations and in particular there is an expression as an
infinite (Fredholm) determinant~\rcw~\rbook. From this starting point
a complete discussion of the long distance behavior was made~\rwmtb~ in
1976. We here reproduce the results of that study.

For $T<T_c$ the two point function has the representation
\eqn\twol{<\sigma_{0,0}\sigma_{M,N}>=
M_{-}^2\exp(-\sum_{n=1}^{\infty}F_{M,N}^{(2n)})}
where and
\eqn\flsum{\eqalign{&F_{M,N}^{(2n)}=
(-1)^n[2z_v(1-z_h^2)]^{2n}(2n)^{-1}(2\pi)^{-4n}\cr
&\int_{-\pi}^{\pi}d\phi_1\cdots\int_{-\pi}^{\pi}
d\phi_{4n}
\prod_{j=1}^{2n}\left( {e^{-iM\phi_{2j-1}-iN\phi_{2j}}
\sin{1\over2}(\phi_{2j-1}-\phi_{2j+1})\over
\Delta(\phi_{2j-1},\phi_{2j})\sin{1\over
2}(\phi_{2j}+\phi_{2j+2})}\right) .}}
Here $\phi_{4n+1}=\phi_1~~\phi_{4n+2}=\phi_2, {\rm Im} \phi_j<0$
and
\eqn\deltadef{\Delta(\phi_{2j-1},\phi_{2j})=(1+z_h^2)(1+z_v^2)-2z_v(1-z_h^2)
\cos\phi_{2j-1}-2z_h(1-z_v^2)\cos\phi_{2j}.}

For $T>T_c$ the corresponding result is (in the more
symmetric form of~\rmwl~\rmwn)
\eqn\fg{<\sigma_{0,0}\sigma_{M,N}>=M_{+}^2X(M,N)
\exp{(-\sum_{n=1}^{\infty}F_{M,N}^{(2n)})}}
where
\eqn\xsum{X(M,N)=\sum_{n=0}^{\infty}X^{(2n+1)}(M,N)}
\eqn\exn{\eqalign{X^{(2n+1)}(M,N)=
(2\pi)^{-4n-2}\int_{-\pi}^{\pi}&d\phi_1
\cdots\int_{-\pi}^{\pi}d\phi_{4n+2}\cr
&\prod_{j=1}^{2n+1}{e^{-i(M-1)\phi_{2j-1}-i(N-1)\phi_{2j}}
\over \Delta(\phi_{2j-1},\phi_{2j})}e^{-i\phi_{4n+1}-i\phi_{4n+2}}\cr
&\prod_{j=1}^{2n}[2iz_v(1-z_h^2)({e^{-i\phi_{2j-1}}-e^{-i\phi_{2j+1}}\over
1-e^{i\phi_{2j}}e^{i\phi_{2j+2}}})]}}
with $\phi_{4n+1}=\phi_1$, $\phi_{4n+2}=\phi_{2}$, Im$\phi_j<0$
and
\eqn\condg{M_+=
[(\sinh 2K^v\sinh K^h)^{-2}-1]^{1/4}}

If the exponentials in ~\twol~ and ~\fg~ are expanded~\rpt~ these series
are seen to be the K{\'a}llen--Lehmann form factor (soliton)
expansions of Part 1 (7.23) where the index $2n~(2n+1)$ represents
the number of particles in the state.

The leading terms of the expansion (of the diagonal correlation)
for $T>T_c$ as $N \rightarrow \infty$ is
\eqn\tgcorr{<\sigma_{0,0}\sigma_{N,N}>\sim{1\over( \pi N )^{1/2}}{k_>^N\over
(1-k_>^2)^{1/4}}+\cdots}
where
\eqn\kg{k_>=\sinh 2K^h \sinh 2K^v.}
The correlation length thus behaves near $T_c$ as
\eqn\corg{\xi^{-1}=\ln k_>\sim |T-T_c|}
and hence the critical exponent $\nu  =1.$

The leading term of the expansion (of the diagonal correlation)
for $T<T_c$ as $N\rightarrow \infty$ is
\eqn\tlcorr{<\sigma_{0,0}\sigma_{N,N}>\sim M^2_-\{1+
{2k_<^{2(N+1)}\over2 \pi N^2(k_<^{-2}-1)^2}+\cdots\}}
where
\eqn\kl{k_<=(\sinh 2K^h \sinh 2 K^v)^{-1}.}
The correlation length thus behaves near $T_c$ as
\eqn\corl{2\xi^{-1}=2\ln k_<\sim2|T-T_c|}
and hence the critical exponent $\nu '=1.$

\bigskip
\newsec{Scaling theory and Painlev{\'e} equations.}

We are now in a position to make contact with the scaling theory of part 1.
First we note that the two scaling laws $2\beta=\nu '(d-2 +\eta)$ and
$d\nu'=2-\alpha '$ are satisfied by $\alpha '=0,~\beta=1/8,~\eta=1/4$ and
$\nu'=1$. We also note the predictions  from $\gamma'=d\nu'-2\beta$ that
$\gamma '=7/4$ and from $\eta=2-d(\delta-1)/(\delta+1)$
that $\delta=15$. We also note that the critical exponents above $T_c$ equal
those below $T_c.$

The scaling limit of the full correlation function may now be  obtained from
the definition
\eqn\scale{G_{\pm}(r)=
\lim_{\rm scaling}M^{-2}_{\pm}<\sigma_{0,0}\sigma_{M,N}>.}
and the results are~\rwmtb~for $T<T_c$
\eqn\gminus{G_{-}(r)=\exp[-\sum_{n=1}^{\infty}\lambda^{2n}f_{2n}(r)]}
with
\eqn\fscale{f_{2n}(r)={(-1)^{n}\over n}\int_1^{\infty}dy_1\cdots
\int_1^{\infty} dy_{2n}
\prod_{j=1}^{2n}[{e^{-ry_j}\over(y_j^2-1)^{1/2}}(y_j+y_{j+1})^{-1}]
\prod_{j=1}^{n}(y_{2j}-1)}
and for $T>T_c$
\eqn\above{G_+(r)=X(r)G_-(r)}
where
\eqn\ex{X(r)=\sum_{n=0}^{\infty}\lambda^{2n+1}x_{2n+1}(r)}
with
\eqn\exsum{\eqalign{x_{2n+1}(r)=(-1)^n\int_1^{\infty}&dy_1\cdots
\int_1^{\infty} dy_{2n+1}\cr
&\prod_{j=1}^{2n+1}{e^{-ry_j}\over (y_j^2-1)^{1/2}}\prod_{j=1}^{2n}
{1\over y_j+y_{j+1}}\prod_{j=1}^n(y_{2j}^2-1)}}
and in the above $\lambda=1/\pi.$

For large $r$ the leading behavior for $T>T_c$ is
\eqn\largerg{G_+(r)\sim{1\over \pi}K_0(r)}
where $K_n(r)$ is the modified Bessel function of order $n.$
The Fourier transform of this is a single pole and confirms the existence
of a particle in the theory.

For large $r$ and $T<T_c$ the leading behavior is
\eqn\largel{G_-(r)\sim1+\pi^{-2}[r^2[K_1(r)-K_0^2(r)]-rK_0(r)K_1(r)
+{1\over2}K_0^2(r)]}
The Fourier transform of this is not a single particle pole but rather
contains a branch cut.
This is consistent with the fact that the excitation spectrum of the transfer
matrix has only even particle excitations for $T<T_c$

But we are able to prove even more. Not only does this scaled dispersion
relation hold but it was shown in 1973~\rbmw~\rmta~\rwmtb ~that
the scaling function has the remarkable property that

\eqn\corpain{G_\pm(r)={1\over 2}[1\mp\eta(r/2)]\eta(r/2)^{-1/2}
\exp{\int_{r/2}^{\infty}d\theta {1\over 4}\theta\eta^{-2}[(1-\eta^2)^2-
(\eta ')^2]}}
where $\eta(\theta)$ satisfies the Painlev{\'e} III equation~\rpain~\rince
\eqn\painleve{\eta''={1\over\eta}(\eta ')^2-{\eta'\over \theta}+\eta
^3-\eta^{-1}}
with the boundary conditions
\eqn\boundary{\eta(\theta)\sim {1-2\lambda}K_0(2\theta)~{\rm as}~
\theta\rightarrow \infty~{\rm where}~~ \lambda=1/\pi.}
If we set
\eqn\psieqn{\eta(\theta)=e^{-\psi(t)}~~{\rm with}~~t=2\theta}
the Painlev{\'e} III equation reduces to the radial sinh-Gordon equation
\eqn\sinhgordon{{d^2 \psi\over dt^2}+{1\over t}{d\psi\over
dt}={1\over 2}\sinh 2\psi}
and the correlation function becomes
\eqn\corpsi{G(r)_{\pm}={\sinh{1\over 2}\psi(r)\choose \cosh{1\over
2}\psi(r)}\exp\{{1\over 4}\int_r^{\infty}ds s[-({d\psi\over
ds})^2+\sinh^2\psi]\}. }
In addition if we set
\eqn\newzeta{\zeta=r{d\ln G_{\pm}\over dr}}
that for both $G_{+}$ and $G_{-}$ we have~\rjm
\eqn\painV{(r\zeta '')^2=4(r\zeta'-\zeta)^2-4(\zeta')^2(r\zeta
'-\zeta)+(\zeta ')^2}
which has an intimate relation with Painlev{\'e} V~\roka--\rokc.

It is of interest to relate these nonlinear differential equations for
$T\neq T_c$ to the previously studied case of $T=T_c.$ To do
this~\rmpb~  we
explicitly introduce a mass scale $m$ into the equation ~\painV~ as
$r=m{\bar r}$ and then let $m\rightarrow 0$. In this limit the
nonlinearity of~\painV~ cancels and one integral can be done to yield
\eqn\lineqn{{\bar r}^2{d^2G\over d{\bar r}^2}+2(c+{1\over 4}){\bar
r}{dG\over d{\bar r}}+[(c+{1\over 4})^2-{1\over 4}]G=0.}
If the integration constant $c$ is chosen to be zero and if we write
${\bar r}=(z z^*)^1/2$ then, considered as an equation in $z,$ the
equation~\lineqn~ is recognized as the second order equation
satisfied by a second level degenerate operator in conformal field
theory~\rbpza~\rbpzb.

Moreover, once these nonlinear differential equations have been found
for the scaling limit it is to be expected that similar results exist
for the full correlations on the lattice as given by the dispersion
relations~\twol~ and ~\fg. In fact there are several such results known.

The first of these generalizations are quadratic difference equations for the
correlation functions on the lattice.
These are most compactly displayed if we note that the correlation
functions above and below $T_c$ are related by what is called a
duality relation. Defining the new constants $K^{v*}$ and $K^{h*}$ by
\eqn\dual{\sinh2K^{v*}\sinh2K^h=1~~{\rm and}~~\sinh2K^{h*}\sinh2K^v=1}
and letting $<\sigma_{0,0}\sigma_{M,N}>^*$ to be the spin--spin
correlation on this dual lattice we have
\eqn\dualratio{{<\sigma_{0,0}\sigma_{M,N}>^*\over
<\sigma_{0,0}\sigma_{M,N}>}=
(\sinh2K^v\sinh2K^h)^{1/2}X(M,N).}
Thus we find the two nonlinear partial difference equations~\rmwl~\rperka~\rmwn
\eqn\partiala{\eqalign{<\sigma_{0,0}\sigma_{M.N}>^2-
<\sigma_{0,0}&\sigma_{M-1,N}><\sigma_{0,0}\sigma_{M+1,N}>\cr
=-\sinh^{-2}2K^v(&<\sigma_{0,0}\sigma_{M,N}>^{*2}\cr
&-<\sigma_{0,0}\sigma_{M,N-1}>^{*}<\sigma_{0,0}\sigma_{M,N+1}>^*)}}
and
\eqn\partialb{\eqalign{<\sigma_{0,0}\sigma_{M,N}>^2-
<\sigma_{0,0}&\sigma_{M,N-1}>
<\sigma_{0,0}\sigma_{M,N+1}>\cr
=-\sinh^{-2}2K^h(&<\sigma_{0,0}\sigma_{M,N}>^{*2}\cr
&-<\sigma_{0,0}\sigma_{M-1,N}>^*<\sigma_{0,0}\sigma_{M+1,N}>^*).}}
At $T=T_c$ we have
$<\sigma_{0,0}\sigma_{M,N}>=<\sigma_{0,0}\sigma_{M,N}>^*$
and these two equations reduce to the remarkable simple result which is
valid except at $M=N=0$
\eqn\toda{\eqalign{\sinh2K^v&(<\sigma_{0,0}\sigma_{M,N}>^2-
<\sigma_{0,0}\sigma_{M-1,N}><\sigma_{0,0}\sigma_{M+1,N}>)\cr
&+\sinh2K^h(<\sigma_{0,0}\sigma_{M,N}>^2-<\sigma_{0.0}\sigma_{M,N-1}>
<\sigma_{0,0}\sigma_{M,N+1}>)=0.}}
This is the discrete (imaginary) time Toda equation~\rtoda~ introduced
by Hirota~\rhir. We may also express the lattice correlations solely in
terms of $X(M,N)$ which generalizes ~\corpain and derive a partial
difference equation for $X(M,N)$ which is a lattice generalization of the
Painlev{\'e} equation~\painleve. For details see~\rmwn.

The second generalization is to  nonlinear
differential equations in the temperature. The simplest example is for
the diagonal correlation function $<\sigma_{0,0}\sigma_{N,N}>$.
Here we define for $T<T_c$
\eqn\tlessa{t=(\sinh2K^v \sinh2K^h)^2}
and
\eqn\tlessb{\sigma(t)=t(t-1){d\over
dt}\ln<\sigma_{0,0}\sigma_{N,N}>-{1\over 4},}
and for $T>T_c$
\eqn\tgreata{t=(\sinh2K^v \sinh 2K^h)^{-2}}
and
\eqn\tgreatb{\sigma(t)=t(t-1){d\over
dt}\ln<\sigma_{0,0}\sigma_{N,N}>-{1\over 4}t}
where in both cases $t>1.$
Then it was found by Jimbo and Miwa~\rjm~ that
\eqn\painsix{\eqalign{(t&(t-1){d^2\sigma\over dt^2})^2\cr
&=N^2[(t-1){d\sigma\over
dt}-\sigma]^2-4{d\sigma\over dt}[(t-1){d\sigma\over dt}-\sigma-{1\over
4}](t{d\sigma\over dt}-\sigma).}}
This is equivalent to the sixth Painlev{\'e} equation~\rjmb. For the
correlations off  the diagonal there also exist systems of such
differential equations. We refer the reader to ~\rjmb~ for details.

But we have not yet succeeded in verifying that the functions $G_{\pm}(r)$
satisfy all the properties of a scaling function. To do this we must show
that as $r\rightarrow 0$ we regain the $T=T_c$ expression
of ~\tccorr~ and show that
$\eta=1/4.$ This was first done in 1977 when it was shown that~\rmtwa~
that as $r\rightarrow 0$
\eqn\etazero{\eta(r/2)=
Bt^{\sigma}\{1-{1\over 16}B^{-2}(1-\sigma)^2r^{2-\sigma}+
O(t^2)\}}
where the two constants $B$ and $\sigma$ are determined in terms of $\lambda$
as
\eqn\sigmalam{\sigma(\lambda)={2\over \pi}{\rm arcsin}(\pi \lambda),~~
B=B(\sigma)=2^{-3\sigma}{\Gamma((1-\sigma)/2)\over \Gamma((1+\sigma)/2)}}
When  $\lambda\rightarrow 1/\pi$ then $\sigma \rightarrow 1$
and the correlation behaves as
\eqn\smallr{G_{\pm}(r)={\rm const}~ r^{-1/4}\{1\pm{1\over2}r[\ln({1\over 8}r)
+\gamma_E]+{1\over 16}r^2+\cdots\}}
where $\gamma_E$ is Euler's constant.

To  complete the verification of scaling theory  we must show that the
constant in ~\smallr~ is the one obtained from the transcendental constant $A$
of~\tccorr. This was done by Tracy~\rtrc~\rbt~ in 1991.
 Thus for the two point
function all the features in the definition of the scaling theory have been
verified. It took 47 years
to achieve this level of understanding.

Similar studies have been made for all the higher multipoint
correlations. Dispersion relations on the lattice have been
found~\rmtwb, sets of holononic partial differential equations have
been found in the scaling limit~\rsmjab,\rsmjb,\rsmjc--\rsmje,\rsmjf,
and quadratic difference equations exist on the lattice for all
temperatures~\rmpw.
These multipoint correlations  satisfy the scaling properties and the
Osterwalder--Schrader axioms~\rpt~\rrptb.
Moreover the S matrix has been computed for
$T>T_c$~\rsmjaa. The scattering  is completely elastic and in the
$n$ particle channel the S matrix is $(-1)^{n(n-1)/2}.$
Thus both for $T$ above and below $T_c$ this Ising field theory constructed
from statistical mechanics satisfies precisely the properties needed for
a relativistic field theory. This is the only interacting field theory
for which this construction has been explicitly carried out.

These expressions of the Ising model correlations in terms of
Painlev{\'e} functions are some of the most beautiful results in
mathematical physics. But what is even more remarkable is the
fact that these are not isolated results but form part of a
very much larger mathematical structure which, though it was first
glimpsed in the Ising model, has a very large range of applicability.

One extension of these results is to the correlation functions of
other specific statistical and field theory models:
the impenetrable Bose~{\rjmms--\rkbi} and Fermi~{\rberka~\rberkb} gases,
  the $XY$
spin chain~{\rperka~\rpcqn~\rlsm--\riiksc}
 and the closely related sine--Gordon field theory at the
decoupling point~\rikt~\rbele.

But the mathematical structure has extensions far beyond these
specific models.
One part of this structure is the theory of isomonodromic
deformations of linear differential equations and their relation to
determinants of singular operators. These determinants are called $\tau$
functions and are mathematical  generalizations of the determinants
of the Ising correlations. One presentation of this theory
is~\rjmu--\rjmc. This treatment not only encompasses the six second
order Painlev{\'e} functions but also establishes the general
principles that derive all higher order equations with the
Painlev{\'e} property.

Another part of the mathematical structure is the connection which
 the theory of random
matrices~\rmehta~ has with Painlev{\'e} functions.
This connection was first
found for the Gaussian ensembles
by Jimbo, Miwa, Mori, and Sato \rjmms~ in 1980 and subsequently
vastly extended to many other ensembles of random matrices~\rbk--\rdysb.
Moreover the random
matrix problems have been connected with 2 dimensional quantum
gravity~\rbk--\rbobr~
and this forms another major connection between statistical mechanics
and quantum field theory.

Finally we note that exactly the Ising correlations given by the PIII
equation have recently been seen to arise in $N=2$ supersymmetric
quantum field theories~\rceva. However in this case  values of the
boundary condition constant $\lambda$ other than $1/\pi$
lead to physical results.
Further discussion of these connections to quantum field theory are
beyond the scope of these lectures but it is surely expected that
there will be more applications and generalizations of the Ising model
results.

It remains to consider the particle interpretation of the Ising field theory.
But now it may appear as if we have encountered a paradox. On the one
hand we have faithfully executed all the steps needed to show that the
theory has interactions. However, on the other hand the actual computations
were done by a formalism that related the Ising model to a free Fermi system.
{}From that point of view the reader may ask what right do I have to call the
theory anything but free.

The resolution of this question is most interesting. We emphasized in the
first lecture in discussing the concept of particle that we really have not
given a definition of a particle as a thing in itself but only in its
interaction with those fields which we chose to call observable. In the
interpretation of the Ising model as a spin model we are certainly conceiving
of the spins being observable by their interactions with the external
magnetic field $H.$ Indeed neutron scattering experiments do exactly that and
the
Fourier transform of the two point spin correlation is what it is claimed
is measured. In terms   of this interpretation and measurement the computations
discussed above are certainly to be interpreted as showing the presence of
strongly interacting bosonic excitations. The construction of the Fermi
operators is very non-local in terms of the spins and therefore will not
couple to the same external fields that the spins do. Indeed it is hard to
conceive of a possible experiment that will couple to these fermions. Thus
the fermion Greens functions which are indeed free and which are indeed used
in the intermediate stages of some methods of computing the Ising
spin correlations are not excited by the possible laboratory experiment.
The observed particles are the interacting bosons and not the free fermions.

In fact this duality between bosons and fermions seen in these
computations is not unique to the
Ising model and in fact is very common in one dimensional quantum systems
and the related two dimensional classical systems. Much as it may seem to
violate naive expectations one local field theory can
often be transformed into another local field theory by means of a non local
transformation. This may indeed be an effect that cannot occur in higher
dimensions (although I do not know of a definitive study) but because the
effect exists we must be very cautious when giving a particle interpretation
to a quantum field theory. In the end it must never
be forgotten that it is the measuring apparatus which is the authority that
tells us which interpretation is physically observed.

 Finally we remark about the $\phi^4$  $\epsilon$ perturbation
theory from d=4. Such an expansion gives results for the critical
exponents of the scaling theory. This expansion
does not converge but is hopefully asymptotic and Borel summable. The
results in two dimensions reproduce the known results with a remarkably good
accuracy~\rzinn. But for field theory  applications these critical
exponent computations are not really what is desired. What we really
want are the
Greens functions. So far there has been little attempt to regain the
Painlev{\'e}
property of the Greens functions or the S matrix from
the $\epsilon$ expansion.

We close with the observation that in the $\epsilon$ expansion it is tacitly
assumed that the particle spectrum in d=4 is the same as in d=3 or d=2.
As long as this is the case there is some hope that the method can succeed.
But if the genuine particle spectra in d=3 or d=2 is not the same then it
is extremely doubtful that the $\epsilon$
perturbation expansion can be powerful enough to give the
Greens functions of the
theory. We will return to this in the next lecture since for $H\neq0$ the
question of how many particles are in the spectrum is a matter of
major concern.

\bigskip
\newsec{The layered Ising model at $H=0$}

The results of the previous sections have been restricted to the homogeneous
Ising model. However the fermionization and combinatorial methods are not
restricted to this case and their initial formulation is valid also for the
completely inhomogeneous lattice~\ising~ where the bonds $E^v(j,k)$ and
$E^h(j,k)$ are completely arbitrary. For this reason many of the
difference equations satisfied by the correlations are valid for
the completely random lattice~\rmpw. However, in order to get complete
results it has been necessary to impose translational invariance in
one direction. Therefore in this final section we will consider the
layered Ising model where $E^v(j,k)=E^v(j)$. We will also consider for
convenience $E^h(j,k)=E^h$ but this specialization can easily be removed
and does not affect any of the physics.

We begin our considerations with the case where the $N$ bonds $E^v(j)$ for
$1\leq j\leq N$ are arbitrarily chosen but then this configuration is repeated
to make up the full lattice. Thus in this case the translational invariance
in the vertical direction is reduced but not totally destroyed. The free
energy of this model has been studied~\rayma~ where it was shown that there is
still  a unique critical temperature determined from
\eqn\rantc{1=\prod_{j=1}^{N}z^{2}_{vc}(j)e^{4E^h/kT_c}.}
At this critical temperature the specific heat still has a logarithmic
divergence but now the amplitude depends on all the $N$ energies $E^v(j).$
Moreover the amplitude depends sensitively on the geometric ordering of
the $N$ bonds. In general all $N!$ permutations of the $N$ distinct bonds give
different amplitudes and furthermore
these amplitudes are less than what would
be obtained is we considered replacing all the vertical bonds by the average
value. In this way we argue that there is a sense in which the
logarithmic divergence can be said to be caused by all the bonds acting
together in a cooperative fashion. The question of greatest interest then
whether the destruction of translational invariance can also destroy the
logarithmic singularity by driving the amplitude of the logarithm to zero.

There are of course many ways to destroy translational invariance. The mildest
way is to have the vertical bonds form a sort of quasi crystal. A particularly
nice example of this is the Fibonacci lattice considered by Tracy~\rtra.
This lattice is defined as follows. Consider a set of sequences $S_n$
of the letters
$A$ and $B$ defined recursively by
\eqn\fib {S_{n+1}=S_n S_{n-1},~~{\rm with}~~S_0=B,~~S_1=A}
For example $S_2=AB,~S_3=ABA$, and $S_4=ABAAB.$ The sequence $S_n$ contains
$F_{n-1}$ A's where $F_n$ are  the Fibonacci numbers defined by
\eqn\fibnum{F_{n+1}=F_n+F_{n-1},~~F_0=F_1=1}
Now consider two values of the energies $E^v=E^v_A$ and $E^v_B$, place them in
the lattice according to the sequence $S_n$ and  repeat the configuration
to build up the complete lattice. In the limit that the sequence length
goes to infinity, so that the translational invariance is destroyed,
 it is found that there is still a single transition
temperature obtained from
\eqn\fibtc{e^{-2E^h/kT_c}=z^{\alpha}_{vAc}z_{vBc}^{\alpha ^2}}
where
\eqn\almo{\alpha ^{-1}=(1+2^{1/2})/2}
and at this temperature the specific heat still has a logarithmic divergence
with the amplitude
\eqn\fibam{A={1\over 4\pi (kT_c)^2}(z^{-1}_{hc}-z_{hc})|{x^2 \ln x^2\over
1-x^2}|
\{2E^h+E^v_A\alpha(z^{-1}_{vAc} -z_{vAc})+
E^v_{B}\alpha^2(z^{-1}_{vBc}-z^v_{Bc})\}^2}
where $x=z_{vBc}/z_{vAc}.$
Thus it is definitely not the case that the mere destruction of
translational invariance is sufficient to destroy the logarithmic singularity
in the specific heat.

However if the lattice is made completely random in the sense that each
vertical bond is chosen  as an independent random variable from a probability
distribution $P(E^v)$ the situation changes drastically~\rmwd,~\rmwc,~\rbma,~
\rbmb,~\rbmc,~\rbook.

The first qualitative change is that now instead of the zeroes of the partition
function pinching the real temperature axis at a point they
 pinch in an entire line
segment that runs from $T^u$ to $T^l$ where $T^{u,l}$ is the critical
temperature that would result if all bonds $E^v$ were replaced by the maximum
(minimum) value allowed by the probability distribution $P(E^v).$ These two
temperatures are called the Griffiths temperatures of the system~\rgri.
There is still a singularity in the free energy at the
temperature we can still denote as $T_c$
 defined by~\rantc~ which is now written as
\eqn\probtc{0=\int_0^{\infty}dE^v P(E^v)\ln(z_{vc}^2e^{4E^h/kT_c})}
but now the singularity is not a logarithmic divergence. There is a new
temperature scale introduced into the problem when $T$
is between the Griffiths temperatures such that near this $T_c$  we
define
\eqn\tempscale{\delta={2\int_0^{\infty}dE^vP(E^v)\ln(z_v^2e^{4E^h/kT})
\over \int_0^{\infty}dE^v P(E^v)[\ln(z_v^2e^{4E^h/kT})]^2}.}
and find that the singular part of the specific heat  is now proportional to
\eqn\singularfree{\int_0^{\infty}d\phi[{\partial ^2\over \partial \delta^2}
\ln K_{\delta}(\phi)-(1+\phi)^{-1}].}
This function does not diverge at $\delta =0$ but instead has
an infinitely differentiable essential singularity.

There are many other new features that occur in the physics of this random
layered Ising model. For example the magnetic susceptibility does not diverge
at $T_c$ but is infinite in an entire range of temperatures around $T_c.$
The conclusion is that the scaling theory developed for the standard
second order phase transition with one length scale has
totally broken down in this random system and that our intuition
must be enlarged.

One interpretation of these computations is that randomness
smoothes out singularities in the specific heat. Therefore,
since the one dimensional randomness of the layered Ising model has already
smoothed out the logarithmic divergence to an infinitely differentiable
essential singularity it can be argued that  making the
system random in the second
direction cannot make the specific heat more singular.
This would argue that the specific heat of the fully random Ising model should
be finite.

This argument has been opposed by Dotsenko and Dotsenko~\rdodo~ who study
a system with $E^v(j,k)$ and $E^h(j,k)$ all chosen as independent
random variables with the probability distribution
\eqn\dodoprob{P(E)=(1-p)\delta(E-E_1)+p\delta(E-E_2).}
If $p$ is considered small and if only the lowest terms in $p$ are kept then it
is found~\rdodo~ that the specific heat has a divergence of the form
$\ln(\ln|T-T_c|^{-1}).$ However the fact that only lowest orders in $p$
 are kept
means that effectively the distinction between the upper and lower Griffiths
temperatures is ignored and the transition seems to be replaced by some
effective transition in which the zeroes still pinch at a point. This would
seem to exclude from consideration the temperature scale inside the
Griffiths temperatures where the actual transition takes place. Hence
it may be argued that this computation~\rdodo~ does not incorporate the new
scales seen in the layered Ising model and is not applicable directly at the
point of singularity itself.

More recently the fully random Ising model has been considered by
Ziegler~{\rzieone--\rziethree } in a field theoretic fashion which does
incorporate the new length scales and explicitly considers the new
behavior between the Griffiths temperatures. These considerations do
lead to a finite specific heat. Furthermore D. Fisher~\rfis~ has
applied renormalization group arguments to the $XY$ random spin chain
related to the layered Ising model and finds that the exponent of the
spontaneous magnetization is $(3-5^{1/2})/2.$
There are still
many open questions in the theory of random Ising models and their study
will continue to enlarge our intuition   of the effects of impurities
on phase transitions.


\bigskip
\bigskip



\vfill

\eject
\listrefs

\vfill\eject

\global\secno=0

\global\meqno=1

\global\subsecno=0

\global\refno=1


%
\def\rhob{{\rho\kern-0.465em \rho}}

\def\eps{\epsilon}

\def\ontopss#1#2#3#4{\raise#4ex \hbox{#1}\mkern-#3mu {#2}}

\setbox\strutbox=\hbox{\vrule height12pt depth5pt width0pt}
\def\tablerule{\noalign{\hrule}}
\def\tr{\tablerule}
\def\strut{\relax\ifmmode\copy\strutbox\else\unhcopy\strutbox\fi}

\nref\rweg{F.J. Wegner, Duality in generalized Ising models and phase
transitions without local order parameters, J. Math. Phys. 12 (1971) 2259.}
\nref\rbdi{R. Balian, J.M. Drouffe and C. Itzykson, Gauge fields  on a lattice
II. Gauge invariant Ising Model, Phys. Rev. D11 (1975) 2098.}
\nref\rwil{K.G. Wilson,Confinement of quarks, Phys. Rev. D10 (1974) 2445.}
\nref\rmw{B.M. McCoy and T.T. Wu, The two-dimensional Ising model in a
magnetic field: Breakup of the cut in the two point function, Phys. Rev. D18
(1978) 1259.}
\nref\rzama{A.B. Zamolodchikov, Integrable field theory from conformal
field theory, Adv. Stud. in Pure Math. 19 (1989) 641.}
\nref\rzamb{A.B. Zamolodchikov, Integrals of the motion and the
S--matrix of the (scaled) $T-T_c$ Ising model with magnetic field, Int. J. Mod.
Phys. A4 (1989) 4235.}
\nref\rmy{B. M. McCoy and M--L. Yan, Gauge invariant correlation functions
for the Ising-gauge Ising-Higgs system in 2--dimensions, Nucl. Phys.
B215[FS7] (1983) 278.}
\nref\rly{T.D. Lee and C.N. Yang, Statistical theory of equations of state and
phase transitions II. Lattice gas and Ising models, Phys. Rev. 87 (1952) 410.}
\nref\rwns{S.O. Warnaar, B. Nienhuis and K.A. Seaton, New construction
of solvable lattice models including an Ising model in a field, Phys.
Rev. Lett. 69 (1992) 710; Int. J. Mod. Phys. B7 (1993) 3727.}
\nref\rbnw{V.V. Bazhanov, B. Nienhius and S.O. Warnaar, Lattice Ising model in
a field: $ E_8$ scattering theory, Phys. Letts. A (in press).}
\nref\rzamc{A.B. Zamolodchikov, private communication}
\nref\rhecht{R. Hecht, Correlation functions for the two-dimensional Ising
model, Phys. Rev. 158, (1967) 557.}
\nref\rmyb{B.M. McCoy and M--L. Yan, Energy density correlation
function of the Ising model in a small magnetic field for $T>T_c$,
Nucl. Phys. B257[FS14] (1985) 303.}

\centerline{\bf Part III. Ising Field Theory at $H\neq 0$}



\vskip 13mm

\centerline{\bf Abstract}
\vskip 4mm
We discuss the relation of the Ising-gauge Ising Higgs theory to the Ising
model in a magnetic field. We then present the methods used to study
the Ising model in a magnetic field and summarize all of
the known results for the field theory limit. These results are then used
to illustrate the phenomena of confinement and to discuss the notion
of the size of a particle.

\vskip 3mm

\newsec{Introduction}

The solution of the Ising model in a magnetic field has been considered an
outstanding problem ever since the solution of the zero field free energy
50 years ago. At first it was considered because of the relevance to
magnetism but in the early 70's it achieved even more importance
when it was recognized~\rweg~\rbdi~ as the simplest non trivial realization
of the lattice gauge theory of Wilson~\rwil.

In the 50 years in which this problem has been investigated there have been
two main lines of investigation. The first is to use our detailed
knowledge of the correlation functions at zero magnetic field as the
basis of an expansion for small values of $H$. This leads~\rmw~
to the picture that
a one parameter family of scaling field theories can be constructed
in the limit
\eqn\hparam{H\rightarrow 0,~T\rightarrow T_c~~{\rm with}~h=
{\rm const}~H/|T-T_c|^{15/8}~
{\rm fixed}.}
The particle content of these field theories depends on the value of
$h$ and for $T<T_c$ and small values of $h$ the spectrum contains a very
large number of particles which become dense as $h\rightarrow 0.$

Much more recently it has been discovered by Zamolodchikov~\rzama~\rzamb~
 that at $T=T_c$
$(h=\infty)$ the field theory limit has the remarkable property that it
is integrable. The spectrum has been computed to contain 8 particles
and the S matrices of the scattering of these particles has been found.

In this lecture we will present these results. In sec. 2 we will present the
connection of the Ising model and lattice gauge theory. In sec. 3 we will
sketch the method of  expansion   for small $H$. In sec.4
we will sketch the method of computations for $T=T_c.$
In sec. 5 we summarize the known results and we conclude in sec. 6 by using
these results to discuss the physics of confinement.
 We conclude in sec. 5 by using
these results to discuss the physics of confinement.

\bigskip
\newsec{Ising--Gauge Ising--Higgs}

In (5.6) of the first lecture we introduced the general lattice gauge theory
of Wilson~\rwil~ for a single scalar field coupled to a gauge field.
However in the subsequent discussion this action was never really discussed
and, in particular, in our discussion of particles and scattering we made
no mention of any particular features connected with gauge theory at all.
But in applications to particle physics gauge theories are of extreme
importance. Therefore if we are to have any genuine understanding  of the
connection between statistical mechanics and quantum field theory we must
consider gauge theories  in detail. If we could not do this we would be in the
position of having two types  of field theory--nongauge theories like the
$n$ vector model that are relevant to statistical physics and gauge
theories which
are relevant to particle physics. It is thus extremely fortunate
that in the early 70's it was discovered by
Wegner~\rweg~ and Balian, Drouffe, and Itzykson~\rbdi~
that there is a connection between the
homogeneous Ising model defined by
\eqn\ising{E=-E\sum_{j,k}(\sigma_{j,k}\sigma_{j+1,k}+
\sigma_{j,k}\sigma_{j,k+1})-H\sum_{j,k}\sigma_{j,k}}
 and the simplest possible lattice
gauge theory of Wilson~\rwil: the Ising--Gauge Ising--Higgs model in
two dimensions specified by the Euclidean action
\eqn\action{\eqalign{S=&-E_g\sum_{j,k=1}^Ls_{j+1/2,k}s_{j+1,k+1/2}
s_{j+1/2,k+1}s_{j,k+1/2}\cr -&
E_h\sum_{j,k=1}^L(\phi_{j,k}s_{j+1/2,k}\phi_{j+1,k}+
\phi_{j,k}s_{j,k+1/2}\phi_{j,k+1}).}}
Here $s_{l,m}=\pm1,\phi_{j,k}=\pm1,$ j(k) labels the row (column)
of the lattice and the variables $s_{l,m}$ are associated with
the links of the lattice and are specified by the location of the
midpoint of the link. In the language of gauge theory $\phi_{j,k}$ is
called a Higgs field and $s_{j,k+1/2}$ is a gauge field. We here will show the
relation of ~\action ~to ~\ising ~following closely ref~\rmy.

The action~\action~ has the local gauge invariance property that if we
consider any particular lattice site $(j,k)$ and replace the four variables
$s$ that lie on the links that end on $(j,k)$ and the variable $\phi_{j,k}$
by their
negatives the action remains unchanged. Therefore in the computation of
expectation values from the general formula
\eqn\corr{<O_j>=Z^{-1}\sum_{\{\phi\}}\sum_{\{s\}}O_je^{-S/kT},~~~~
Z=\sum_{\{\phi\}}\sum_{\{s\}}e^{-S/kT}}
only operators $O_j$ invariant under the gauge transformation
give non vanishing
expectation values. We will be particularly interested in the correlations
\eqn\higghiggs{<\phi_{j,k}s_{j,k+1/2}\phi_{j,k+1}~
\phi_{j',k'}s_{j',k'+1/2}\phi_{j',k+1/2}>}
which we abbreviate as $<H_{j,k}H_{j',k'}>$ and
call a Higgs-Higgs correlation and
\eqn\plapla{\eqalign{<s_{j+1/2,k}s_{j+1,k+1/2}&s_{j+1/2,k+1}s_{j,k+1/2}\cr
&s_{j'+1/2,k'}s_{j'+1,k'+1/2}s_{j'+1/2,k'+1}s_{j',k+1/2}>}}
 which we abbreviate as $<P_{j,k}P_{j'k'}>$
and call a plaquette-plaquette correlation. We note that the correlations of
two gauge variables $<s_{j,k+1/2} s_{j',k'+1/2}>$ and of two Higgs variables
$<\phi_{j,k}\phi_{j',k'}>$ vanish identically because they are not gauge
invariant.

Consider first the partition function $Z$ of ~\corr. Because of the local
gauge invariance the sum over the $\{\phi\}$ is trivially done and thus
\eqn\part{Z=2^{L^2}\sum e^{-S'/kT}}
where   $S'$ is obtained from $S$ by setting all $\phi_{j,k}=1.$ Calling
$E_i/kT=K_i$ and using the fact that $s_{j,k+1/2}$ and $\phi_{j,k}$
take on only the values $\pm1$ we now write
\eqn\fours{e^{K_gssss}=\cosh K_g(1+z_gssss)~~{\rm and}~~e^{K_h s}=\cosh K_h
(1+z_hs)}
where $z_{g,h}=\tanh K_{g,h}$. Thus
\eqn\parts{Z=(2\cosh K_g\cosh^2 K_h)^{L^2}\sum_{\{s\}}\prod_{j,k}(1+z_gssss)
(1+z_hs)}
where the indices  on $s$ have been suppressed. We may now expand the
products and do the sum over all $s=\pm1$ and obtain
\eqn\partgraph{Z=(8\cosh K_g\cosh^2 K_h)^{L^2}\sum_{\rm graphs}z_g^Az_g^P}
where the sum is over all closed (intersecting) polygons on the lattice, $A$
is the interior area of the polygons and $P$ is the length of the perimeter
of the polygons.

To see the relation with~\ising~ define the variable $\sigma_{j,k}$ to be
minus one if the
plaquette whose center is at $j+1/2,k+1/2$ is  not used   in a term
in ~\partgraph~ and equal to one if it is used. Then
\eqn\adef{A={1\over 2}\sum_{j,k}(1-\sigma_{j,k}),~P_v={1\over 2}\sum_{j,k}
(1-\sigma_{j,k}\sigma_{j,k+1})~
{\rm and}~P_h={1\over 2}\sum_{j,k}(1-\sigma_{j,k}\sigma_{j+1,k})}
with $P=P_h+P_v.$ Thus
\eqn\newising{\eqalign{Z=(8 \cosh K_g\cosh^2 K_h)^{L^2}&\sum_{\sigma=\pm1}
\exp\{{1\over 2}\ln z_g\sum_{j,k}(1-\sigma_{j,k})\cr
+&{1\over 2}\ln z_h\sum_{j,k}
[(1-\sigma_{j,k}\sigma_{j,k+1})+(1-\sigma_{j,k}\sigma_{j+1,k})]\}\cr}}
which is clearly proportional to the homogeneous Ising model~\ising~ with $
E^v(j,k)=E^h(j,k)=E$ if
\eqn\ident{H/kT={1\over 2}\ln z_g~{\rm and}~E/kT={1\over 2}\ln z_h.}

The important point to  note in this derivation is that it goes through
word for word if we let $z_g$ and $z_h$ vary from site to site on the lattice.
This may now be used to calculate correlation functions. Consider, for example
$<H_0H_R>.$ By definition
\eqn\hihi{<H_0H_R>=2^{L^2}Z^{-1}\sum_{\{s\}}s_{0,1/2}s_{M,N+1/2}e^{-S'/kT}.}
{}From $e^{-S'/kT}$ we extract the factor $e^{K_hs_{0,1/2}}e^{K_hs_{M,N+1/2}}$
and write
\eqn\more{\eqalign{s_{0,1/2}s_{M,N+1/2}e^{K_hs_{0,1/2}}&e^{K_hs_{M.N+1/2}}\cr
=&\cosh^2 K_h s_{0,1/2}s_{M,N+1/2}(1+z_h s_{0,1/2})(1+z_hs_{M,N+1/2})\cr
=&\cosh^2 K_h z_h^2(1+z_h^{-1}s_{0,1/2})(1+z_h^{-1}s_{M,N+1/2})\cr}}
where $s^2=1$ has been used. Thus we find
\eqn\hihhiz{<H_0H_R>=z_g^2Z'/Z}
where the partition function $Z'$ is identical with $Z$ except that the
bonds on $(0,1/2)$ and $(M,N+1/2)$ are replaced by $z_h^{-1}.$ The argument
leading to ~\newising~ may now be made  in an identical fashion and
we find that $Z'$ is given by~\newising~ with the difference that the
coefficient of $(1-\sigma_{-1,0}\sigma_{0,0})$ and $(1-\sigma_{M-1,N}
\sigma_{M,N})$ is ${1\over 2}\ln z_g^{-1}$ instead of ${1\over 2}\ln z_g$. Thus
we can write
\eqn\morehihi{\eqalign{<H_0H_r>
=&z^2_h<\exp[-\ln z_g(1-\sigma_{-1,0}\sigma_{0,0})
\exp[-\ln z_g(1-\sigma_{m-1,n}\sigma_{m,n})]>\cr
=&<\exp[\ln z_h\sigma_{-1,0}\sigma_{0,0}]\exp[\ln z_h \sigma_{M-1}\sigma_{M,N}
]>\cr
=&<(\cosh \ln z_g + \sinh \ln z_h \sigma_{-1,0}\sigma_{0,0})
(\cosh \ln z_h + \sinh \ln z_h \sigma_{M-1,N}\sigma_{M,N})>.}}
Therefore we find the desired result
\eqn\hihires{<H_{0}H_R>-<H_0>^2=\sinh^2( 2E/kT) (<\sigma_{-1,0}\sigma_{0,0}
\sigma_{M-1,N}\sigma_{M,N}>-<\sigma_{-1,0}\sigma_{0,0}>^2).}
where the correlations on the right are for the Ising model specified by
{}~\ident.

In an identical fashion we find that
\eqn\pp{<P_0P_R>-<P_0>^2=\sinh ^2(2H/kT) (<\sigma_{0,0}\sigma_{M,N}>
-<\sigma_{0,0}>^2).}

The correlation of the plaquettes~\pp~ is thus the same as the spin-spin
correlations considered in general in the previous lecture.
But the Higgs-Higgs correlation~\hihires~ involves
an operator we did not explicitly consider in
part 1. This operator $\sigma_{j,k}\sigma_{j,k+1}$ is denoted
by $\epsilon ^h _{j,k}$ and
\eqn\eps{\sigma_{j,k}\sigma_{j,k+1}=\epsilon_{j,k}^h~~({\rm and}~~
\sigma_{j,k}\sigma_{j+1,k}=\epsilon_{j,k}^v)}
and is distinguished from $\sigma_{j,k}$ in that it is even under
spin reversal.
For the complete description of the gauge theory
$\epsilon_{j,k}^{h,v}$ must be considered in some sense
to be on an equal footing with the original operator $\sigma_{j,k}$

There are many further questions that need to be addressed about the physical
interpretation of these gauge invariant correlations, but we will defer
further remarks until sec.5  after we have summarized our current understanding
of the Ising correlation functions.
\bigskip
\newsec{Methods of computation for small magnetic field}

The most naive procedure to study the Ising model for $H\neq 0$ is to use the
results for the correlation functions at $H=0$ discussed in the
previous lecture and to expand the Greens functions as a series in $H$. Thus
if we denote the interaction energy ~\ising~ as
\eqn\hising{E=E_0-H\sum\sigma_{j,k}}
we can study the two point function
\eqn\twoh{<\sigma_{0,0}\sigma_{M,N}>_H=Z(H)^{-1}\sum_{\sigma}\sigma_{0,0}
\sigma_{M,N}e^{(-E_0/kT+H/kT)\sum_{j,k}\sigma_{j,k})}}
in a power series in $H$ as
\eqn\expand{\eqalign{&<\sigma_{0,0}\sigma_{M,N}>_H-M(H)^2\cr
&=\sum_{n=0}^{\infty}{1\over n!}(H/kT)^n\sum_{M_i,N_i}<\sigma_{0,0}\sigma_{M,N}
\sigma_{M_1,N_1}\cdots\sigma_{M_n,N_n}>_{H=0}^c}}
where the superscript $c$ indicates that only the connected
part of the correlation is used. To be specific
consider $T<T_c.$ Then the correlation functions at $H=0$  have the
representation
\eqn\escale{<\sigma_{0,0}\sigma_{M,N}\sigma_{m_1,N_1}\cdots\sigma_{M_n,N_n}>=M_-^{2+n}
e^{F_{n+2}}}
where $M_-$ is the spontaneous magnetization. We may now take the scaling
limit by using for each of the coordinates in ~\expand~
 the scaling variables of part 1
\eqn\variables{m=A({\pi\over 2})^{-1}M|T-T_c|~{\rm and}~n=A(0)^{-1}N|T-T_c|.}
Then if
\eqn\hscale{HM_-(T)A({\pi\over 2})A(0)/|T-T_c|^2\sim {\rm const}~ H
|T-T_c|^{-15/8}=h}
is held fixed the scaled Greens function is given as
\eqn\sgreen{G^c_2(0,{\vec r};h)=\sum_{n=0}^{\infty}{1 \over n!}h^n
\int d{\vec r}_1\cdots d{\vec r}_nG^c_{n+2}(0,{\vec r},{\vec r_1}
\cdots{\vec r_n}).}
The scaling parameter $h$ of~\hscale~ is the parameter referred to
in ~\hparam. It can vary from $0$ to $\infty$ and plays the
role of an adjustable coupling constant in the gauge theory.

The  stable particles of the theory are obtained as the poles in the
two point function which are obtained in coordinate space for large $r$ as
\eqn\poles{G_2^c(r)=\sum_l a_l(h)K_0[(2+k_l(h))r)]\sim\pi^{1/2}r^{-1/2} e^{-2r}
\sum_la_{l}(h)e^{-rk_l(h)}.}
 We thus can compute the particle spectrum for small   $h$ by using the large
$r$ expansion of the multi point Greens functions in the cluster expansion.
Orient the vector ${\vec  r}$ to point along the $y$ axis.
Then the multi point
Greens functions have a remarkable string property that the correlation is
concentrated about the $y$ axis in the following form
\eqn\greenstring{G^c_{n+2}(0,{\vec r}_1,\cdots,{\vec r}_n)
\sim r^{-2}e^{-2r}h_{n+2}({m_i\over r},{n_i\over r})}
where $m_i/r$ and $n_i/r^{1/2}$ are of order one. Inserting this form
into~\sgreen~ we see that each integral over ${\vec r}_i$ gives a factor of
$r^{3/2}$ and hence for large $r$
\eqn\moregreen{G_2^c(r;h)\sim r^{-2}e^{-2r}\sum_{n=0}^{\infty}{1\over n!}
(r^{3/2}h)^n c_n}
where
\eqn\cn{c_n=\int_0^{1}\{d {\bar m}_i\}\int_{-\infty}^{\infty}\{d {\bar n}_i\}
h_{n+2}({\bar m}_i,{\bar n}_i).}
Thus we see that $r$ always appears in the combination $hr^{3/2}$. Therefore
the
only way for the form~\poles~ to be possible is if
\eqn\form{k_l(h)=h^{2/3}\lambda_l~~{\rm and}~~a_l(h)=ha_l.}
To determine the numbers $\lambda_l$ and $a_l$
the detailed form for the scaling
functions must be used. When this is done it is found that $a_l$ is independent
of $l$ and that $\lambda_l$ is obtained as the solutions of
\eqn\result{J_{1/3}({\lambda^{3/2}\over 3})+
J_{-1/3}({\lambda^{3/2}\over 3}) =0}
where $J_n(z)$ is the Bessel function of order $n.$

This expansion method can also be applied to correlations for spin
for $T>T_c$ and again the scaling parameter $h$ of ~\hparam~ is a one parameter
adjustable coupling constant for the gauge theory. Thus to be complete
in addition to the value of $h$ we need to specify whether the
scaling theory is constructed from above or below $T_c.$ From this discussion
might appear that there could be a singularity at $h=\infty ~(T=T_c,~H>0)$
since it can
be approached from two different directions. However the theorem of
Lee and Yang~\rly~ says that the only singularities in the real $H-T$ plane
are at $H=0$ for $T<T_c$. Consequently the field theory limits must
join analytically at $h=\infty$

Finally the expansion method can  be used to study
the correlations for $\epsilon_{j,k}.$
For this case, of course, the correlations at $H=0$ of arbitrary numbers of
$\sigma$'s and $\epsilon$'s must be first evaluated by the methods of the
second lecture. We will defer the presentation of results to section 5 where
we summarize all results needed to discuss confinement.
\bigskip
\newsec{Method of computation for $T=T_c$ and $H\neq 0$}

When $T=T_c$ and $H=0$ the methods of the preceding section break down
completely because $h=\infty.$ It is thus extremely useful that for this
case it was found in 1988 by Zamolodchikov\rzama~\rzamb~ that
results may be obtained
by a completely different method which is quite unlike any method I have
discussed previously in these lectures.

In principle you might try to extend the perturbation techniques of the
previous section to $T=T_c$ by using the exact $T=T_c$ correlation
functions in the cluster expansion. Furthermore you might pass to the continuum
limit by using the large $R$ form of the correlations and symbolically
writing
\eqn\zamham{E_{Ising}=E_{T=T_c}-H\int\sigma(x)d^2x.}
Unfortunately the integrals in the cluster expansion
are badly divergent and without further definition the procedure is
meaningless.

However, the basic reason that the Ising model  is solvable at $H=0$ is
that it has an infinite number of conservation  laws which we exploited when
we related the Ising model to free fermions. What Zamolodchikov recognized~
\rzama~\rzamb~ is that if $T=T_c$ and if the continuum
limit is taken then there is a sense in which ~\zamham~ preserves some,
but not all,
of the conservation laws of $H=0$ for $H\neq 0.$

This observation is both remarkable and mysterious. The mystery is that no one
has found these conservation laws for the original lattice definition~\ising~
of the Ising model in the sense that no one has found a family of
commuting transfer matrices $T(u)$ which contains the Ising model with
$H\neq 0.$ Apparently Zamolodchikov has discovered that there is some more
general concept of partial integrability that does not
demand that all states on the
finite lattice be of the desired form. He has managed to throw away those
states which do not obey the conservation laws.

The second departure Zamolodchikov makes from our previous methods is
that he implements the conservation laws by directly assuming that the states
of the system have a particle interpretation in Minkowski space
 and that these particle states
scatter with an S matrix that depends on the rapidity variable $\theta _k$
defined by
\eqn\rapidity{p^0+p^1=m_ke^{\theta _k},~~~p^0-p^1=m_ke^{-\theta _k}}
where $p^0$ is the energy, $p^1$ is the momentum, and $m_k$ is the mass
of the $k$th particle. He then supplements
the basic equations of crossing
\eqn\cross{S_{ab}(\theta)=S_{ab}(i\pi-\theta)}
and unitarity
\eqn\unit{S_{ab}(\theta)S_{ab}(-\theta)=1}
with the equations for a factorizible S matrix and the
requirement that the poles in the S matrix elements be at the positions
determined
by the masses of the other particles in the problem.
Then when the S matrix elements are
constrained by the conservation laws the S matrix can be computed and it is
found that there are 8 particles with the masses
\eqn\mass{\eqalign{m_1&=m,~~m_2=2m\cos{\pi\over 5},
{}~~m_3=2m\cos{\pi\over 30},\cr
m_4=&2m_2\cos{7\pi\over 30},~~m_5=2m_2\cos{2 \pi \over 15},
{}~~m_6=2m_2\cos{\pi\over 30},\cr
m_7=&4m_2\cos{\pi\over 5}\cos{7\pi\over 30},~~m_8=4m_2\cos{\pi \over 5}
\cos{2\pi\over 15}.}}
These masses have the property that they are given in terms of
the components $S_i$ of the Perron-Frobenius vector of the Cartan matrix of the
Lie algebra $E_8$ as $m_i/m_j=S_i/S_j$

More recently a lattice model has been found~\rwns~ that also seems to be
in the same universality class as the critical Ising model in a magnetic
field. It is called the dilute $A_3$ model and the variable
has 3 states per site. The model depends on a parameter we may call $H$ and
the free energy behaves as $H\sim H^{1+1/15}$ as $H\rightarrow 0$ which is
precisely the behavior of the critical Ising model. Furthermore the
thermodynamics of the associated quantum spin chain have been computed very
recently~\rbnw~ and from this the mass spectrum~\mass.

{}From these computations it may be said that the mass spectrum of the Ising
field theory at $T=T_c$ has been computed. Clearly it is most desirable to
obtain
this result directly for the original Ising interaction instead of using
this indirect argument.

Finally it is most interesting to remark that only the masses $m_1$,
$m_2$ and $m_3$  are less than $2m_2$. Therefore it might have been expected
that the particles of mass $m_4-m_8$ would be unstable. This instability cannot
occur in an integrable model defined by commuting transfer matrices since
it is rigorously known that production and decay are forbidden by the
infinite conservation laws. However it seems highly likely that any
deviation of $h$ from $\infty$ will destroy the
infinite number of conservation laws and
hence the particles lying above the threshold of decay into 2 particles
of mass $m_1$ should become unstable. It is thus very tempting to see if
there is some way to perturb this model in this  small non integrable
fashion to obtain the first model computation of particle decay in a 2
dimensional model system.

\newsec{Summary of results}
\bigskip

We may now summarize the results known for the scaled Ising model in a
magnetic field. The most basic of these results in given in graphical
form in Fig. 1 where we schematically indicate the masses and
threshold singularities for the spectrum as we go from a) $H=0,T>T_c$ to
b) $H\sim 0,T>T_c$ to c) $H>0,T=T_c$ to d) $H\sim 0,T<T_c$ and finally to
e) $H=0,T<T_c.$ Clearly as we move along the curve from 1) to 5)
we pull more and
more masses out from the two particle threshold and eventually
at $H=0,T<T_c$ we get an infinite number of poles which become a cut.

If the spin spin correlation couples to all the particles then this gives
the singularity structure of the scaled two point function $G_{\pm}$. However
there is some indication~{\rzamc } that not all particles couple to the two
point function (at least at $T=T_c$). It is surely important to analytical
study this phenomenon.

To complete the summary we need the corresponding results for the density
density correlation $<\epsilon_{0,0}\epsilon_{M,N}>-<\epsilon_{0,0}>^2.$
At $H=0$ it is known ~\rhecht~  for both $T>T_c$ and $T<T_c$ that
\eqn\scaledden{D_{\pm}(r)={\rm const }~\lim_{\rm scaling}|T-T_c|^{-2}
(<\epsilon_{0,0}\epsilon_{M,N}>-<\epsilon_{0,0}>^2)=K_1^2(r)-K_0^2(r).}
Thus $D(r)$ couples only to 2 particle states and not to any higher
multi particle states. As limiting cases we have
\eqn\denl{D_{\pm}(r)\sim{\pi \over 2}r^{-2}e^{-2r}
{}~~{\rm as}~~r\rightarrow \infty}
and
\eqn\dens{D_{\pm}(r)\sim r^{-2}~~{\rm as}~~r\rightarrow 0.}

Finally for $h$ small it has been shown~\rmyb~ that for large $r$
\eqn\denhpos{D_{\pm}(r,h)\sim {\rm const}~ G_{\pm}(r,h).}
However this
equality is only valid for the leading singularities and in general there
is no reason to suppose that $\epsilon$ and $\sigma$ couple to the same
particles in the spectrum.
\newsec{Confinement in the Ising-gauge theory}

\bigskip

We may now combine the results of the previous sections to discuss
confinement in this most simple of all gauge theories coupled to matter.
In previous talks at this summer school the test
for confinement was whether    a
large Wilson loop behaves with an area or a perimeter law. Unfortunatly this
criterion is only useful for either pure gauge theories or for those for which
the matter fields are treated as infinitely heavy or static.
As soon as matter fields are dynamic the large Wilson loops all follow a
perimeter law and thus do not distinguish between confined and
unconfined phases. Consequently we are forced to discuss confinement in
terms of the scaled plaquette--plaquette correlation which we call
\eqn\scaledpp{P_{\pm}(r,h)=\lim_{\rm scaling}
\sinh^{-2}(2H/kT)M_{\pm}^{-2}(T)(<P_0P_R>-<P_0>^2)=G_{\pm}(r,h)}
and the scaled Higgs--Higgs correlation which we call
\eqn\scaledhh{\eqalign{H_{\pm}(r,h)&=\lim_{\rm scaling}{\rm const}
\sinh^{-2}(2E/kT)
|T-T_c|^{-2}(<H_0H_r>-<H_0>^{2})\cr
&=D_{\pm}(r,h).}}

Consider first the Higgs-Higgs correlation $H_{\pm}(r,h)$.
At some naive level this operator creates a pair of Higgs
particles connected by a gauge string at one point and destroys them
at some other point. The correlation function is thus a measure of
the propagation. Consider first $H=0,T>T_c$. Then from the
results of sec. 5 on $D_+(r,0)$
we see that $H_+(r,h)$ couples only to two particle states and not to
one particle states. This is interpreted as saying that two free particles are
created each of mass $m(=1).$

Next consider $h>0$ with $T>T_c$. Now from sec. 5 $H_{+}(r,h)$ couples
to a single particle state of mass $m$. This is clearly interpreted as
saying that the two Higgs particles are confined by the gauge string to
form a single meson of mass $m$. This change in total mass
from $2m$ to $m$ is a very
strong confinement. As we continue around the circle
towards $H=0,T<T_c$ the confining potential becomes weaker and weaker
and more excited states (mesons) are produced. The number of confined
mesons increases until $H=0,T<T_c$ where the weak confining potential
disappears altogether and a two particle state of unconfined Higgs
particles each of mass $m$ is regained.

In terms of the Ising model magnetic interpretation what is happening
is  for $T<T_c$ and weak magnetic field $H$ is that the field
 creates a weak linear confining
potential between two kink states. Assuming that the two kink states obey
Schroedinger's equation we must solve that equation with a linear potential.
The eigenvalues of this equation are exactly the Bessel function
spectrum we obtained by the cluster expansion. Of course in the strongly
confining region such a simple picture is not possible.

This picture of confinement in the weak coupling regime suggests another
phenomenon. For the low lying mesons in the weak coupling regime the
wave function is going to be nonzero over a large region (which spreads
out to $\infty$ as $h\rightarrow 0$). This certainly
suggests that there will be
a size associated with these weakly confined mesons which can become quite
large. Consider a     hypothetical experiment which detects the
scattering of the single meson   in the strongly confining regime. The
size grows constantly as $h\rightarrow 0$ for $T<T_c$. But for
a finite size detector the meson will eventually become larger than the
detector and hence it will be undetectable. This seems a good
illustration of the care that must be made in interpreting the results of
quantum field theory computations.

Now consider the other gauge invariant correlation $P_{\pm}(r,h)$. This
has no Higgs fields  in it and thus seems to couple not to Higgs but only to
the gauge   field. The question now arises if $P_{\pm}(r,h)$ is coupled to
anything other than mesons and if such excitations could possibly be
identified with excitations of the gauge field (sometimes called glueballs). If
indeed there are poles in $P_{\pm}(r,h)$ which are not poles of $H_{\pm}(r,h)$
it would seem that there are excitations of the gauge field separate from
the mesons of confined Higgs particles.

In conclusion I draw the readers attention to the fact that this gauge theory
interpretation  does not seem quite to be the same as the particle
interpretation used by Zamolodchikov in his bootstrap computation. In
order to fully understand scattering     in Zamolodchikov's  interpretation
one would like to find in terms of the Ising model operator $\sigma$
operators which
create each of the eight particles separately. These
are presumably the interpolating fields which are much discussed in
quantum field theory and their construction should
shed light on the vague concept of the size or extension of
particles.


\bigskip
\bigskip



{\bf Figure Captions}

Fig.1. The masses and threshold singularities of the Ising model with
a scaled magnetic field $h.$

a) $H=0,T>T_c$

b) $H\sim 0,T>T_c$

c) $H>0,T=T_c$

d) $H\sim 0,T<T_c$

e) $H=0,T<T_c.$
\vfill

\eject
\listrefs

\vfill\eject

\global\secno=0

\global\meqno=1

\global\subsecno=0

\global\refno=1


%
\def\rhob{{\rho\kern-0.465em \rho}}

\def\eps{\epsilon}

\def\ontopss#1#2#3#4{\raise#4ex \hbox{#1}\mkern-#3mu {#2}}

\setbox\strutbox=\hbox{\vrule height12pt depth5pt width0pt}
\def\tablerule{\noalign{\hrule}}
\def\tr{\tablerule}
\def\strut{\relax\ifmmode\copy\strutbox\else\unhcopy\strutbox\fi}

\nref\rbax{R.J. Baxter, One-dimensional anisotropic Heisenberg chain,
Phys. Rev. Letts. 26 (1971) 834.}
\nref\rperk{J.H.H. Perk, Equations of motion for the transverse
correlations of the one-dimensional XY-model at finite temperatures,
Phys. Lett. 79A (1980) 1.}
\nref\rpcqn{ J.H.H. Perk, H.W. Capel, G.R.W. Quispel and F.W. Nijhoff,
Finite--temperature correlations of the transverse Ising chain in a
transverse field, Physica 123A (1984) 1.}
\nref\rnm{T. Niemeijer, Some exact calculations on a chain of spins 1/2
 Physica 36 (1967) 377.}
\nref\rsjl{A. Sur, D. Jasnow and I.J. Lowe, Spin dynamics for the
one-dimensional XY model at infinite temperatures,
Phys. Rev. B 12 (1975) 3845.}
\nref\rbj{U. Brandt and K. Jacoby, The transverse correlation function
of anisotropic X--Y chains: Exact results at $T=\infty,$
Z. Physik B26 (1977) 245.}
\nref\rcp{H.W. Capel and J.H.H. Perk, Autocorrelation function of the
X-component of the magnetization in the one-dimensional XY model,
Physica 87A (1977) 211.}
\nref\rpc{J.H.H. Perk and H.W. Capel, Time-dependent XX-correlation
functions  in the one-dimensional XY-model,
Physica 89A (1977) 265.}
\nref\rmps{B.M. McCoy, J.H.H. Perk and R.E. Shrock, Time dependent
correlation functions of the transverse Ising chain at the critical
magnetic field, Nucl. Phys. B220[FS8] (1983) 35.}
\nref\rmpsb{B. M. McCoy, J.H.H. Perk and R.E. Shrock, Correlation
functions of the transverse Ising chain at the critical temperature
for large temporal and spatial separations, Nucl. Phys. B220[FS8]
(1983) 269.}
\nref\riiks{A.R. Its, A.G. Izergin, V.E. Korepin, and N.A. Slavnov,
Temperature correlations of quantum spins, Phys. Rev. Letts. 70 (1993) 1704.}
\nref\rsl{A. Sur and I.J. Lowe, Time dependent autocorrelation function for a
linear Heisenberg chain at infinite temperature,
Phys. Rev. B11 (1975) 1980.}
\nref\rrmp{J.M.R. Roldan, B.M. McCoy and J.H.H. Perk, Dynamic spin correlation
functions of the XYZ chain at infinite temperature: A study based on moments,
Physica 136A (1986) 255}
\nref\rbl{M. B{\" o}hm and H. Leschke, Dynamic spin-pair correlations
in a Heisenberg chain at infinite temperature based on an extended short-time
expansion, J. Phys. A 25 (1992) 1043.}
\nref\rpg{O. Platz and R.G. Gordon, Rigorous bounds for time dependent
spin correlation functions, Phys. Rev. B7 (1973) 4764; and Rigorous
bounds for time-dependent correlation functions, Phys. Rev.
Letts. 30 (1973) 264.}
\nref\rbltwo{M. B{\" o}hm and H. Leschke, Dynamical aspects of spin chains
at infinite temperature for different spin quantum numbers, Physica A 199
(1993) 116.}
\nref\rpoil{D. Poilblanc, T. Ziman, J. Bellissard, F. Mila and G. Montambaux,
Poisson {\it vs.} GOE statistics in integrable and non-integrable quantum
mechanics, Europhys. Lett. 23 (1993) 537.}
\nref\rwig{E.P. Wigner, Results and theory of resonance absorption, in
Gatlinberg Conference on neutron physics by time of flight, 1956, Oak
Ridge Natl. Lab. Report ORNL--2309 (1957) 59.}
\nref\rmehta{M.L. Mehta, {\it Random Matrices}, second ed.
Academic Press, 1991 }
\nref\rhkd{S. Howes, L.P. Kadanoff and M. den Nijs, Quantum model for
commensurate incommensurate transitions, Nucl. Phys.
B215[FS7] (1983) 169.}
\nref\rampty{H. Au-Yang, B.M. McCoy, J.H.H. Perk, S. Tang and M--L.
Yan, Commuting transfer matrices in the chiral Potts models: Solutions
of the star-triangle equations with genus $>1$, Phys. Lett. A123 (1987) 219.}
\nref\rbpa{R. J. Baxter, J.H.H. Perk and H. Au-Yang, New solutions of
the star triangle relations for the chiral Potts model, Phys. Lett.
A128 (1988) 138.}
\nref\rayp{H. Au--Yang and J.H.H. Perk, Onsager's star--triangle
equation: Master key to integrability, Adv. Stud. in Pure Math. 19
(1989) 57.}
\nref\rpf{O. Perron, Math. Ann. 64 (1907) 248; G. Frobenius, S. B.
Deutsch. Akad. Wiss. Berlin. Math--Nat. Kl (1912) 456; see also F.R.
Gantmacher, {\it Matrix Theory}, Chelsea Publishing Co (1959), vol.2.}
\nref\rbaxa{R.J. Baxter, The superintegrable chiral Potts model,
Phys. Letts. A133 (1988) 185.}
\nref\ramp{G. Albertini, B.M. McCoy, J.H.H. Perk, Eigenvalue spectrum
of the superintegrable chiral Potts model, Adv. in Pure Math. 19
(1989) 1.}
\nref\rdkm{S. Dasmahapatra, R. Kedem and B.M. McCoy, Spectrum and
completeness of the three-state superintegrable chiral Potts model,
Nucl. Phys. B396 (1993) 506.}
\centerline{\bf Part IV. Quantum Statistics and the Chiral Potts Model}



\vskip 13mm

\centerline{\bf Abstract}
\vskip 4mm
We extend the considerations of the previous lectures from classical
to quantum statistical mechanics.
We discuss diffusion in quantum spin chains and the relation this
has  to random matrices.
We conclude with a discussion of the chiral Potts model as
an example of a system where
the physics in Euclidean and Minkowski space can be very different.

\vskip 3mm

\newsec{Introduction}

Thus far in these lectures I have discussed problems in which
classical statistical
mechanics and quantum field theory are very closely related. This very
close relation has come about because in the systems considered the models
have had the feature that it is possible to make the analytic continuation from
the Euclidean space where classical statistical mechanics is naturally
interpreted to Minkowski space where quantum field theory is most naturally
interpreted. However there are many situations in physics where these analytic
continuations can not be made and where it is more appropriate to directly
consider quantum statistical mechanics in Minkowski space to begin with.
Consequently in this final lecture I want to briefly sketch some of these
problems in quantum many body statistics. In contrast to the previous lectures
I will make no attempt to pretend that these remarks will be self contained
and I will confine myself as far as possible to sketching problems and
results to the exclusion of explaining methods.

In sec.~2 I  will return to the question of diffusion raised in the first
lecture and explain in some detail what is known. In sec.~3 I will
briefly discuss what is known about the relation of diffusion and
the theory of random matrices.
I will
conclude in sec.~4 by discussing
the chiral Potts model which was introduced
and using it
to discuss  situations where Euclidean and Minkowski physics are different.

\newsec{Quantum diffusion}
\bigskip

In quantum statistics expectation values are computed from
\eqn\expac{<O_j>=Z^{-1}
\Tr O_je^{-H/kT},~~Z=\Tr e^{-H/kT}}
where $H$ is a quantum Hamiltonian. This formulation is directly in Minkowski
space and thus is extremely natural to consider time dependent correlations
such a
\eqn\tcorr{<A(t)B(0)>=Z^{-1}\Tr e^{-iHt}Ae^{iHt}Be^{-H/kT}.}
It is clear that the study of these correlation functions involves a knowledge
of both the spectrum of $H$ and the matrix elements of the operators $A$ and
$B$
in the basis of the eigenstates of $H$.

Let us first consider the spectrum of $H$. In our previous discussion
of quantum field theory
and statistical mechanics we considered the
excitations which are of order one above the ground
state in the thermodynamic limit. More precisely we considered
systems where in dimension $d$ the limit
\eqn\groung{\lim_{L\rightarrow \infty}E_{GS}/L^d~~{\rm exists}}
and also where
\eqn\order{\lim_{L\rightarrow \infty}(E_{ex}-E_{GS})~~{\rm exists}.}
If in addition we have the property
\eqn\qpspec{\lim_{L\rightarrow \infty}(E_{ex}-E_{GS})=
\sum_{\alpha=1}^N \sum_{j_{\alpha},{\rm rules}}^{m_\alpha}
e_{\alpha}(P^{\alpha}_{j_{\alpha}})}
where $N$ is the number of species of particles, $m_{\alpha}$ is the
number of particles in the state, and $e_{\alpha}(p)$ are called the
single particle energy levels we say that the spectrum is of the quasiparticle
form.

It is generally assumed that (but I know of no proof) that most quantum
Hamiltonians have this quasi-particle form for the order one excitations. This
is usually sufficient to study the system in the limit that $T\rightarrow 0.$
But if we are interested in quantum statistics at $T>0$ this quasi-particle
discussion of the order one low lying excitations is definitely not
 good enough.
In fact for any $T>0$ these order one excitations above the ground
state, even though they
form a Hilbert space built on these single particle excitations, are not
sufficient to describe the physics of the system. Indeed these
order one excitations form a set of measure zero in the total (non separable)
space of all states. In quantum statistics we are thus forced immediately to
deal with the fact that there may be phenomena in nature that are not
particle like. The aim of this section is to give some insight into the
study of these phenomena. This insight will be maximized if we get as far away
from the zero temperature quasi-particle
regime as possible. Therefore we will focus on  the opposite situation and set
$T=\infty.$

To be concrete we will concentrate on the cases where the most exact
information is known:

 1) The various special cases of the nearest neighbor
XYZ model
\eqn\hxyz{H_{xyz}=-\sum_{j=1}^{L}(J^xS_j^xS_{j+1}^x+J^yS_j^yS_{j+1}^y+
J^zS_j^zS_{j+1}^z+HS^z_j)}
where $S^l_j$ are the three spin $S$ rotation matrices
on the site $j$ and in particular
\eqn\shalf{S^l_j={1\over 2}\sigma_j^l~~~{\rm for}~S={1\over2}}
where $\sigma^l_j$ are the three Pauli spin matrices on the site j.
When $J^x=J^y$ this is called the XXZ model and we write
\eqn\hxxz{H_{xxz}=-2J\sum_{j=1}^L(S^x_jS^x_{j+1}+S^y_jS^y_{j+1}+\Delta
S^z_j S^z_{j+1})-H\sum_{j=1}^LS_j^z}
and when $J^x=
J^y=J^z$ ($\Delta=1$) this is the spin $S$ Heisenberg magnet.

2) The nearest neighbor plus next nearest neighbor Heisenberg chain
\eqn\npnn{H_{nnn}=-J_1\sum_{j=1}^{L}{\vec S}_j\cdot{\vec
S}_{j+1}-J_2\sum_{j=1}^{L}
{\vec S}_j\cdot{\vec S}_{j+2}.}

The first, and  perhaps surprising, fact about these quantum
spin chains at $T=\infty$ is that they do not reduce to a classical limit.
Moreover the setting of $T=\infty$ has not even simplified the problem. Indeed
for the XYZ chain at $T=0$ with  $S=1/2$ and $H=0$ a large number of exact
results have
been  computed
 since  Baxter~\rbax~ discovered in 1971 that the system
is integrable. In contrast at $T=\infty$ the only exact results for
spin correlations are
at $J^z=0.$ Even the integrability  of the $S={1\over 2}$ XYZ chain
has not been exploited to any great extent for $T>0.$

At $J^z=0$ with $S=1/2$ on the other hand
the Hamiltonian~\hxyz~ is called the XY model and
the time dependent correlation functions have been studied in great detail. At
all temperatures the correlations of $\sigma^x$ and $\sigma^y$
 are known to satisfy partial
difference-differential equations of the Toda type~\rperk~\rpcqn.
 Many asymptotic results are
known for large space and time separations~\rnm--\riiks.
These results all stem, ultimately,
from the fact that the quasi-particle form for the excitations~\qpspec~ is
not just valid for the low lying excitations in the $L\rightarrow \infty$ limit
but is in fact valid for all finite lattices. Thus the spectrum is of the free
Fermi form form all temperatures.

{}From the large number of results available on the XY model we will restrict
our
attention here to $T=\infty$ and to the special case of ~\hxxz~with
$\Delta=0$ and $H=0$.
For this case it has been found~\rnm~ that
\eqn\sz{<\sigma^z_m(t)\sigma_n^z(0)>=[J_{|m-n|}(2Jt)]^2}
where $J_n(t)$ is the Bessel function of order n, and~\rsjl--\rpc
\eqn\sx{<\sigma_m^x(t)\sigma_n^x(0)>=\delta_{n,m}e^{-J^2t^2}}

The correlation function $C^z(n,t)=<\sigma_0^z(0)\sigma_n^z((t)>$
satisfies the  conservation law
\eqn\cons{\sum_{n=-\infty}^{\infty}C^z(n,t)=1}
and therefore can be compared with the spin diffusion forms of part 1. From the
exact result~\sz~ we find as $t\rightarrow \infty$
\eqn\szlt{C^z(0,t)={1\over \pi J t}\cos^2(2Jt-{\pi\over 4})+\cdots}
and we find that the exact result for the spatial variance $\sigma (t)^2$ is
\eqn\var{\sigma^2(t)=\sum_{n-\infty}^{\infty}C^z(n,t)n^2=
\sum_{n-\infty}^{\infty}
[J_n(2Jt)]^2 n^2=2(Jt)^2.}
Both of these results are in contradiction with the spin diffusion forms for
$t\rightarrow \infty$  of
\eqn\diff{C^z(0,t)\sim~(4\pi Dt)^{-1/2},~~\sigma^2(t)\sim2Dt.}

It is of course not particularly surprising that a system whose spectrum
is exactly that of free fermions does not have a diffusive character at
infinite temperature. The more interesting question is what is the behavior of
$C^z(n,t)$ for the spin $1/2$ XXZ model~\hxxz~(with $H=0$).
Here for $\Delta \neq 0$ the energy spectrum is definitely
not that of a free Fermi system and the excitations do scatter.
On the other hand this system is
integrable and the $n$ body scattering is factorizible into a series of 2 body
scatterings. Therefore any initial momentum distribution of $n$
excitations will be
preserved in time. This would lead to the conjecture that even though the
system is not free diffusion still will not occur.

It is of interest to confirm or deny this
conjecture  of no spin diffusion in the spin 1/2 XXZ model
on the basis of an exact computation. Indeed for the isotropic Heisenberg
case $\Delta =1$ this problem has been investigated for the last 20
years~\rsl--\rbl.
However no one has found a way to turn the intuition of the argument
given above into a computation. Instead what has been done is to expand
the correlation function into a series in $t$ as
\eqn\expansion{C^z(n,t)=
\sum_{l=0}^{\infty}{(-1)^l\over (2l)!}M^z_{2l}(n)t^{2l}}
where by directly expanding the exponentials in~\tcorr~we have
\eqn\moments{M^z_{2l}(n)=\lim_{L\rightarrow \infty}2^{-L} \Tr
\{[\underbrace{[H,[H,\cdots,[H,\sigma_0^z]]\cdots]]}_{2l~ \rm times}
\sigma_n\}}
and we make use of the identity
\eqn\ident{\eqalign{\Tr
&\{\underbrace{[A,[A,\cdots,[A,B]\cdots]]}_{2l~ \rm times}C\}\cr
&=(-1)^{l}\Tr\{(\underbrace{[A,[A,\cdots,[A,B]\cdots]]}_{l~ \rm times})
(\underbrace{[A,[A,\cdots,[A,C]\cdots]]}_{l ~\rm times})\}.}}
The longest of these series for the Heisenberg chain~\rbl~ have been
obtained on the computer for terms up to $t^{30}.$

The technical question is now to determine the $t\rightarrow \infty$ behavior
of the infinite system from these truncated power series expansions. The
most naive thing to do is to just use the first 30 terms of~\expansion~
 as they stand. This, however, does not give a very good extrapolation and much
better results are obtained for the autocorrelation function $C^z(0,t)$
if we use the
exact piece of information that the Fourier transform of this function is
nonnegative. The finite truncation of the series~\expansion, however, does not
have a nonnegative Fourier transform. However there is a lovely theory
of Tchebycheff bounds~\rrmp~\rpg~ which uses this positivity of the Fourier
transform
and the finite number of coefficients $M^z_{2l}(0)$ to
produce exact upper and lower bounds for $C^z(0,t).$ We give the bounds from
{}~\rbl~ in Fig. 1.

It is obvious from these bounds that there are oscillations in the correlation
function which are reminiscent of the oscillations in the case $J^z=0.$
It is also obvious that the asymptotic regime has not yet been reached.
Similarly
if the spatial variance is studied~\rbl~one finds that the asymptotic regime
is also not obtained from the first 30 terms. Accordingly none of the
authors who have studied the problem have been able to make a definitive
statement about whether or not the spin diffusion form has been confirmed or
denied.

These short time series expansions have been extended to several of the
other one dimensional spin chains~\rbltwo~ which are not integrable in the
sense of commuting transfer matrices. These systems are the following
special cases of ~\hxxz~with $H=0$ and the given maximum
computed power of $t$:

1) $S=1$, $\Delta=0$ to order $t^{22},$

2) $S=1$, $\Delta=1$ to order $t^{18},$

3) $S=\infty$, $\Delta=0$ to order $t^{18},$

4) $S=\infty$, $\Delta=1$ to order $t^{16}.$

In all of these cases the maximum order is much smaller that the
$t^{30}$ which
was computed for the $S=1/2$ Heisenberg magnet and as is to be expected
these series are not  long enough to allow any conclusions
to be made about their long time behavior. Consequently the present status of
the short time series studies of the one dimensional spin chains is that
there is no system for which series are known of sufficient length to
determine whether or not the spin diffusion forms hold nor can a difference
between integrable versus non integrable be detected.

\newsec{Random Matrices}

We  have now discussed all of the results which are available for the
time dependent correlation functions at infinite temperature for quantum spin
models. Thus we have exhausted all the results for which a direct comparison
with the theory of spin diffusion can be made. However, as was stated at
the outset these time dependent spin correlation functions depend both on the
spectrum of $H$ and the matrix elements of the spin operators. It is therefore
natural to split the problem into two parts and to study the spectrum
of $H$ by itself and to attempt to find a way to characterize the energy
levels which are not low lying levels of order one. Indeed it seems most
plausible that it should be possible to determine whether or not a system
has diffusion solely on the basis of the eigenvalue spectrum of $H$.

Quite recently a very interesting study has been made~\rpoil~of the spin 1/2
Heisenberg chain with nearest neighbor and next
nearest neighbor interactions~\npnn. For a chain of length 20 the eigenvalues
have been numerically computed for states of fixed total spin and fixed
momentum. This leads to a large number of levels for which
$E_{max}-E_{min}=O(L)$ as $L\rightarrow \infty.$ As noted earlier the
levels that have $E_{ex}-E_{min}$ and $E_{max}-E_{ex}$ of order 1 as $L
\rightarrow \infty$ are expected to have the quasi-particle
form~\qpspec~ and these are excluded from the data. The spacing of the
remaining  levels is computed and in any region of order one the average
spacing is normalized to one (where we note that this average spacing varies
slowly as we go through the entire distribution of eigenvalues.)

In Fig. 2 we plot the results of such a study for the spin $1/2$
chains ~\hxxz~ and ~\npnn~ with 16
sites, momentum $P=2\pi/16$, $S^z=1$ and $H=0.$ Here we plot $P(s)$, the
probability of  the normalized
level spacing $s$ which is defined as the
true level spacing $S$ divided by the
local mean level spacing $D$. We consider the following 4
cases

a) $H_{nnn}$ for $J_1=1$ and $J_2=0.5.$ This model is not integrable.

b) $H_{xxz}$ for $\Delta=1.$ This model is integrable.

c) $H_{xxz}$ for $\Delta=0.05.$ This is also integrable.

d) $H_{xxz}$ for $\Delta=0.$ This model is not only integrable but the
spectrum is that of completely free fermions for which the quasi
particle property ~\qpspec~holds on the finite lattice without the need
to take the thermodynamic limit.

These results have a very remarkable property. For the integrable
cases shown in Fig.~2b and 2c
the  normalized level spacing statistics follows the Poisson distribution
\eqn\pois{P(s)=e^{-s}.}
 The
probability is maximum at zero spacing and decreases monotonically. But if
$J_2$ is sufficiently large $(J_2/J_1=.5)$ the normalized
level spacing statistics of case a)
have a completely different character. The probability now vanishes at zero
level spacing and has a single maximum. A more refined characterization
is to note that the spacing distribution is well fit by the distribution
function of the ensemble of random matrices described by the
Gaussian Orthogonal Ensemble which is numerically well approximated by
the surmise of Wigner~\rwig
\eqn\wig{P(s)={\pi s\over 2}\exp(-{\pi\over 4}s^2).}
(For a further discussion of this random
matrix problem the reader is referred to ~\rmehta). In ~\rpoil~ it is
suggested that this difference between Poisson and GOE distinguishes the
integrable from the non integrable case. However, we see in Fig. 2d for
case d) that if the spectrum is totally free that the spectrum is not
even Poisson but has an enormous degeneracy of levels at zero spacing.

\newsec{Chiral Potts model}

In the discussion of quantum gravity in the first lecture of this
series it was emphasized that even though there has been much work on
Euclidean quantum gravity that in fact there are many problems with
such a formulation. Furthermore in this summer school in the lectures
on string theory and quantum gravity the necessity of considering
gravity directly in Minkowski space has been strongly emphasized.
Consequently I wish to conclude these lectures on the relation between
quantum field theory and statistical mechanics with a discussion of a
model where the physics in Minkowski space  and Euclidean space are
very different and not connected by a Wick rotation. This is the
Chiral Potts model which was introduced as a classical
statistical mechanical model in Part 1 of this lecture series as the
classical interaction energy
\eqn\cp{E=-\sum_{j,k}^L\sum_{n=1}^{N-1}\{E^v_n(\sigma_{j,k}\sigma_{j,k+1}^*)^n
+E_n^v(\sigma_{j,k}\sigma_{j+1,k}^*)^n\}}
where $\sigma _{j,k}^N=1.$ The Boltzmann weights of this model are
given in terms of the $E_j^{v,h}$ as
\eqn\bold{W^{v,h}(n)=\exp{1\over kT}\sum_{j=1}^{N-1}E_j^{v,h}}
where $\omega =e^{2\pi i/N}$ and
the transfer matrix is given in terms of these Boltzmann weights as
\eqn\tran{T_{\{l\},\{l'\}}=\prod_{j=1}^lW^v(l_j-l_j')W^h(l_j-l_{j+1}')}
The model has been found to be integrable~\rampty--\rayp~
if the Boltzmann weights are restricted to
lie on the manifold~\rbpa~
\eqn\wh{{W^h_{p,q}(n)\over
W^h_{p,q}(0)}=\prod_{j=1}^n({d_pb_q-a_pc_q\omega^j\over
b_pd_q-c_pa_q\omega ^j})}
\eqn\wv{{W^v_{p,q}(n)\over W^v_{p,q}(0)}=\prod_{j=1}^n({\omega a_p
d_q-d_p a_q\omega^j\over c_p b_q-b_p c_q \omega^j})}
where $a_p,b_p,c_p,d_p$ and $a_q,b_q,c_q,d_q$ are restricted to lie on
the generalized elliptic curve
\eqn\curve{a^N+\lambda b^N=\lambda ' d^N,~~~\lambda a^N+b^N=\lambda '
c^N}
with
\eqn\lamp{\lambda '=(1-\lambda ^2)^{1/2}.}
By integrability we mean the commutation relation
\eqn\ttcom{[T_{p,q},T_{p,q'}]=0.}
The curve ~\curve~ has genus $g=N^3-2N^2+1.$ It is important to note
that this integrable model will only describe a physical two
dimensional classical model if the Boltzmann weights are real and
positive. This will be the case if the points $p$ and $q$ are restricted
to
\eqn\real{a^*c=\omega^{1/2}bd,~~|a|=|d|,~~{\rm and}~~|b|=|c|.}

There is also a one dimensional quantum spin chain which corresponds
to this two dimensional model in a fashion similar to the way the XY
model corresponds to the Ising model, namely the Hamiltonian
introduced by Howes, Kadanoff and den Nijs~\rhkd~ for the special case $N=3$
\eqn\hcp{H_{cp}=-\sum_{j=1}^L\sum_{n=1}^{N-1}\{{\bar \alpha}_n(X_j)^n
+\alpha_n(Z_jZ_{j+1}^{\dagger})^n\}}
where we use
\eqn\xd{X_j=I_N\otimes\cdots\underbrace{X}_{{\rm site}~j}\cdots\otimes I_N}
\eqn\zd{Z_j=I_N\otimes\cdots\underbrace{Z}_{{\rm site}~j}\cdots\otimes I_N,}
$I_N$ is the $N\times N$ identity matrix, the elements of the
$N\times N$ matrices $X$ and $Z$ are
\eqn\xnn{X_{l,m}=\delta_{l,m+1}~~({\rm mod}~N)}
and
\eqn\znn{Z_{l,m}=\delta_{l,m}\omega^{l-1},}
and
\eqn\al{\alpha_n=\exp[i(2n-N)\phi/N]/\sin(\pi n/N)}
\eqn\albar{{\bar \alpha}_n=\lambda \exp[i(2n-N){\bar
\phi}/N]/\sin(\pi n/N).}
When the parameters $\phi,{\bar \phi}$ and $\lambda$ are related by
\eqn\restriction{\cos\phi=\lambda\cos{\bar \phi}}
the Hamiltonian is obtained from the classical model
obtained by considering the limit $p\rightarrow q$
\eqn\htrel{T_{p,q}=1(1+{\rm const})+uH_{cp}+O(u^2)}
where $u$ measures the deviation of $p$ from $q$.

The quantum spin Hamiltonian~\hcp~ has several properties:

1) $H_{cp}$ is hermitian for $\phi,{\bar \phi}$ and$\lambda$ real;

2) $H_{cp}$ commutes with the spin rotation operator $\prod_{j=1}^L X_j;$

3) $H_{cp}$ is translationally invariant;

4) Even when $H_{cp}$ is hermitian the matrix elements are real only
for $\phi={\bar \phi}=0.$

{}From 1) it follows that the eigenvalues of $H_{cp}$ are real and thus
that the Hamiltonian is a physical energy operator, from 2) it follows
that the eigenvalues of the spin rotation operator $e^{2\pi i Q/N}$
for $Q=0,1,\cdots N-1$ are good quantum numbers and from 3) it follows
that the eigenvalues of the translation operator $e^{iP}$ with $P={2
\pi k/L}$ with $k=0,1,\cdots L-1$ give the total momentum of the
system. These are standard properties which are shared with many other
of the systems discussed in this set of lectures. However property 4
is new and leads to features we have not seen before:

1) When the matrix elements are complex the system is not time
reversal invariant

2) For ${\phi \neq 0}$ the system is not parity invariant. This is
the reason the model is called chiral;

3) The Perron-Frobenius theorem~\rpf~cannot be invoked and thus ground
state level crossing may occur.

The Perron-Frobenius theorem has in fact already played a major role
in the connection between statistical mechanics and quantum field
theory. It says that if the off diagonal matrix elements of a matrix
are positive then the state with the largest eigenvalue is non
degenerate. For any statistical mechanical system with real interaction
energies the Boltzmann weights are clearly real and positive. Therefore
the Perron-Frobenius theorem says that the eigenstate of the
transfer matrix with the maximum eigenvalue (from which we obtained
the free energy) is never degenerate no matter how many parameters are
varied.

The integrable two dimensional classical model had real positive
Boltzmann weights when~\real~holds
and thus on this manifold the theory of critical phenomena and the
relation with Euclidean field theory developed in the previous 3
lectures will apply. However this manifold is not the same as the
manifold where $\phi, {\bar \phi}$ and $\lambda$ are real. Therefore the
manifold where the Hamiltonian is Hermitian is not the same as the
manifold where the statistical system is physical. Thus there is no
reason to believe that the physics of the Minkowski space quantum
Hamiltonian will be related to the Euclidean statistical system.

In fact the physics of the statistical and quantum system are very
different for this model. This is most vividly seen in explicit
computations done for the special case of~\hcp~called superintegrable where
$\phi={\bar \phi}=\pi/2$. Here it is found that the ground state
energy is~\rbaxa~\ramp~
\eqn\gronde{E_{GS}/L=-(1+\lambda)\sum_{n=1}^{N-1}F(-{1\over 2},{n\over
N};1;{4\lambda\over (1+\lambda)^2})}
where $F(a,b;c;z)$ is the hypergeometric function. This ground state
energy is only singular at $\lambda =1$ and the previous
discussions of the connection between statistical mechanics and
quantum field theory might lead us to expect that the model had a mass
gap which vanished only at $\lambda=1.$

However this expectation turns out not to be the case. The excitation
spectrum of the superintegrable case has been explicitly computed. It
is found to be of the quasi particle form ~\qpspec~with the fermionic
exclusion rule
\eqn\rulef{P^{\alpha}_j\neq P^{\alpha}_k~~{\rm for}~~j\neq k}
 and for $N=3$ there are two
distinct types of quasi particle energies~\ramp~{\rdkm} which in the sector
$Q=0$
are
\eqn\er{e_r(P^r)=2|1-\lambda|+{3\over \pi}\int_1^{|{1-\lambda\over
1-\lambda}|^{2/3}}dt({\omega v_r\over \omega t v_r-1}+{\omega ^2
v_r\over \omega^2 t v_r -1})[{4\lambda\over t^3-1}-(1-\lambda)^2]^{1/2}}
where
\eqn\pr{e^{-iP^r}={1-\omega^2 v_r\over 1-\omega v_r}}
with $-\infty<v_r<\infty$ and
\eqn\prange{0<P^r<2 \pi}
and
\eqn\etwos{e_{2s}(P^{2s})=4|1-\lambda|+{3\over
\pi}\int_1^{|{1+\lambda\over 1-\lambda}|^{2/3}}dt
{v_{2s}[4(v_{2s}t)^2-v_{2s}t+1]\over (v_{2s}t)^3+1}[{4\lambda\over
t^3-1}-(1-\lambda)^2]^{1/2}}
where
\eqn\pts{e^{-iP^{2s}}={1-e^{-\pi i/3}v_{2s}\over 1-e^{i \pi /3}v_{2s}}}
with $v_{2s}>0$ and
\eqn\tsrange{{ 2\pi\over 3}<P^{2s}<2\pi.}
Furthermore there is a conservation law on the number of excitations
$m_{r}$ and $m_{2s}$
\eqn\mrestrictions{m_r+2m_{2s}= 0~~({\rm mod}3).}

There are two very important features to be noted in this spectrum.
First the momentum range~\tsrange~ is not the full Brillouin zone of
0--$2\pi$.
This is perhaps not completely in accord with the usual intuition
about particle excitations but in fact does not violate any principle.

However the most important feature of the excitation energies is the
fact the $e_r(P)$ is not always positive. In fact explicit numerical
evaluation of the integral in~\er~ shows that if
\eqn\negrange{.9013<\lambda<1/.9013\cdots}
then there is a region of momentum P where $e_r(P)$ is negative. Thus
when ~\negrange~ holds there has been level crossing and the state for
which the eigenvalue ~\gronde~ has been computed is no longer the
ground state of the system. Therefore a phase transition has taken
place because the ground state levels have crossed and this transition
does not
reflect itself in a singularity in the ground state energy~\gronde.
This sort of phase transition is outside the scope of the critical
phenomena discussed on the first lecture.

Level crossing transitions such as the one considered here  can potentially
occur whenever they are not excluded by the Perron-Frobenius theorem.
This is precisely the sort of physics which the chiral Potts model was
originally invented to study~\rhkd. Moreover in the chiral Potts model
 it can be
shown that the correlation functions do not have a Lorentz invariant
form and that space and time occur in a very asymmetric fashion.
The full study of these effects is beyond the scope of this lecture
and, indeed, many of the properties which are known for the $N=2$ case,
which is the Ising model, have not yet been computed. However it is the
opinion of the author that at least for the superintegrable case most
of the computations done for the Ising model can be extended to this
model and that when this is done we will have greatly increased our
understanding of the physics of level crossing transitions.
\bigskip
\newsec{Conclusion}

I now have come to the end of this lecture series and I hope that I
have fulfilled my mandate of explaining how statistical mechanics and
quantum field theory are connected. Indeed, I hope that I have
persuaded the reader that far from being the completely separated
fields of research which they were considered to be 30 years ago that
they are in fact different aspects of exactly the same subject.
 You might try to
make a distinction by saying that statistical mechanics is more
general because it allows the explicit consideration of vacua which
are not rotationally or Lorentz invariant and it allows for diffusion.
But on the other hand since cosmology deals with elevated temperatures
and compactified string theory certainly breaks Lorentz invariance of
the underlying 10 or 26 dimensional space even this distinction seems
to be only of marginal importance.

 It is very impressive
that modern day  mathematical physics embraces fields as
far separated as algebraic geometry and the theory of critical
phenomena; that gauge theory and the theory of magnets are related:
that random matrices and two dimensional quantum gravity lead to the
same type of nonlinear differential equations that characterized the
Ising model correlation functions.
In the beginning of this century some of the deepest insights into the
emerging science of quantum mechanics came from statistical mechanics.
I hope that in this series of lectures I have demonstrated
that at the end of the century this connection of the two subjects
continues to provide profound insights into the deepest problems of physics.

\bigskip

\noindent
{\bf Acknowledgements}

\smallskip
The author is very pleased to thank Prof.~V.V. Bazhanov for the
opportunity to participate in the Sixth Annual Theoretical
Physics Summer School
of the Australian National University and to thank Prof.~R.J. Baxter
for hospitality extended at the Australian National University.
The author is deeply indebted to Prof.~C. Itzykson,
Dr.~E. Melzer, Prof.~J.H.H. Perk, Prof.~C.A. Tracy,
and Prof.~A.B. Zamolodchikov
for suggestions on the preparation of this set of lectures.
The
author is also pleased to acknowledge the help of W. Orrick in
computing the distributions for Fig. 2 of part IV.
This work is partially supported by the National Science
Foundation under grant DMR--9106648.
\bigskip
\noindent
{\bf Figure Captions}

Fig. 1. The bounds on the autocorrelation function
$<\sigma^z_0(t)\sigma^z_0(0)>$ at $T=\infty$ of the Heisenberg spin
chain as obtained by B{\"o}hm and Leschke in~\rbl.

Fig. 2 Plot of the histogram of the probability of the normalized
level spacing $s=S/D$ for various spin chains with 16 sites, momentum
$P={2\pi\over 16}$ and $<\sigma^z>=1$. The following cases are
considered:

a) $H_{nnn}$~\npnn~ with $J_1=1$ and $J_2=0.5.$
The dashed line is the Poisson distribution~\pois~ and the dotted
line is the distribution function of the Gaussian Orthogonal
Ensemble~\wig.

b) $H_{xxz}$~\hxxz~ with $\Delta=1$ and $H=0.$ The dashed line is the
Poisson distribution~\pois.

c) $H_{xxz}$~\hxxz~with $\Delta=0.05$ and $H=0.$ The dashed line is the
Poisson distribution ~\pois.

d) $H_{xxz}$~\hxxz~with $\Delta=0$ and $H=0.$ The dashed line is the
Poisson distribution ~\pois.
\vfill

\eject
\listrefs

\vfill\eject

\bye
\end